\documentclass[11pt]{article}
\usepackage{epsfig}
\usepackage{graphicx}
\usepackage{amssymb} 
\usepackage{here}
\usepackage{eucal}
\evensidemargin=\oddsidemargin

\newcommand{\beq}{\begin{equation}}
\newcommand{\eeq}{\end{equation}}
\newcommand{\bq}{\begin{equation}}
\newcommand{\eq}{\end{equation}}
\newcommand{\ba}{\begin{array}}
\newcommand{\ea}{\end{array}}
\newcommand{\beqa}{\begin{eqnarray}}
\newcommand{\eeqa}{\end{eqnarray}}

\newcommand{\NAIVE}{na{\"\i}ve~}
\newcommand{\X}{\chi}
\newcommand{\nuL}{\nu^{~}_L}

\def\fbar{\bar{f}}
\def\ep{\epsilon}

\def\ah{\widehat{a}}
\def\A{{\cal A}}
\def\C{{\cal C}}
\def\O{{\cal O}}
\def\D{{\cal D}}
\def\R{{\cal R}}
\def\T{{\cal T}}
\def\TT{{\cal T}}

\def\V{{\cal V}}

\def\End{\end{document}}

\def\sq{\sqrt{2}}

\def\to{\rightarrow}

\def\dis{\displaystyle}
\def\f{\frac}
\def\ov{\overline}

\def\[{\left[}
\def\]{\right]}
\def\({\left(}
\def\){\right)}

\def\G{{\cal G}}

\def\l{{\ell}}

\def\C{\mathcal C}
\def\U1EM{U(1)_{\rm em}}
\def\O{\mathcal O}
\def\R{\mathcal R}

\def\leqq{\leqslant}
\def\geqq{\geqslant}
\def\lan{\langle}
\def\ran{\rangle}

\def\sb{\sin\beta}

\def\ssb{s_{\beta}^{~}}
\def\ssbb{s_{\beta}^{2}}
\def\ccb{c_{\beta}^{~}}
\def\ccbb{c_{\beta}^{2}}
\def\ssa{s_{\alpha}^{~}}
\def\ssaa{s_{\alpha}^{2}}
\def\cca{c_{\alpha}^{~}}
\def\ccaa{c_{\alpha}^{2}}

\def\d{\delta}

\def\[{\left[}
\def\]{\right]}
\def\dis{\displaystyle}

\def\cut{\Lambda}

\def\RE{\Re\mathfrak{e}}
\def\IM{\Im\mathfrak{m}}
\def\MP{M_{\rm Pl}}
\def\MI{\mathfrak{m}_{\rm I}^{~}}
\def\MII{\mathfrak{m}_{\rm I}^{2}}
\def\B{\mathfrak{B}}
\def\nb{\overline{n}}
\def\nL{\nu_L^{~}}

\def\nLp{\nu_{L+}^{~}}
\def\nLm{\nu_{L-}^{~}}
\def\nLpm{\nu_{L\pm}^{~}}
\def\thisday{September, 2004}
\begin{document}
\setcounter{footnote}{1}   

\begin{titlepage}
\title{
\vspace*{-16mm}
\begin{flushright}
{\large hep-ph/0409131} \hfill
{\small UT-HEP-04-12}   \\
\end{flushright} 
\vspace*{18mm}
        {\bf    Scales of Fermion Mass Generation } \\ 
        {\bf    and Electroweak Symmetry Breaking      } \\[1cm]
}

\author{
{\large {\sc Duane A. Dicus}}\,\footnote{
Electronic address: dicus@physics.utexas.edu}
  \,~~and~~\,
{\large {\sc 
Hong-Jian He}}\,\footnote{
Electronic address: hjhe@physics.utexas.edu}
\\[5mm]
Center for Particle Physics and Department of Physics, \\[2mm]
University of Texas at Austin, Texas 78712, USA }

\date{(\,\thisday\,)}

\maketitle

\vspace*{1.2cm}

\begin{abstract}
\vspace*{3mm}
\noindent 
The scale of mass generation for fermions (including neutrinos)
and the scale for electroweak symmetry breaking (EWSB) 
can be bounded from above by the unitarity of scattering 
involving longitudinal weak gauge bosons
or their corresponding would-be Goldstone bosons.
Including the exact $n$-body phase space
we analyze the \,$2\to n$\, ($n\geqq 2$) processes
for the fermion-(anti)fermion scattering into multiple gauge boson
final states.
Contrary to na{\"\i}ve energy power counting,
we demonstrate that as $n$ becomes large,
the competition between an increasing energy factor
and a phase-space suppression leads to a {\it strong new
upper bound} on the scale of fermion mass generation at 
a finite value $n=n_s$, which is {\it independent of the EWSB
scale,}  $v = (\sqrt{2}G_F)^{-\f{1}{2}}$.
For quarks, leptons and Majorana neutrinos,
the strongest \,$2\to n$\, limits 
range from about $3$\,TeV  to $130-170$\,TeV 
(with $2\lesssim n_s \lesssim 24$), 
depending on the measured fermion masses. 
Strikingly, given the tiny neutrino masses as constrained 
by the neutrino oscillations, neutrinoless double-beta decays and 
astrophysical observations,
the unitarity violation of \,$\nL\nL\to nW_L^a$\, scattering 
actually occurs at a scale no higher than  $\sim\! 170$\,TeV.
Implications for various mechanisms of neutrino mass generation
are analyzed.  On the other hand, 
for the $\,2\to n\,$ pure Goldstone-boson scattering,
we find that the decreasing phase space factor always dominates over 
the growing overall energy factor when $n$ becomes large, 
so that the best unitarity bound on the scale of EWSB
remains at \,$n=2$\,.
\\[3mm]
PACS number(s): 12.15.Ff, 14.60.Pq, 11.80.-m, 12.60.-i
\hfill   [\,Phys. Rev. D\,{\bf 71} (2005) 093009\,]
\end{abstract}

\thispagestyle{empty}  
\end{titlepage}

%%%%%%%%%%%%%%%%%%%%%%%%%%%%%%%%%%%%%%%%%%%%%%%%%%%%%%%%%%%%%%%
\newpage
\baselineskip 18.5pt 
\setcounter{footnote}{0}
\renewcommand{\thefootnote}{\arabic{footnote}}
\setcounter{page}{2}

%\vspace*{1.5mm}
\noindent
{\large {\bf CONTENTS}}  \hfill Page$^{\#}$
\\[4mm]
{{\bf 1. Introduction}}  \dotfill  3
\\[3mm]
{{\bf 2. Scales of Mass Generation vs.~Unitarity Bounds}}
                         \dotfill  7
\\
%\hspace*{1em}
{{\bf 2.1.} {\tt Defining the Scale for Mass Generation}}
                         \dotfill  7
\\
%\hspace*{1em}
{{\bf 2.2.} {\tt High Energy Scattering 
                              and  Power Counting of Energy}}
                         \dotfill  9
\\
%\hspace*{1em}
{{\bf 2.3.} {\tt General Unitarity Condition and the Customary 
                ${\mbox {\boldmath  $2\to 2$}}$ Limits}}
                         \dotfill  13
\\
%\hspace*{2em}
{\bf 2.3.1.} {\tt General  $2\to n$
            Unitarity Condition}
                         \dotfill  13
\\
%\hspace*{2em}
{\bf 2.3.2.} {\tt Customary Limits from $2\to 2$
            Scattering}
                         \dotfill  16
\\[3mm]
{{\bf 3. Challenge from ${\mbox {\boldmath  $2\to n$}}$
Inelastic Scattering}}
                         \dotfill  19
\\
%\hspace*{1em}
{\bf 3.1.} {\tt Puzzle:~Energy Power Counting 
                             vs.~Kinematic Condition}
                         \dotfill  19
\\
%\hspace*{1em}
{\bf 3.2.} {\tt Resolution:~Increasing Power of Energy  
                vs.~Phase Space Suppression}
                         \dotfill  20
\\[3mm] 
{\bf 4. 
Scales of Mass Generation for Quarks and Leptons}
                         \dotfill  26
\\
%\hspace*{1em}
{\bf 4.1.} {\tt Unitarity Bounds on 
the Mass Generation Scales from
${\mbox {\boldmath $f\bar{f}\to n\pi^a$}}$}
                         \dotfill  26
\\
%\hspace*{1em}
{\bf 4.2.} 
{\tt Model-Dependent Effects from the EWSB Sector}
                         \dotfill  30
\\[3mm]
{\bf 5. Scale of Mass Generation for Majorana Neutrinos}
                         \dotfill  38
\\
%\hspace*{1em}
{\bf 5.1.} 
{\tt Majorana Masses and Neutrino-Neutrino Scattering}
                         \dotfill  38
\\
%\hspace*{1em}
{\bf 5.2.} 
{\tt  Unitarity Violation vs.~Mechanisms 
      for Majorana Mass Generation}
                         \dotfill  42
\\
%\hspace*{2em}
{\bf 5.2.1.} {\tt Unitarity Violation vs.~Seesaw 
                  and Radiative Mechanisms}
                         \dotfill  43
\\
%\hspace*{2em}
{\bf 5.2.2.} 
{\tt Scale of Mass Generation for Majorana Neutrinos\\ 
\hspace*{10mm}  in Supersymmetric Theories}
                        \dotfill  46
\\
%\hspace*{3em}
{\tt 5.2.2A.\,Scale of Mass Generation for 
                  Majorana Neutrinos via MSSM Seesaw}
                        \dotfill  46
\\
%\hspace*{3em}
{\tt 5.2.2B.\,Unitarity Violation Scale 
                  vs.~SUSY Breaking Induced Neutrino Masses}
                        \dotfill  49
\\[3mm]
{\bf 6. On the Scale of Electroweak Symmetry Breaking}
                    \dotfill  51
\\[3mm]
{\bf 7. Conclusions}
                    \dotfill  55
\\[3mm]
{\bf Acknowledgments}   
                    \dotfill  58
\\[3mm]
{\bf Appendix A:  Derivation of $\mathbf{n}$-Body Phase Space}
                     \dotfill  59
\\[1.2mm]
{\bf Appendix B:  Refined Unitarity Conditions}
                     \dotfill  60
\\[3mm]
{\bf References}     \dotfill  62

\newpage
\section{\hspace*{-6mm}.$\!$
Introduction}

Despite the astonishing success of the standard model \cite{SM1,SM2} 
for electroweak and strong interactions over the past thirty years, 
the origin and scales of mass generation remain, perhaps, the 
greatest mystery. 
The standard model (SM) contains %19 
nineteen free input parameters in total, 
among which %12 
twelve can be expressed in terms of {\it unpredicted} 
mass-eigenvalues (two weak gauge boson masses, one Higgs mass, 
six quark masses and three lepton masses) 
with one good equality\footnote{
Here $v=\(\sqrt{2}G_F\)^{-\f{1}{2}}\simeq 246\,$GeV is 
determined by the Fermi constant $G_F$.}
\,$\dis\f{m_t}{\,v/{\sqrt{2}}\,} \simeq 1$\, and a huge hierarchy
\,$\dis\f{m_e}{m_t} \simeq 3\times 10^{-6}$\, 
as fixed by experimental data.   
There are four additional 
inputs (three mixing angles and one {\tt CP}-phase) in
the Cabibbo-Kobayashi-Maskawa (CKM) matrix \cite{CKM}
coming from diagonalizing the quark mass matrices, 
so that sixteen out of the nineteen free parameters\footnote{
The remaining three parameters are 
two gauge couplings $(\alpha_s,\,\alpha_{\rm em})$ 
and one strong {\tt CP} phase.}
are due to our lack of knowledge about the origin of mass generation.
The recent exciting neutrino oscillation experiments 
\cite{atm,sol,K2K,CHOOZ,Kam} 
further point to three extra neutrino masses\footnote{For  Majorana 
neutrinos there are three mixing angles\,\cite{MNS} 
and three {\tt CP} phases
in the leptonic sector \cite{nu-rev}.
The global analysis\,\cite{G4nu} of all the current oscillation data 
strongly disfavors an extra singlet sterile neutrino.
The upcoming Fermilab MiniBooNE experiment\,\cite{BooNE} will resolve 
the controversy with LSND\,\cite{LSND}. 
} 
at the scale of \,$\sim 0.05$\,eV,\footnote{
The recent astrophysics analyses of the 
WMAP (Wilkinson Microwave Anisotropy Probe) data\,\cite{WMAP},
in conjunction with the 
2dF Galaxy Redshift Survey\,\cite{2dF}, give an impressive 95\%\,C.L.
upper bound,  
\,$\,\sum_j m_{\nu j} \leqq 1.01$\,eV \cite{WMAP-2dF}.
Combining the WMAP, SDSS (Sloan Digital Sky Survey)
and Lyman-$\alpha$ Forest data results in
a better limit of \,$\,\sum_j m_{\nu j} \leqq 0.65$\,eV
(95\%\,C.L.) \cite{Hannest}.}\,
giving another huge hierarchy: 
\,$\dis\f{m_\nu}{m_t} \sim 3\times 10^{-13}$.\,
In the SM, a single hypothetical Higgs boson \cite{higgs} is assumed
to generate masses for all gauge bosons, quarks and leptons, which,
however,  leaves all mass scales and their hierarchies fixed by empirical
inputs, without real explanation. Furthermore, such a Higgs boson cannot
generate neutrino masses without losing renormalizability 
(given the observed SM particle spectrum) \cite{weinberg5} 
or adding extra new particles 
(to retain the renormalizability) \cite{nu-seesaw,nu-rad}.
Various interesting extensions of the SM have been sought, ranging from
compositeness \cite{DSB,seesaw,LH}, 
low or high scale supersymmetry \cite{SUSY,SUSYhigh}, 
grand unification \cite{GUT}, 
to extra dimensions and deconstructions \cite{exd,Decons}, 
and string/M theories \cite{string}. 
Each of them overcomes some unsatisfactory aspects of 
the SM while adding many new unknowns. 
%either at higher scales or in its own.
Among so many kinds of possible new physics, there is 
no doubt that understanding the mass generation holds the greatest
promise to unravel immediate new clues beyond the SM.
Fortunately, on-going and upcoming high energy experiments 
will actively explore the scales of mass generation (and the associated
mixings) for weak gauge bosons, quarks, leptons and neutrinos, either
directly or indirectly, in this decade.

While we still seem a long way from finding an imagined 
theory of everything with only a single input parameter 
to explain the world, 
we  wish to ask a much more modest ``bottom-up'' question:
How do we use  high energy scattering processes as a fundemental
means to probe the scales of mass generation
in a {\it model-independent way}\,?  What is the energy scale at 
which the new physics underlying the relevant
mass generation has to show up?  In particular, 
fermion mass generation can be independent of the electroweak
symmetry breaking (EWSB) that generates the gauge boson masses of
the $W^\pm$ and $Z^0$; since the fermion and gauge boson masses 
need not originate from the vacuum
expectation value (VEV) of the same Higgs doublet (as assumed
in the SM)\footnote{For instance, 
in the Top-color models\,\cite{Hill-topc,DSB},
the top-condensate Higgs (which gives the observed large top-quark mass)
only changes the EWSB vacuum expectation value ($v\simeq 246$\,GeV) 
by a few percent and thus has little to do with the EWSB\@. 
In the traditional dynamical models, the EWSB is caused by technicolor
(TC) gauge interactions\,\cite{xTC,DSB}, while the fermion masses
are generated by technifermion condensates via additional 
extended technicolor (ETC)\,\cite{xETC,DSB}, 
independent of EWSB.}, 
we would hope to probe the scales for
fermion mass generation without any {\it ad hoc} theory
assumption.
The SM gauge symmetry forbids all bare mass terms for gauge
bosons and fermions. 
If any observed particle mass {\it is put in by hand} 
in the SM Lagrangian, the theory is no longer renormalizable and we 
expect the high energy scattering amplitudes involving this particle to
grow with energy and violate unitarity at a certain scale
above which the new physics has to show up. 
In this way the unitarity of the $S$-matrix puts a 
{\it universal upper bound} on the scale of 
the corresponding mass generation.

In the standard model EWSB sector for generating the 
$W/Z$ masses, once the neutral physical Higgs boson is removed,
the $2\to 2$ quasi-elastic scattering amplitudes of longitudinally
polarized $W/Z$'s exhibit the worst 
high energy behavior \cite{Ebad1,DM,Ebad2}, 
proportional to the square of the c.m. energy
$\sqrt{s}$\,,\, leading to partial wave 
unitarity violation at a scale  
\beq
\label{eq:UBw22}
E_W^\star ~\simeq~ \sqrt{8\pi}v ~\simeq~ 1.2\,{\rm TeV} \,,
\eq
similar to the classic analyses of the upper bound on the mass 
of the Higgs boson\,\cite{DM,LQT,Vel} where tree-unitarity is violated
at a critical Higgs mass 
$M_H^{\rm crit}\simeq 
 \sqrt{\f{16\pi}{3}}v\simeq 1$\,TeV~\footnote{Subsequent 
studies and applications may be found, for instance, 
in Refs.\,\cite{CG,LW,MVW,DJL} and references therein.}.\,
Eq.\,(\ref{eq:UBw22}) puts a model-independent upper limit on the 
scale of EWSB and justifies TeV energy scale for 
the Large Hadron Collider (LHC) at CERN. 
Appelquist and Chanowitz derived an analogous upper 
bound on the scale of  fermion mass generation by considering
the $2\to 2$ inelastic channels of fermion-antifermion scattering
$f_\pm\bar{f}_\pm \to V_L^{a_1}V_L^{a_2}$ ($V^a=W^\pm,Z^0$)\,\cite{AC},
\beq
\label{eq:UBf22}
E_f^\star ~\simeq~ \dis  \f{8\pi v^2}{~\sqrt{N_c}\,m_f~} \,,
\eeq
where $N_c = 3\,(1)$ for quarks (leptons).
This bound indicates that the upper limit on
the scale of fermion mass generation is proportional to the
inverse power of the fermion mass itself and is thus an independent
bound from the customary unitarity bound on the EWSB scale in
Eq.\,(\ref{eq:UBw22}). Also, for all the SM fermions (except the
top quark), the bound (\ref{eq:UBf22})
is much weaker than Eq.\,(\ref{eq:UBw22}).
However, this traditional wisdom was recently challenged 
by an interesting study\,\cite{scott}, where it was noted 
that for inelastic scattering with a multi-gauge-boson final state,
$f\fbar \to nV_L^a$ ($n>2$), the $n$-body phase space
integration contributes a nontrivial energy factor $s^{n-2}$ 
to further enhance the cross section 
(in addition to the energy dependence of $s^1$ 
from the squared amplitude),
so that the unitarity bound goes like
\beq
\label{eq:UBfscot}
E_f^\star ~\sim~ \dis v\(\f{v}{m_f}\)^{\f{1}{n-1}}
\longrightarrow~ v\,,~~~~~({\rm for}\,~n\to\,{\rm large})\,,
\eeq
which could be pushed arbitrarily close to the weak scale $v$ and
thus become independent of the fermion mass 
$m_f$ for {\it large enough $n$.} 
From this, Ref.\,\cite{scott} deduced  
that {\it the fermion-antifermion
scattering into weak bosons does not reveal an independent
new scale for  fermion mass generation.}
This is very counter-intuititive  since
the following {\it kinematic condition} must hold for 
any $2\to n$ scattering with $n$
weak gauge bosons in the final state, 
\beq
\label{eq:KC}
\dis
\sqrt{s} ~\,>~\, n M_{W(Z)} \,\simeq\, \dis\f{n}{3}v ~,
~~~\Longrightarrow~~~
\f{E^\star}{v} ~\,>~\, \f{n}{3} ~.
\eeq
This shows that as $n$ becomes large, the required c.m. energy
$\sqrt{s}$\, has to be arbitrarily above the weak scale $v$,
so that the unitarity violation scale, \,$\sqrt{s}=E^\star$,\,
has to {\it grow at least linearly with $n$}, instead of 
decreasing to \,$v$\,,\, contrary to the 
power counting argument in Eq.\,(\ref{eq:UBfscot}).

The observation in (\ref{eq:KC}) is completely
general, valid for any $2\to n$ process including both
\,$f\bar{f}\to nV_L^a$\, and \,$V_L^{a_1}V_L^{a_2}\to nV_L^a$.\,
Thus, a similar contradiction may possibly occur for the 
$V_LV_L\to nV_L$ scattering. We note that
the $n$-body phase space is the same as before but 
the squared amplitude now behaves as \,$s^2/v^{2n}$\,,\,
so by \NAIVE power-counting the 
inelastic unitarity bound goes as,
\beq
\label{eq:UBwus}
E_W^\star \,\sim\, \dis v\({C}_0\)^{1\over{2n}}
~\longrightarrow~ v\,,~~~~~({\rm for}\,~n\to\,{\rm large})\,,
\eeq
where \,${C}_0 > 1$\, is a dimensionless constant determined by
the phase-space and the square of the $S$-matrix element. 
For \,$n=2$,\, we know from Eq.\,(\ref{eq:UBw22}) that 
$\,C_0\simeq \(\sqrt{8\pi}\)^4\simeq 632$\,.\,
Hence, as $n$ becomes large, we would naively
expect $E_W^\star$ to approach \,$v\simeq 246$\,GeV,\, 
significantly below the 
traditional $2\to 2$ bound \,$E^\star_W \simeq 1.2$\,TeV\,
in Eq.\,(\ref{eq:UBw22}). If this argument 
holds, we would deduce the new physics scale associated with 
EWSB to be less than $\sim\! 250$\,GeV, 
which could manifest itself  
as a light Higgs boson or something similar 
and might already be partly accessible
at the on-going Tevatron Run-2.

It is the above general observation based upon the kinematic
condition (\ref{eq:KC}) that leads us to believe that the
energy power counting arguments in 
Eqs.\,(\ref{eq:UBfscot}) and (\ref{eq:UBwus}) cannot be true
and a new resolution must be sought.
Therefore, in this study, we systematically and quantitatively analyze 
the $2\to n$ inelastic channels, with $n>2$, for both 
\,$f\bar{f}\to nV_L$\, and \,$V_LV_L\to nV_L$\, scattering.
Our findings show that the exact $n$-body phase space 
contributes a non-trivial {\it dimensionless} factor which 
{\it sufficiently suppresses} 
the effect of the energy growth given by 
Eqs.\,(\ref{eq:UBfscot}) and (\ref{eq:UBwus}).
As a consequence, the kinematic condition  (\ref{eq:KC}) is 
satisfied for any large $n$ value, and the scale for 
fermion mass-generation is proven to be independent of the
EWSB scale.

Section\,2 establishes the necessary concepts 
and formalism, and reviews the customary
unitarity bounds from $2 \to 2$ scattering.  We first
define the scale of mass generation in connection with the
unitarity violation scale.  We then discuss the equivalence
theorem, give a general power counting formula and apply it to
count the energy dependence of the relevant scattering amplitudes
for our later analyses.
We present a systematical derivation of 
the general unitarity condition for $2 \to n$ 
scattering, extending that of Ref.\,\cite{scott}.
We then summarize the customary unitarity limits 
from \,$2 \to 2$\,
scattering for the scale of EWSB and the scales of 
mass generation for Dirac fermions and Majorana neutrinos.

In Sec.\,3 we return to the new puzzle presented around 
Eqs.\,(\ref{eq:UBfscot})-(\ref{eq:KC}) 
and resolve it by using an exact expression
for the n-body phase space.  
With estimated amplitudes for the
\,$2 \to n$\, scattering, we analyze the 
unitarity violation scales for the scatterings
\,$\xi_1 \xi_2 \to n~V^{a}_{L}$\, where the \,$\xi$\, denotes quarks,
leptons, Majorana neutrinos, or longitudinal weak bosons.

In Sec.\,4,~5, and 6 we further quantify the  
estimates in Sec.\,3 for quarks and leptons,
for Majorana neutrinos, and for EWSB, respectively.  
In Sec.\,4 we first compute
the scattering amplitude for \,$f\bar{f}\to nV_L^a \,(n\pi^a)$\,
from the universal contact interaction.  
This gives slightly weakened limits
for the unitarity violation, i.e., the estimates of Sec.\,3 are
relaxed by a factor of 2 or so.  We then 
examine the effect of including model-dependent EWSB contributions 
and note that the effects decrease 
with the increase of $n$, and, in particular,
we explicitly show that for a SM (or SM-like) EWSB sector the effect 
on \,$2\to 3$\, scattering is to enhance the bounds slightly.
Thus, we find that for all light SM fermions 
(except possibly the top quark) the \,$2\to n$\, unitarity limits 
on the scale of fermion mass generation become the
strongest at a value \,$n=n_s > 2$\, and are substantially tighter
than the classic Appelquist-Chanowitz bound \cite{AC}.

In Sec.\,5, we present a quantitative calculation of the scattering
amplitude for Majorana neutrinos \,$\nL\nL\to nV_L^a$\,.
This relaxes the estimate in Sec.\,3 by a factor of about 
\,$1.3-1.6$\,.\,
Thus we demonstrate that, without assuming\footnote{For
models where neutrino mass generation is
independent of the EWSB sector, see, {\it e.g.,}  
Refs.\,\cite{nudy,nudy1}.} {\it a priori}
that the same SM Higgs doublet used for
EWSB participates in  neutrino mass generation,
the unitarity of the scattering 
\,$\nL\nL\to nV_L^a$\, puts an upper bound 
on the Majorana neutrino mass generation at a scale substantially
below \,$M_{\rm GUT}^{~}$\,,\footnote{$M_{\rm GUT}^{~}$
stands for the energy scale of the 
grand unification (GUT)\,\cite{GUT}.}\,
around $130-170$\,TeV, with \,$n_s\approx 20-24$\,.\, 
We also discuss the implications of 
such low-scale unitarity violation 
for various mechanisms of Majorana neutrino mass generation.

Finally, in Sec.\,6, we explicitly analyze the scattering
\,$V_L^{a_1}V_L^{a_2}\to nV_L^a$\, 
($\pi^{a_1}\pi^{a_2}\to n\pi^a$),
with \,$n\leqq 4$\,,\, and show that as $n$ increases
the upper bound on the EWSB scale is not driven towards the 
scale \,$v\simeq 246$\,GeV\,
[cf. Eq.\,(\ref{eq:UBwus})], rather, it becomes weaker so that 
the best constraint still occurs at \,$n_s =2$, in agreement
with Sec.\,3.\,
Including our earlier general estimates 
from Sec.\,3 for large $n$ values, 
we conjecture that this feature  holds
for any value of \,$n\geqq 2$\,.\,

We conclude in Sec.\,7 and derive the 
expression for n-body phase space in Appendix\,A.
For completeness, in Appendix\,B we also refine  
the unitarity conditions of Sec.\,2.3.1.

\vspace*{5mm}
\section{\hspace*{-6mm}.$\!$
Scales of Mass Generation vs.~Unitarity Bounds}
\vspace*{3mm}

In this section, we start by introducing necessary concepts
and formalism for the gauge and fermion mass terms and for 
the $2\to n$ ($n\geq 2$) unitarity limit. Then we briefly review
the traditional \,$2\to 2$\, unitarity bounds derived for  EWSB and
fermion (neutrino) mass generation.

\vspace*{4mm}
\subsection{\hspace*{-5mm}.$\!$
Defining the Scale for Mass Generation}
\vspace*{2mm}

It turns out that the SM Lagrangian, with the physical Higgs boson
removed (often called the Higgsless SM) and with all observed particle
masses put in by hand, 
exhibits a nonlinearly realized gauge
symmetry \cite{CCWZ,weinberg79,ABL}, under which
the three would-be Goldstone bosons $\{\pi^a\}$ \cite{GB}
can be formulated as
\beq
\label{eq:U}
U ~=~ \exp \[ i\pi^a\tau^a/v \]
\eeq
which transforms, 
under the SM gauge group
$\G=SU(2)_L\otimes U(1)_Y$ as,
\beq
U \,~\rightarrow~\, U' \,=\, g_L^{~} U g_Y^{\dagger}\,, 
\eeq
where
$~
g_L^{~} = \exp [-i\theta_L^a\tau^a/2]$\, and
$~
g_Y^{~} = \exp [-i\theta_Y\tau^3/2]$\,.
Thus the weak gauge boson mass terms,
$M_W^2W^{+\mu}W^-_\mu+\f{1}{2}M_Z^2Z^\mu Z_\mu$,
are given by the dimension-2 nonlinear operator,
\beq
\label{eq:V-mass}
{\cal L}_V ~=~ \dis\f{v^2}{4}{\rm Tr}
               \[\(D_\mu U\)\(D^\mu U\)^\dag\]\,,
\eeq
where 
~$\dis D_\mu U=\partial_\mu U+i\f{g}{2}W^a_\mu{\tau^a}U
                       -i\f{g'}{2}UB_\mu{\tau^3}$\,.\,

Now consider a pair of generic SM fermions $(f,\,f')$ 
which form a left-handed $SU(2)_L$ weak doublet
$F_L=(f_L,\,f'_L)^T$ (with hypercharge $Y_L$) 
and two right-handed weak singlets
$f_R$ and $f'_R$ (with hypercharges $Y_R$ and $Y_R'$).
The electric charges of $(f,\,f')$ are  given by 
$\,Q_f = \f{1}{2}+Y_L=Y_R$\, and 
\,$Q_{f'}=-\f{1}{2}+Y_L=Y'_{R}$\,,\,
respectively.
We can write down the following gauge-invariant nonlinear
operator of dimension-3,
\beq
\label{eq:f-mass}
{\cal L}_f ~=~ \dis
-m_f\ov{F_L}U\(\ba{c} 1 \\ 0 \ea\)f_R
-m_{f'}\ov{F_L}U\(\ba{c} 0 \\ 1 \ea\)f'_R \,+\, {\rm H.c.} \,,
\eeq 
which contains the fermion bare mass-terms
\,$- m_f\ov{f}f - m_{f'}\ov{f'}f'$\,.
%[Here we have used an identity 
%$\,i\tau^2 U^\ast\(1,\,0\)^T  =U\(0,\,1\)^T\,$\, 
%to simplify the second term on the right-hand side
%of Eq.\,(\ref{eq:f-mass}).]

The SM has no (light) right-handed singlet 
neutrino $\nu_R^{~}$, so that
the tiny masses for the active neutrinos ($\nu_L$) indicated by the
neutrino oscillations can be naturally of Majorana-type, i.e., 
\,$\dis -\f{1}{2}{m_\nu^{ij}}\,\nu^T_{Li} \widehat{C} \nu_{Lj}^{~} 
       + {\rm H.c.}$\,,\, 
which can be formulated in a gauge-invariant manner, 
\beqa
\label{eq:nu-mass}
{\cal L}_\nu &=& \dis
-\f{{\cal C}_{ij}}{\cut} {L^{\alpha}_i}^T \widehat{C}L^{\beta}_j
 \Phi^{\alpha'}\Phi^{\beta'}
\epsilon^{\alpha\alpha'} \epsilon^{\beta\beta'} \,+\, {\rm H.c.}
\nonumber
\\[2mm]
&=& \dis
-\f{1}{2}\,{\cal C}_{ij}\f{v^2}{\cut} 
   {L^{\alpha}_i}^T \widehat{C}L^{\beta}_j
 \ov{\Phi}^{\alpha'}\ov{\Phi}^{\beta'}
\epsilon^{\alpha\alpha'} \epsilon^{\beta\beta'} \,+\, {\rm H.c.}
\eeqa
where we choose \,$F_{Lj}=L_j$ ($j=1,2,3$)\,
as the left-handed lepton doublet,
\,$\widehat{C}=i\gamma^2\gamma^0$\, 
is the charge-conjugation operator, and 
\,$\Phi =\dis U\(0,\,{v}/{\sqrt{2}}\)^T
       \equiv \f{v}{\sqrt{2}\,}\ov{\Phi}
\,$.\,
Thus, the tiny neutrino Majorana masses are given by
\beq
\label{eq:mij-nu}
m_\nu^{ij} ~=~ \dis {\cal C}_{ij}\f{v^2}{\cut}
\eeq
and naturally suppressed by
\,$v^2/\cut \lll v$\, in the seesaw mechanism\,\cite{nu-seesaw} 
or by the loop-induced factor 
\,${\cal C}_{ij}\lll 1$\, (which also includes small ratios of 
lepton masses to the TeV scale)
in the radiative mechanism\,\cite{nu-rad}. 
If $\Phi$ is replaced by the linear SM-Higgs-doublet $H$, 
Eq.\,(\ref{eq:nu-mass}) recovers 
the traditional dimension-5 operator of Weinberg\,\cite{weinberg5},
implying the {\it assumption} that 
the Higgs doublet $H$ (responsible for the EWSB)
already participates in the neutrino mass generation.
However, like other SM fermions, the mass generation for neutrinos
needs {\it not}  originate from the {\it same} Higgs doublet as
EWSB\,\cite{nudy,nudy1},\,    
so we will not make the {\it ad hoc} assumption of
\,$\Phi = H$\,.\, 
Instead, we will systematically investigate the most generic neutrino
mass operator  (\ref{eq:nu-mass}) which, in the unitary gauge, 
is just the bare Majorana mass term 
\,$\dis -\f{1}{2}\nu_L^T m_\nu \widehat{C}\nu_L^{~} + {\rm H.c.}$

We see that despite the fact 
that the SM gauge symmetry forbids all bare mass terms
for the weak gauge bosons and fermions, 
they can nonetheless be formulated in a gauge-invariant way 
via the nonlinear realization and thus are necessarily
{\it nonrenormalizable.} 
Hence, the $S$-matrix elements of the
high energy scattering involving
these massive particles will unavoidably violate unitarity at a certain
energy scale at which the new physics responsible for the mass generation
has to recover the proper unitarization.
Generically, we can define the 
scale $\cut_x$ for generating a mass $m_x$ 
{\it to be the minimal energy above which the bare mass term
for $m_x$ has to be replaced by a renormalizable interaction}
(adding at least one new physical
degree of freedom to the experimentally observed particle spectrum).
The unitarity violation scale $E^\star$ provides an {\it upper bound}
on such a scale \,$\cut_x$\,,\, i.e.,
\beq
\label{eq:cutx-UB}
\cut_x ~\leqq~ E^\star  \,,
\eeq
and is thus a {\it  conservative and model-independent} estimate of
the mass generation scale \,$\cut_x$\,.

Generally speaking, 
the scale for  EWSB and mass generation of \,$W^\pm /Z^0$\,
is naturally around 
$\O(v)$ [cf. eq.\,(\ref{eq:UBw22})], 
but the mass generation scales 
for the SM fermions are expected to be much higher (except,
perhaps, for the  top quark).  To study the unitarity bound on
the fermion mass 
we can classify the theories into two types, depending
on the nature of  EWSB: (i) EWSB is generated 
in a weakly coupled scenario with a fundamental Higgs boson(s);
(ii) EWSB is induced by dynamical symmetry breaking
(with a composite Higgs scalar(s) or no Higgs boson).
In either case, we can study the scale of the fermion
mass generation by examining how the
(nonrenormalizable) bare fermion mass term leads to 
unitarity violation in high energy scattering. For 
type-(i), without assuming the EWSB Higgs
boson(s) to also couple to the fermions, we can derive the
unitarity violation bounds from the scattering
\,$f\bar{f} \to n V_L^a$\, (which is well-defined above
$\O(1)$\,TeV and below the unitarity violation scale).    For 
type-(ii), due to the dynamical EWSB, 
the longitudinal gauge boson $V_L^a$ and its Goldstone boson 
$\pi^a$  are composite and cannot be treated as local
fields at energy scales $\gg \O(1)$\,TeV. Thus, the
scattering \,$f\bar{f} \to n V_L^a$\, no longer provides
a reliable, model-independent bound on the scales of
fermion mass generation. 
Instead, we may consider the scattering 
\,$f\bar{f} \to (f\bar{f})^n$\, from  
effective higher dimensional multi-fermion operators.

\vspace*{4mm}
\subsection{\hspace*{-5mm}.$\!$
High Energy Scattering and Power Counting of Energy
}
\vspace*{2mm}

According to the systematic power counting analysis
for electroweak gauge theories\,\cite{He-p1,He-p2},
high energy scattering of longitudinally polarized 
weak gauge bosons $V_L^a$ 
exhibits the worst high energy behavior
and thus is expected to provide the strongest unitarity bounds.
This is easy to understand intuitively 
by noting that in the $S$-matrix 
element each external fermion field contributes an energy 
factor of \,$E^{1\over2}$\, while an external scalar field gives 
no additional energy enhancement. However, for scattering
involving longitudinally polarized weak bosons, each external
field $V_L^a$ gives rise to a factor of \,$E^1$\, 
due to its longitudinal polarization vector
\beq
\label{eq:VL-pol}
\ep^\mu_L(k) = \dis\f{1}{M_a}\(|\vec k|,~ k^0\vec{k}/|\vec{k}|\)
=\f{k^\mu}{M_a} + v^\mu(k)\,,~~~~~~
v^\mu(k)=\O\(\f{M_a}{E}\) ,
\eeq
where \,$M_a = M_{W,Z}$\, and the term 
\,$\O\({M_a}/{E}\)$\, is suppressed 
in the high energy regime \,$E \gg M_a$\,.\, 
The term $k^\mu/M_a$ leads to a potentially
bad high energy behavior for the scattering amplitude and is thus 
the most sensitive to the unitarity violation. 
But, the actual energy-dependence of the $V_L$-amplitude is much
more involved due to the various intricate energy cancellations.
Consider, for instance,
$V_LV_L\to V_LV_L$ scattering at tree-level.
Na{\"\i}ve power counting shows that its amplitude has both $\O(E^4)$ 
and $\O(E^2)$ individual terms, but the $\O(E^4)$ terms always cancel 
out due to the generic Yang-Mills gauge structure and the 
$\O(E^2)$ terms can further cancel down to $\O(E^0)$ 
if a physical Higgs boson $h^0$ is present.\footnote{It was shown
recently\,\cite{C5DU,HeDPF04,Csaki}
that in the compactified (or deconstructed) higher dimensional Higgsless 
Yang-Mills theories, such $\O(E^2)$ terms are canceled (or suppressed) 
due to the presence of spin-1 Kaluza-Klein (KK) gauge bosons 
from a geometric Higgs mechanism.} 
Without such a Higgs scalar $h^0$, the
$(W^\pm,\,Z^0)$ mass terms may be formulated as the nonlinear dimension-2
operator (\ref{eq:V-mass})
which is nonrenormalizable and gives rise to {\it non-canceled}
\,$\O(E^2)$\, terms without respecting to unitarity. 
This puts an upper bound on the EWSB scale \cite{DM,LQT,CG,MVW,DJL}.

The key for carrying out a faithful energy power counting 
for a scattering amplitude with $V_L^a$'s 
is to note\,\cite{He-p1,He-p2} 
that the actual $E$-dependence is given by 
in the corresponding amplitude 
involving would-be Goldstone boson $\pi^a$'s.
These two scattering amplitudes are quantitatively connected via 
the Electroweak Equivalence Theorem (ET)\footnote{A 
KK equivalence theorem
(KK-ET)\,\cite{C5DU,HeDPF04} has recently been constructed 
for the geometric Higgs mechanism in  
compactified higher dimensional Yang-Mills theories on
general 5D-backgrounds (including warped space).
This is valid independent of whether the zero-mode gauge bosons 
are massless or not.}
\cite{Ebad2,LQT,CG,Eold,HKL,HKY94,HK96,xet},
\beq
\label{eq:ET}
\ba{l}
\dis
\TT \[V_L^{a_1},\,\cdots,\,V_L^{a_n};\Phi_{\rm phys}\] ~=~
C_{\rm mod}\,
\TT \[-i\pi^{a_1},\,\cdots,\,-i\pi^{a_n};\Phi_{\rm phys}\] 
~+~ B \,,
\\[2mm]
\hspace*{3mm}
B \equiv\dis 
  \sum_{\ell=1}^n \[
  C^{a_{\ell+1}}\cdots C^{a_n} 
  \TT\[v^{a_1},\cdots,v^{a_n},
       -i\pi^{a_{\ell+1}},\cdots,-i\pi^{a_n};\Phi_{\rm phys}\]
  + \!{\rm permutations} \]
\\[5mm]
\hspace*{7mm}\dis
  = \O\!\(M_W/E_j\)\!-{\rm suppressed} ,
\ea
\eeq
where $\,v^a\equiv v^{\mu}V_\mu^a\,$ and
\,$\Phi_{\rm phys}$ denotes possible amputated
external physical fields\,\cite{HKY94,He-p1}\footnote{This 
ET relation is deeply rooted in the underlying Higgs mechanism
with spontaneous gauge symmetry breaking, 
even if a fundamental Higgs boson does not exists. 
As explicitly shown 
in the second paper of Ref.\,\cite{HKL} [cf. its Eqs.\,(88)-(95)] 
and in Ref.\,\cite{He-p2}, 
the ET (\ref{eq:ET}) is derived from an amputated 
Slavnov-Taylor identity 
$~\TT\[V_S^{a_1}+iC^{a_1}\pi^{a_1},\cdots ,
     V_S^{a_n}+iC^{a_n}\pi^{a_n}, \Phi_{\rm phys}\] = 0
~$, which directly reflects the fact that 
{\it the would-be Goldstone boson
$\pi^a$ and the unphysical scalar component of the vector
field $V_S^a=\ep^\mu_S V^a_\mu$ 
($\ep_S^\mu=k^\mu/M_a$, $M_a=M_{W,Z}$)
are ``confined'' together so that no net effect from them can be
observed in the physical $S$-matrix elements -- a quantitative
formulation of the general Higgs mechanism at the $S$-matrix level
(independent of whether a Higgs boson actually exists).} 
}.
The $B$-term on the RHS of the first equation is
suppressed by $\,\O(M_W/E_j)\,$ relative to the leading
term and its precise form\,\cite{HKY94,He-p1}  
is given by the second equation
in (\ref{eq:ET}).
The constant modification factor 
\,$C_{\rm mod} 
   = C^{a_1} \cdots C^{a_n} 
   = 1+\O({\rm loop})
$\, 
is generated from radiative corrections
and can be exactly simplified to unity under 
realistic renormalization schemes\,\cite{HKL,HKY94,HK96}, 
but will not affect the power counting analysis. 
Since the Goldstone fields $\pi^a$ have no extra polarization
vector like (\ref{eq:VL-pol}), there is no non-trivial 
$E$-cancellation in the $\pi^a$-amplitude [the right-hand side
of Eq.\,(\ref{eq:ET})] so that the $E$-power counting can directly
apply.\footnote{In the case that the graviton or its Kaluza-Klein
states interact with the longitudinal $V_L^a$'s via the energy-momentum 
tensor\,\cite{PRL-KK-WZ}, the $E$-cancellation associated with
the $V_L^a$-polarizations again manifests in the
corresponding amplitude involving the Goldstone boson $\pi^a$'s
due to the ET (\ref{eq:ET}).}\,
It is shown\,\cite{He-p1,He-p2} that such a power counting 
method can be systematically constructed 
(\`{a} la Weinberg\,\cite{weinberg79}) for any given scattering
amplitude $\TT$ and its $E$-power dependence \,$\D_E$\, is, for
\,$E\gg M_{W,Z},\,m_f$\,, 
\beq
\label{eq:DE}
\dis
{\cal D}_E ~=~ 2+2L-\sum_j \V_j\(2-d_j-\f{1}{2}f_j\) -e_v^{~} \,,
\eeq
where $\V_j$ is the number of vertices of type $j$ which contain
$d_j\,(\geqq\! 0)$ derivatives and 
$f_j\,(\geqq\! 0)$ fermionic lines,
$L$ is the number of loops,
and $e_v^{~}\,(\geqq\!0)$ is the number of external gauge fields 
$v^a=V^a_\mu v^\mu$ appeared in an $S$-matrix element 
[cf. (\ref{eq:VL-pol}) for definition of $v^\mu$].
From (\ref{eq:DE}), it can be proven\,\cite{He-p1,He-p2} 
that, for any $n\geqq 2$, 
\beq
\D_E \,~ \leqq ~\, 2+2L ~,
\eeq
where the equality holds when the amplitude
contains only Goldstone boson vertices.

%%%%%%%%%%%%%%%%%%%%%%%% Fig.1 %%%%%%%%%%%%%%%%%%%%%%%%%%%%%%%%%%%%%%%%
\begin{figure}[t]
\label{fig:ff-npi}
\begin{center}
\vspace*{-5mm}
\hspace*{-8mm}
\includegraphics[width=17.5cm,height=12cm]{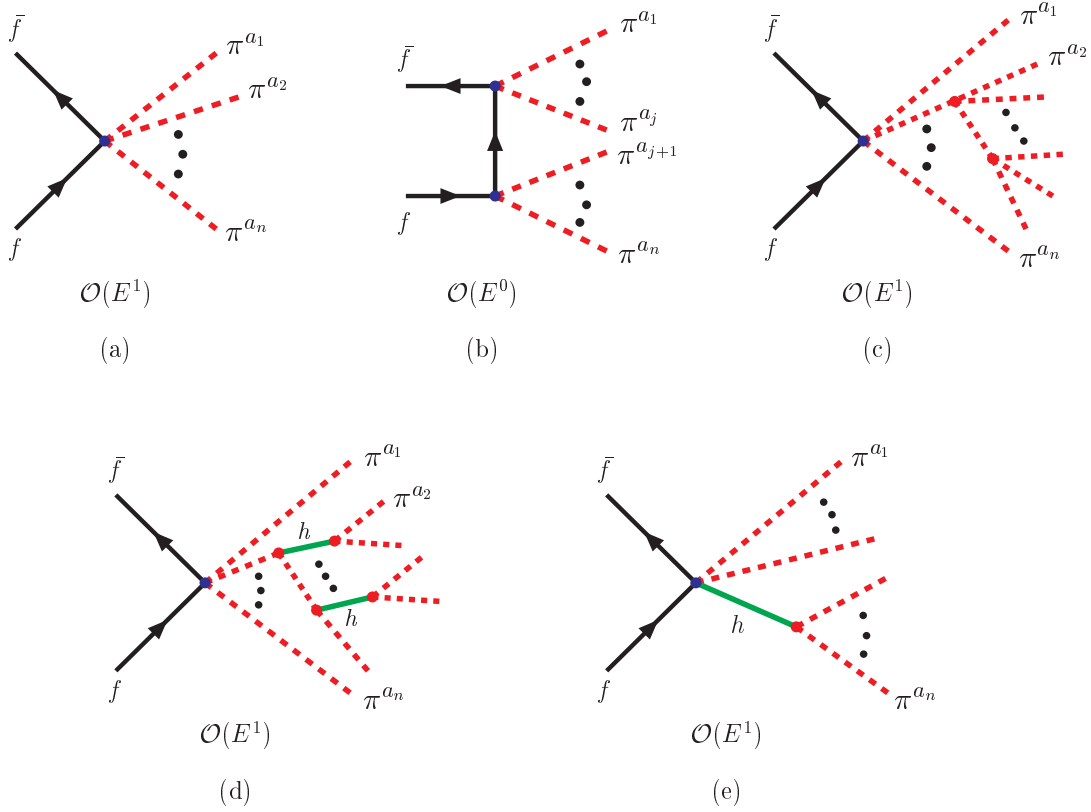}
\vspace*{-10mm}
\caption{Typical contributions to $f\bar f \to n\pi^a$
scattering: 
(a) the leading ``contact'' diagram of $\O(E^1)$;
(b) the sub-leading ``$t(u)$-channel'' type diagrams of 
$\O(E^0)$  which can be ignored for the unitarity analysis;
(c) the ``hybrid''  contact diagrams ($n\geqq 3$) including Goldstone
self-interactions from the EWSB sector;
(d) the ``hybrid''  contact diagrams ($n\geqq 3$) including Goldstone
interactions with the EWSB quanta (illustrated here by the SM Higgs
boson $h^0$  as the simplest example of EWSB);
(e) the ``hybrid''  contact diagrams due to
the {\it assumption} of
Yukawa interactions between the fermions and the SM Higgs boson $h^0$,
as the simplest example of fermion mass generation.
}
\vspace*{-0.5cm}
\end{center}
\end{figure}
%%%%%%%%%%%%%%%%%%%%%%%%%%%%%%%%%%%%%%%%%%%%%%%%%%%%%%%%%%%%%%%%%%%%%%

In the following sections,
we will systematically analyze unitarity for the
\,$2\to n$ ($n> 2$)\, scattering amplitudes
\,$f\bar{f},\nL\nL\to nV_L^a$\, and 
\,$V_L^{a_1}V_L^{a_2}\to nV_L^a\,$.\,  According to the ET
discussed above, to extract the leading energy behavior,
we can analyze the corresponding Goldstone boson amplitudes
\,$f\bar{f},\nL\nL\to n\pi^a$\, and 
\,$\pi^{a_1}\pi^{a_2}\to n\pi^a$\,,\,
where the energy power counting can be safely carried
out [cf. Eq.\,(\ref{eq:DE})].
At the tree level ($L=0$),  using the rule (\ref{eq:DE}),
and the interactions (\ref{eq:f-mass})
(which shows that the coupling of two fermions to any number of
$\pi^a$ has no derivatives) and (\ref{eq:V-mass})
(which shows that the self-coupling of $\pi^a$\,'s 
always has two derivatives), 
we can thus deduce, for \,$E \gg M_{W,Z}, m_f^{~}$\,, 
\beq
\label{eq:Torder}
\ba{lclcl}
\dis
\TT [f\bar{f},\nL\nL \to nV_L^a] &\simeq &
\TT [f\bar{f},\nL\nL \to n\pi^a] &=& 
    \dis\O\(1\)\f{\,m_{f,\nu}^{~}\,}{v^n}E \,,
\\[4mm]
\dis
\TT [V_L^{a_1}V_L^{a_2} \to nV_L^a] &\simeq & 
\TT [\pi^{a_1}\pi^{a_2} \to n\pi^a] &=& 
    \dis\O\(1\)\f{\,E^2\,}{\,v^n\,} \,,
\ea
\eeq
where the dimensionality of \,$\TT$\, is always given by
~$\D_{\TT}=4-(2+n)=2-n$~ and is independent of whether
the external fields are  bosonic or fermionic. 
[It will be shown from the exact calculation in Sec.\,6
that the leading contribution to the second amplitude
in Eq.\,(\ref{eq:Torder}) is nonvanishing only for 
$n({\rm even})=2,4,6,\cdots$.]
Eq.\,(\ref{eq:Torder})
shows that the $E$-power dependence of $\TT$ is actually
independent of the number of particles in the final state.
Therefore, we note that any conceptual change from including
$n$-body final state reactions ($n> 2$) has to arise from the 
$n$-body {\it phase space integration} which will be
systematically analyzed in the following sections.

%\vspace*{5mm}
In Fig.\,1, we illustrate the relevant 
contributions to the scattering 
\,$f\bar{f} \to n\pi^a$.\,  Fig.\,1(a) is 
the leading contribution
of $\O(E^1)$ from the contact fermion-Goldstone-boson 
interaction, solely reflecting the  nonzero bare fermion
mass term in the unitary gauge. Fig.\,1(b) 
is the sub-leading $t/u$-channel
type of contributions of $\O(E^0)$, and can be ignored for
deriving the unitarity bound.   Figs.\,1(c-e)
denote the ``hybrid'' contact graphs due to the presence
of Goldstone self-interactions from
Eq.\,(2.8) and the exchange of associated EWSB quanta (such
as the SM Higgs).   The EWSB sector is unitarized at
the lower scale of $\O(1)$\,TeV, but such contributions in
Fig.\,1(c) and Fig.\,1(d-e) can be at most of $\O(E^1)$ in general,
as we counted above in Eq.\,(\ref{eq:Torder}).
Unlike the model-independent leading ``contact'' contributions in 
Fig.\,1(a), the additions in Fig.\,1(c-e) rely
on the details of  EWSB and are thus highly 
{\it model-dependent}.
But, this does not affect our general estimate in
Sec.\,3 which only makes use of the above generic 
power counting results 
in Eq.\,(\ref{eq:Torder}). For a more precise analysis, 
we will first estimate the unitarity bounds by
quantitatively computing the universal model-independent
contribution of Fig.\,1(a) in Sec.\,4.1. 
Then, in Sec.\,4.2, we further
examine the effect of including the type of contributions in
Figs.\,1(c-e) by specifying an EWSB sector.
We will show that such model-dependent contributions
only affect the unitarity bound by factors of \,$\O(1)$,\,
so that our unitarity limits on the scales of 
fermion mass generation are robust.

\vspace*{4mm}
\subsection{\hspace*{-5mm}.$\!$
General Unitarity Condition and the
Customary ${\mbox {\boldmath  $2\to 2$}}$ Limits}
\vspace*{2mm}

For the convenience of later analysis and discussion,
we give a unified derivation of 
the unitarity condition on elastic and inelastic scattering
including general spins for the particles and
a consistent treatment of identical particles in the final state, 
which extends the derivation of Ref.\,\cite{scott}.
Then we briefly review the customary unitarity limits
for various elastic and inelastic $2\to 2$ scattering processes.

%\vspace*{2mm}
\subsubsection{\hspace*{-5mm}.$\!$
General ${\mbox {\boldmath  $2\to n$}}$
Unitarity Condition}
%\vspace*{2mm}

The unitarity bound
originates from the unitarity condition of
the $S$-matrix, ${\cal S}^\dag {\cal S}=1$, which, with the definition
\,${\cal S}=1+i\,\TT\,$,\, can be expressed as,
%
%\beqa
%\label{eq:TUC}
\,$\TT^{\,\dag}\TT = 2\,\Im\mathfrak{m}\,\TT$\,.\,
%\eeqa
%
Taking the matrix element of both sides of this relation
between identical  2-body states and inserting 
a complete set of intermediate states into its 
left-hand side, one arrives at
\beqa
\label{eq:TUC1}
\dis\int_{{\rm PS}_2}
\left|\TT_{\rm el}[2\to 2]\right|^2  +
\sum_n\int_{{\rm PS}_n} 
\left|\TT_{\rm inel}[2\to n]\right|^2
&=&
2\,\IM\,\TT_{\rm el}[2\to 2]  \,,
\eeqa 
where on the left-hand side 
$\int_{{\rm PS}_n}$ denotes the $n$-body phase space
integration (including the symmetric factors
from the possible identical particles in the final state)
and, in the second term, the summation is over all 
inelastic channels ($n\geqq 2$).
The right-hand-side is evaluated in the forward direction.
For the $2\to 2$ elastic channel, 
we make the partial wave expansion for the scattering
amplitude
\beq
\label{eq:T22pdj}
\ba{l}
\dis
\TT_{\rm el}[2\to 2] ~=~ 16\pi\, e^{i(\nu-\nu')\varphi}\sum_{j} 
(2j+1) d_{\nu'\nu}^j (\cos\!\theta)\,a^{\rm el}_j \,,
\\[4mm]
\dis
a^{\rm el}_j ~=~ \f{1}{32\pi}\, e^{i(\nu'-\nu )\varphi}
\int^1_{-1}d\cos\!\theta \,
d^j_{\nu'\nu}(\cos\theta) \,\TT_{\rm el}[2\to 2]  \,,
\ea
\eeq
where \,$\theta$\, and \,$\varphi$\, are the scattering angles, 
and  \,$(\nu,\,\nu')=(\nu_1-\nu_2,\,\nu_3-\nu_4)$\,   
with \,$(\nu_1,\,\nu_2)$\, and \,$(\nu_3,\,\nu_4)$\, 
the helicity indices
of the incoming and outgoing particles, respectively.
The $d$-functions satisfy the orthonormal condition, 
~$\dis\int_{-1}^1 \! dx\, d^{j}_{\nu'\nu} (x)\,d^{j'}_{\nu'\nu}(x)
  =\f{2\,\delta_{jj'}}{2j+1}$\,,\, and for the special case
of \,$\nu=\nu'=0$\,,\, we have
\,$d^{j}_{00}(x)=P_j(x)$\,,\, with \,$P_j(x)$\, the 
Legendre polynomials.
So, for the incoming and outgoing particles of the same
helicities, \,$\nu_1=\nu_2$\, and  \,$\nu_3=\nu_4$\,,\,
(\ref{eq:T22pdj}) reduces to
\beq
\label{eq:T22p}
\ba{l}
\dis
\TT_{\rm el}[2\to 2] ~=~ 16\pi\sum_{j} 
(2j+1)P_j (\cos\!\theta)\,a^{\rm el}_j \,,
\\[4mm]
\dis
a^{\rm el}_j ~=~ \f{1}{32\pi} \int^1_{-1}d\cos\!\theta \,
P_j(\cos\!\theta)\, \TT_{\rm el}[2\to 2]  \,.
\ea
\eeq
With (\ref{eq:T22pdj}),
it is straightforward to compute
\beqa
\label{eq:SE22}
\dis\int_{{\rm PS}_2}
\left|\TT_{\rm el}[2\to 2]\right|^2
~=~ \dis\f{\,32\pi\,}{\varrho_{\rm e}^{~}} 
  \sum_{j} (2j+1)|a^{\rm el}_j|^2 \,,
\eeqa
where the symmetry factor 
$\varrho_{\rm e}^{~}$ equals $1!\,(2!)$ if the 
2-body final state consists of non-identical (identical) 
particles\footnote{Throughout our derivation the factors
due to the proper counting of final state identical particles
will solely originate from the 
{\it phase space integration of the $n$-body final state}
 in the total cross section.}.
In (\ref{eq:SE22}) and the rest of this paper, 
we consider the mass of any initial/final state particle to be 
much smaller than the c.m.\ energy, \,$m_j^2\ll s$\,, 
which is justified for the scattering processes to be analyzed
in Sec.\,3-6.\footnote{For completeness, in Appendix\,B 
we give the precise formula of 
$\int_{{\rm PS}_2}
 \left|\TT_{\rm el}[2\to 2]\right|^2$ 
and the refined unitarity conditions by including 
the possible masses of initial/final state.}\,
Using Eqs.\,(\ref{eq:T22pdj}) and (\ref{eq:SE22}),
the unitarity condition (\ref{eq:TUC1}) becomes
\beq
\label{eq:TUC2}
\dis
\sum_{j} (2j+1) \f{1}{\varrho_{\rm e}^{~}}
\[ \f{\varrho_{\rm e}^2}{4} - \( {\RE}\,a^{\rm el}_j\)^2
  -\( {\IM}\,a^{\rm el}_j - \f{\varrho_{\rm e}^{~}}{2}\)^2  
\] 
\,=~ \f{1}{\,32\pi}\sum_n
\int_{{\rm PS}_n}\left|\TT_{\rm inel}[2\to n]\right|^2
~\geqq 0  \,\,,
\eeq
Because the right-hand side of (\ref{eq:TUC2}) is non-negative,
we find a consistent solution, for each \,$j$\,,
\beq
\label{eq:al-ucircle}
\dis
   \( {\RE}\,a^{\rm el}_j\)^2
  +\( {\IM}\,a^{\rm el}_j - \f{\varrho_{\rm e}^{~}}{2}\)^2 
~\leqq~ \f{\varrho_{\rm e}^2}{4} \,,
\eeq
which results in
\beq
\label{eq:al-bound}
\ba{rcc}
\dis\left|{\RE}\,a^{\rm el}_j\right| 
   &\leqq&  \dis\f{\,\varrho_{e}^{~}\,}{2} \,,
\\[3mm]
\dis\left|\,a^{\rm el}_j\right| &\leqq&  \dis\varrho_{e}^{~} \,.
\ea
\eeq
When the 2-body final state consists of non-identical
particles ($\varrho_{\rm e}^{~} =1$), these conditions
reduce to the familiar form,
$~\dis |{\RE}\,a^{\rm el}_j| \,\leqq\, \f{1}{2} \,$\, 
and 
$~|a_j^{\rm el}| \leqq 1$\,.\, 
With (\ref{eq:al-bound}) and defining the elastic cross section  
$\,\sigma_{\rm el} \equiv \dis\sum_j\sigma^{\rm el}_j\,$,\,
we deduce a bound on $\,\sigma_{\rm el}\,$ in 
its $j$-th partial wave, from Eq.\,(\ref{eq:SE22}),
\beq
\label{eq:UC-SE2}
\sigma^{\rm el}_j[2\to 2] ~\leqq~ 
\dis\f{\,16\pi (2j+1)\varrho_{\rm e}^{~}\,}{s} ~,
\eeq
which, for $\varrho_{\rm e} =1$,\,
agrees with Eq.\,(38.47) of Ref.\,\cite{PDG}. 
Finally, by assuming the dominance of \,$j=0$\, contribution
to the elastic channel,
an upper bound on the inelastic cross section
$\,\sigma_{\rm inel}[2\to n]$\, ($n\geqq 2$)\, can be derived from 
Eq.\,(\ref{eq:TUC2}),
\beq
\label{eq:UC-IEn}
\sigma_{\rm inel}[2\to n] ~\leqq~ \dis\f{\,4\pi\varrho_e\,}{s} ~.
\eeq
Here the identical particle factor 
$\varrho_e$ is determined by the final state of the 
{\it elastic} channel 
(which has the same initial state as the inelastic channel)
as indicated by the subscript of \,$\varrho_e$\,.\,
For a given initial state of the inelastic channel, 
as long as we can find
at least one elastic channel (with the same initial state) in which the
final state particles are non-identical\footnote{For some 
types of  initial states in a
given model, the final state with identical particles may be automatically
forbidden due to the particular charge of this initial state or the
absence of certain couplings.},\,
we can then set \,$\varrho_e=1$\,
in the above {\it inelastic} bound  (\ref{eq:UC-IEn}) to get the
optimal constraint.
This mild requirement is satisfied for  
all the inelastic channels of 
$\,f\ov{f},f\ov{f'}\to nV_L^a\,$.\,
Before concluding this subsection, let us derive a partial
wave unitarity condition for the $2\to 2$ {\it inelastic} channel
[parallel to (\ref{eq:al-bound})]. 
Using the same partial wave expansion formula (\ref{eq:T22p})
for the inelastic amplitude, we can compute
\beqa
\label{eq:INE22}   %{eq:SE22}
\dis
\int_{{\rm PS}_2}
\left|\TT_{\rm inel}[2\to 2]\right|^2
~=~ \dis\f{32\pi}{\varrho_{\rm i}^{~}} \sum_j (2j+1)
   |a^{\rm inel}_j|^2 \,,
\eeqa
where \,$\varrho_{\rm i}^{~}=1!\,(2!)$\, 
for an inelastic final state being
non-identical (identical) particles.
Substituting this back into the RHS of (\ref{eq:TUC2}),
we deduce
\beq
\label{eq:inel22a}
\dis
\sum_j (2j+1) 
\left\{ 
   \f{\varrho_{\rm e}^{~}}{4} - \f{1}{\varrho_{\rm e}^{~}}
   \[\( {\RE}\,a^{\rm el}_j\)^2
    +\( {\IM}\,a^{\rm el}_j - \f{\varrho_{\rm e}^{~}}{2}\)^2\]  
\right\}
~>~ \sum_j (2j+1)\f{1}{\varrho_{\rm i}^{~}}
  | a_j^{\rm inel} |^2 ~,
\eeq
so that, for each \,$j$\,,
\beqa
\label{eq:a0-inel}
|a_j^{\rm inel}| &<& \dis
\f{\,\sqrt{ \varrho_{\rm i}^{~}
            \varrho_{\rm e}^{~}\,}\,}{2} \,.
\eeqa
Interestingly, the optimal bound for a
given $2\to 2$ inelastic channel is realized 
when we can set $\varrho_e^{~}=1$
for a corresponding elastic channel (which shares the same
initial state), i.e.,
$|a_j^{\rm inel}| < \sqrt{\varrho_{\rm i}^{~}}/2$\,. 
We see that this or (\ref{eq:a0-inel})
takes a {\it different}
form from the elastic partial wave condition (\ref{eq:al-bound})
when final state consists of identical particles or 
when $\IM\,a_j \neq 0$\,.

%\newpage
\vspace*{4mm}
\subsubsection{\hspace*{-5mm}.$\!$
Customary Limits from ${\mbox {\boldmath  $2\to 2$}}$
Scattering
}
\vspace*{3mm}

\noindent
%\underline
{\bf 2.3.2A.}   
{\tt $2\to 2$ Unitarity Bound for the EWSB Scale}
\vspace*{3mm}

The unitarity violation 
from the quasi-elastic scattering 
\,$V^{a_1}_LV^{a_2}_L\to V^{a_3}_LV^{a_4}_L$\, was originally
studied by Dicus-Mathur/Lee-Quigg-Thacker (DM/LQT) \cite{DM,LQT}
for constraining the Higgs boson mass in the SM.
According to the ET, its leading high energy behavior 
is represented by the corresponding Goldstone
boson scattering 
$\,\pi^{a_1}\pi^{a_2}\to \pi^{a_3}\pi^{a_4}$,\,
and can be directly computed from the Higgsless nonlinear Lagrangian
(\ref{eq:V-mass}) as 
\beq
\ba{lll}
\TT[\pi^0\pi^0\to\pi^0\pi^0] = 0\,,
&
\dis\TT[\pi^0\pi^0\to\pi^+\pi^-] = \f{s}{v^2}\,,
&
\dis\TT[\pi^+\pi^-\to \pi^0\pi^0] = \f{s}{v^2}\,,
\\[2mm]
\dis\TT[\pi^0\pi^\pm\to\pi^0\pi^\pm ] = \f{t}{v^2}\,,
&
\dis\TT[\pi^\pm\pi^\pm\to\pi^\pm\pi^\pm] = -\f{s}{v^2}\,,~~
&
\dis\TT[\pi^+\pi^-\to\pi^+\pi^-] = -\f{u}{v^2}\,,~~
\ea
\eeq
where $\,t\simeq -\f{1}{2}s\(1-\cos\theta\)\,$  and
      $\,u\simeq -\f{1}{2}s\(1+\cos\theta\)\,$.\, 
This is usually called the low energy theorem\,\cite{CGG} and
may be reexpressed in terms of isospin amplitudes
$\TT [I]$ \cite{Don},
\beq
\ba{l}
\dis
\TT[0] ~=~ 3\A(s,t,u)+\A(t,s,u)+\A(u,t,s)
       ~=~ \f{\,3s+t+u\,}{v^2} \,,
\\[2mm]
\dis
\TT[1] ~=~ \A(t,s,u) - \A(u,t,s) ~=~ \f{\,t-u\,}{v^2} \,,
\\[3.5mm]
\dis
\TT[2] ~=~ \A(t,s,u)+\A(u,t,s) ~=~ -\f{\,t+u\,}{v^2} \,,
\ea
\eeq
where ~$\A(s,t,u) = \dis\f{\,s\,}{\,v^2\,}$\,,\, and the
relevant identical particle factors are not included.
Considering the isospin-singlet channel and including the identical
particle factor $\,\varrho_{\rm e}=2!\,$ in the first inequality
of Eq.\,(\ref{eq:al-bound}), we see that the unitarity condition
$~\left|\RE\,a^0_0\right| \leqq \dis\f{\varrho_{\rm e}}{2}~$
leads to an optimal limit, 
$\sqrt{s}\leqq E^\star_W$ with
$\,
E^\star_W ~=~ \sqrt{8\pi}\,v ~\simeq~ 1.2\,{\rm TeV}\,.
$\,
Another derivation is to define the normalized 
isospin-singlet state (including identical particle factors),
$\,\left|0\right>=\dis\f{1}{\sqrt{6}}
   \[2\left|\pi^+\pi^-\right> +
      \left|\pi^0\pi^0\right>\]\,$, and compute the
scattering amplitude in the singlet-channel,
\beq
\label{eq:Amp22-T0}
\TT[0] ~=~ \dis
\f{1}{6}\[4\TT[+-,+-]+4\TT[+-,00]+\TT[00,00]\]
\,=\, \f{\,2(s-u)\,}{3v^2} \,.
\eeq
Thus, from (\ref{eq:Amp22-T0})
the usual unitarity condition of the $s$-wave amplitude 
($\,\left|\RE\,a^0_0\right| \leqq \dis{1}/{2}\,$)
imposes an optimal bound for the isospin singlet-channel,
\beq
\label{eq:EWSB-UB22}
E^\star_W ~=~ \sqrt{8\pi}\,v ~\simeq~ 1.2\,{\rm TeV}\,.
\eeq
which we mentioned earlier in Eq.\,(\ref{eq:UBw22}).

\vspace*{7mm}
\noindent
{\bf  2.3.2B.}   %${\mbox {\boldmath  $2\to 2$}}$
{\tt $2\to 2$ Unitarity Bound on the Scale of Fermion Mass Generation}
\vspace*{3mm}

Appelquist and Chanowitz (AC) \cite{AC} 
considered the scattering 
$f_\pm\bar{f}_\pm\to V_L^{a_1}V_L^{a_2}$ \cite{FFVV} with
fermion bare mass terms\footnote{Their equivalent gauge-invariant
nonlinear formulation is given in Eq.\,(\ref{eq:f-mass}).},
and derived the asymptotic amplitude,
\beq
\TT[f_\pm\bar{f}_\pm\to W^+_LW^-_L] 
~=~ \pm \f{m_f}{\,v^2\,}\sqrt{s} \,,
\eeq
where the subscripts for $f_{\pm}\bar{f}_\pm$
indicate the fermion helicity $\,\dis\pm\f{1}{2}$\,.\,
For the color-singlet initial state, this gives the $s$-wave
unitarity limit 
~$\sqrt{s}\leqq E^\star \simeq \dis\f{8\pi v^2}{\sqrt{N_c}\,m_f}$\,,\,
as we mentioned earlier in Eq.\,(\ref{eq:UBf22}).
Defining the spin-singlet combination for the initial state
fermions, 
$~\dis\f{1}{\sqrt{2}}
  \[|f_+\bar{f}_+\rangle - |f_-\bar{f}_-\rangle\]$\,,\,
we may make this bound slightly tighter,
\beq
\label{eq:ACx}
E^\star_f ~\simeq~ \dis\f{8\pi v^2}{\sqrt{2N_c}\,m_f} \,.
\eeq 
One can further define a normalized isospin-singlet state
for the out-going gauge bosons with the combination\,\cite{MVW}
~$\dis\f{1}{\sqrt{6}\,}
\[ 2|W^+_LW^-_L\ran + |Z_L^0Z_L^0\ran\]$\,,\,
%\sum_{a=1}^3\left|V_L^a V_L^a\right\rangle$\,,\, 
and thus slightly reduce the bound (\ref{eq:ACx}) by an additional
factor of \,$\dis\sqrt{\f{2}{3}}\simeq {1}/{1.22}$\,,\,
i.e.,\footnote{For 
clarity of the following analyses with a
multiple $V_L^a$ ($\pi^a$) final state, 
we will ignore such minor improvements 
and thus will use (\ref{eq:ACx}) for comparison.}\, 
$\,\dis E_f^{\star}\simeq 
   \f{8\pi v^2}{\,\sqrt{3N_c}m_f^{~}\,}\,$.\,

\vspace*{7mm}
\noindent
{\bf 2.3.2C.}    
{\tt $2\to 2$ 
Unitarity Bound on the Scale of Majorana Neutrino Mass Generation}
\vspace*{3mm}

The Majorana neutrino scattering 
\,$\nu_{L\pm}^{~}\nu_{L\pm}^{~}\to V_L^{a_1}V_L^{a_2}$\, 
was considered recently by
Maltoni-Niczyporuk-Willenbrock (MNW)\,\cite{scott,scott-PRL}
who derived an upper bound on the scale of the neutrino 
mass generation by using the Weinberg dimension-5
operator\,\cite{weinberg5} with the usual SM Higgs doublet.
For the inelastic scattering 
\,$  \f{1}{\sqrt{2}}\[|\nu_{L+}^{~}\nu_{L+}^{~}\rangle-
               |\nu_{L-}^{~}\nu_{L-}^{~}\rangle\]
     \to \f{1}{\sqrt{2}}|Z_LZ_L \ran$ or 
        $\f{1}{\sqrt{2}} |hh \ran$,\,
one obtains, from the condition (\ref{eq:a0-inel})
($\varrho_{\rm e}=2$),\footnote{In this paper we will not add
extra identical particle normalization factors for the {\it initial} 
state as they are irrelevant to the calculation of the 
cross section and unitarity bound (cf. Sec.\,2.3.1).
For convenience, we may do so in the amplitude calculation only 
for final state identical particles as far as the total
cross section and unitarity bound are concerned (which also means
that in the condition (\ref{eq:a0-inel}) we will accordingly remove 
the final state identical particle factor 
$\varrho_{\rm i}^{~}$ and still retain $\varrho_{\rm e}$), 
but we keep in mind that this is inappropriate when computing the 
differential cross section.}
\beq
\label{eq:nuB22}
E^\star_\nu ~\simeq~ \dis\f{~4\sqrt{2}\pi v^2\,}{m_\nu} \,,
\eeq
which can be improved to
$~E^\star_\nu \,\simeq\, \dis \f{4\pi v^2}{m_\nu}~$
by using a mixed final state
$\,\f{1}{2} |Z_LZ_L+hh\rangle$ \cite{scott}.

In our analysis, the bare neutrino Majorana mass term 
arises from the nonlinear operator (\ref{eq:nu-mass})
{\it without} assuming
the participation of the SM Higgs boson. With this, 
the scattering processes
\,$\f{1}{\sqrt{2}}\[|\nu_{L+}^{~}\nu_{L+}^{~}\rangle-
           |\nu_{L-}^{~}\nu_{L-}^{~}\rangle\]
 \to |W^+_LW^-_L \rangle,\, \f{1}{\sqrt{2}}|Z_LZ_L\rangle$\,
and 
\,$\f{1}{\sqrt{2}}\[|\nu_{L+}^{~}\nu_{L+}^{~}\rangle-
           |\nu_{L-}^{~}\nu_{L-}^{~}\rangle\]
 \to |\pi^+\pi^- \rangle,\, \f{1}{\sqrt{2}}|\pi^0\pi^0\rangle$\,
can be computed. The helicity amplitudes are
\beq
\label{eq:nunu-VV}
\ba{l}
\dis\T\[\nLpm\nLpm \to W^+_LW^-_L \]
\,\simeq\, -\T\[\nLpm\nLpm \to \pi^+\pi^-\]
\,\simeq\,  \dis\pm \f{2m_\nu}{v^2}\sqrt{s} ~,
\\[4.5mm]
\dis\T\[\nLpm\nLpm \to\f{1}{\sqrt{2}} Z_L Z_L \]
\,\simeq\, -\T\[\nLpm\nLpm \to \f{1}{\sqrt{2}}\pi^0\pi^0\]
\,\simeq\, \dis\pm \f{2\sqrt{2}m_\nu}{v^2}\sqrt{s} ~.
\ea
\eeq
Thus, for a spin-singlet initial state, the
scattering 
\,$\f{1}{\sqrt{2}}\[|\nLp\nLp\rangle-|\nLm\nLm\rangle\]
 \to \f{1}{\sqrt{2}}|Z_LZ_L\rangle$\,
leads to a unitarity violation limit 
\beq
\label{eq:nuB22x}
E^\star_\nu ~\simeq~ \dis\f{~2\sqrt{2}\,\pi v^2\,}{m_\nu} \,.
\eeq
The reason that this limit is stronger than
(\ref{eq:nuB22}) by a factor of $2$ is the absence
of the SM Higgs boson in our nonlinear operator (\ref{eq:nu-mass}).
This is due to the fact 
that the $s$-channel SM-Higgs exchange arising 
from the usual dimension-5 Weinberg operator 
will partially cancel 
with the gauge amplitude in the unitary gauge
and push the unitarity violation to a slightly higher scale.
%\footnote{We will return to  this issue in Sec.\,6.}
We also note that with the isospin-singlet 
combination for the final state gauge bosons, 
~$\dis\f{1}{\sqrt{6}\,}\[ 2|W^+_LW^-_L\ran + |Z_L^0Z_L^0\ran\]$\,,\,
the bound (\ref{eq:nuB22x}) may be further improved by 
factor of \,$\dis\f{\sqrt{3}\,}{2}\simeq {1}/{1.15}$\,.\,
With the typical neutrino mass  $m_\nu \sim 0.05$\,eV,
the limit (\ref{eq:nuB22x}) [or (\ref{eq:nuB22})] 
gives,
\,$E^\star_{\nu} \sim 10^{16}\,{\rm GeV}$,\,
which is right at the grand unification (GUT) scale.

\newpage
%\vspace*{5mm}
\section{\hspace*{-6mm}.$\!$
Challenge from ${\mbox {\boldmath  $2\to n$}}$
Inelastic Scattering}
\vspace*{4mm}

\subsection{\hspace*{-5mm}.$\!$
Puzzle: Energy Power Counting vs.~Kinematic Condition}
\vspace*{3mm}

According to the power counting result 
Eq.\,(\ref{eq:Torder}),  the $E$-power
dependence of the scattering amplitudes
for $f\bar{f},\nL\nL \to nV_L^a$ and 
\,$V_L^{a_1}V_L^{a_2}\to nV_L^a$\, ($n\geqq 2$)\, 
is independent of the number of $V_L^a$'s in the final
state. This means that any conceptual change for $n>2$ 
has to come from the $n$-body phase space.
Indeed, it is straightforward to count the $E$-power
dependence from the $n$-body phase space integration,
\beq
\label{eq:PSn}
\ba{ll}
\dis\f{1}{\,{\cal J}_{\rm in}\,}
\int_{{\rm PS}_n} 
& 
\dis =~ \f{1}{\,\varrho\,{\cal J}_{\rm in}\,}\int
\f{d^3k_1\cdots d^3k_n}{2E_1\cdots 2E_n}
(2\pi)^{4-3n}\,\delta^{(4)}\!\(p_1+p_2-\sum_{j=1}^nk_j\)
\\[7mm]
&~ \dis\sim~ s^{n-3} ~=~ E^{2(n-3)} \,, 
\ea
\eeq
where ~$ {\cal J}_{\rm in} = 4\[(p_1\cdot p_2)^2-(m_1m_2)^2\]^{1\over2}
\simeq 2s$~ is the flux factor of the initial state,
\,$\varrho$\, is the symmetry factor from possible identical
particles in the final state, and we define 
\,$E\equiv \sqrt{s}$\,.\,  Eq.\,(\ref{eq:PSn}) shows that the
phase space contributes a nontrivial energy power factor
to the cross section which grows with $n$.

Combining the powers of energy for the scattering amplitude
in Eq.\,(\ref{eq:Torder}) and for the $n$-body phase space
integration in Eq.\,(\ref{eq:PSn}), we arrive at the
following \NAIVE power counting estimate, including all
{\it dimensionful} parameters in the cross section, 
\beq
\label{eq:SE-naive}
\dis
\sigma[2\to n] ~~\propto~~ 
\dis\f{1}{s}\(\f{m_f^{~}}{v}\)^{2(2-\d)}
            \(\f{s}{v^2}\)^{n -2 + \d} \,,
~~~~~~~(n\geqq 2)\,,
\eeq
where ~$\d =1$~ for 
~$f\bar{f},\nL\nL\to nV_L^a$ (as considered in \cite{scott})~
and ~$\d =2$~ for 
~$V_L^{a_1}V_L^{a_2}\to nV_L^a$\,.\,
For simplicity, we have used \,$m_f^{~}$\, to denote the mass of either
a Dirac fermion $f$ or a Majorana neutrino $\nL$.
With the general condition  (\ref{eq:UC-IEn})
the \NAIVE estimate (\ref{eq:SE-naive}) results in the
unitarity limit,
\beq
\label{eq:PB-naive}
\dis
\widetilde{E}^\star \,~\sim~\,
v\[C_0
   \(\f{v}{m_f}\)^{2(2-\d)}\]^{\,\f{1}{2(n-2+\d)}\,} 
\longrightarrow\, v\,,
~~~~~~({\rm for}~n\to\,{\rm large})\,,
\eeq
where $C_0$ is a dimensionless constant 
undetermined by the power counting.
From (\ref{eq:PB-naive}), we see that for \,$\d=1$,\, 
it gives the bound (\ref{eq:UBfscot}) 
(which is the conclusion of Ref.\,\cite{scott}),
while for \,$\d=2$,\, it results in the bound (\ref{eq:UBwus}). 
It is clear that if the above \NAIVE power counting estimate
could really hold, then one is led to a striking conclusion that
all unitarity bounds from  multiple $V_L$-production
would approach the same scale \,$v\simeq 246$\,GeV\, 
for arbitrarily large $n$,
and thus such scattering processes would reveal {\it no new
scale} (other than $v$) for the mass generation. 
As we have observed, even for
the EWSB scale, this would imply a significant reduction
of the customary \,$2\to 2$\, limit (\ref{eq:UBw22}) 
[or (\ref{eq:EWSB-UB22})].

The puzzle is sharpened 
when we consider the general kinematic condition for all
multiple $V_L$-production processes,
\beq
\label{eq:KCx}
\sqrt{s} ~\,>~\, n M_{W(Z)} \,\simeq\, \dis\f{n}{3}v \,,
~~~~\longrightarrow~~~~~
{E^\star} ~\,>~\, v\,\f{n}{3} \,,
\eeq
which must hold for all these processes as we advertized
in the Introduction.  When $n$ becomes large, this condition requires
the unitarity violation scale \,$\sqrt{s}=E^\star$\, to {\it grow
at least linearly with \,$n$\,,}\, rather than arbitrarily approaching
the scale \,$v$\,.\, This unavoidably contradicts to the conclusion
drawn from the \NAIVE power counting estimate (\ref{eq:PB-naive}). 
It is this observation that leads us to believe
that the condition  (\ref{eq:PB-naive}) cannot really hold and
a deeper resolution must exist.

\vspace*{2mm}
\subsection{\hspace*{-5mm}.$\!$
Resolution: 
Increasing Power of Energy vs.~Phase Space Suppression}
\vspace*{2mm}

At this point it appears clear that
in order to be consistent with the kinematic condition
(\ref{eq:KCx}) there must be additional $n$-dependent
{\it dimensionless} factors in  the {\it exact} 
$n$-body phase space integration that  sufficiently 
suppress the $E$-power enhancement in Eq.\,(\ref{eq:PSn}).

When the possible angular dependence is ignored in the squared
amplitude,
\,$\left|\TT\right|^2$,\,
the cross section factorizes into the squared amplitude
times the phase space, so the $n$-body phase space 
integration can be done exactly (cf. Appendix\,A) and 
is given by
\beq
\label{eq:PSnexact}
\ba{ll}
\dis\f{1}{\,{\cal J}_{\rm in}\,}
\int_{{\rm PS}_n} 
& 
\dis =\,\f{1}{\,\varrho\,{\cal J}_{\rm in}\,}\int
\f{d^3k_1\cdots d^3k_n}{2E_1\cdots 2E_n}
(2\pi)^{4-3n}\,
\delta^{(4)}\!\(p^{~}_1+p^{~}_2-\sum_{j=1}^nk_j\)
\\[5mm]
& \dis =~
\f{s^{n-3}}{~2^{4(n-1)}\pi^{2n-3}\,(n-1)!(n-2)!\,\varrho~} \,.
\ea
\eeq
For the processes \,$f\bar{f},\nL\nL \to nV_L^a$\, and
\,$V_L^{a_1}V_L^{a_2}\to nV_L^a$\,,\, using Eq.\,(\ref{eq:Torder})
we can generically write
the form of the squared amplitudes as,
\beq
\label{eq:T2c}
\left|{\cal T}\right|^2 ~=~ c^{~}_0(\theta_j)\,
\(2N_c\)^{2-\d}
\dis\f{~m_f^{2(2-\d)}s^{\d}~}{v^{2n}} \,,
\eeq 
where \,$\d$\, is defined below Eq.\,(\ref {eq:SE-naive}) and
\,$c^{~}_0(\theta_j)$\, is a process-dependent, dimensionless coefficient 
with possible dependence on the scattering
angles ($\theta_j$)\,\footnote{It turns out that for
the scattering \,$f\bar{f},\nL\nL \to nV_L^a$\,, 
the universal leading contributions are given by the ``contact'' 
Feynman diagrams represented in Fig.\,1(a) where the
$c^{~}_0$ has no angular dependence.}.\,
Here, the factor $\,2N_c\,$ comes from the color-singlet channel
of the initial state fermion pair with a proper helicity combination
[cf. Eq.\,(\ref{eq:in})],  and $\,N_c=3\,(1)\,$  
for quarks (leptons or neutrinos).
Thus, we can have a {\it realistic estimate} for the scattering
cross section,
\beqa
\label{eq:PCnsigma}
\sigma &=& \dis
\f{C_0\(2N_c\)^{2-\d}
}{~2^{4(n-1)}\pi^{2n-3}\,(n-1)!(n-2)!~}
\f{1}{\,s\,}
\(\f{\,\sqrt{s}\,}{v}\)^{2(n-2+\d)}
\(\f{\,m_f\,}{v}\)^{2(2-\d)}
\,,
\eeqa
where $\,C_0\equiv \ov{c}^{~}_0/\varrho$\,
and \,$\ov{c}^{~}_0$\, is a constant factor originating from the
coefficient $c^{~}_0$ in the squared amplitude (\ref{eq:T2c}) 
and includes the possible modification to the 
$n$-body phase space integration (\ref{eq:PSnexact}) 
(if $c^{~}_0$ has nontrivial angular dependence).
Substituting (\ref{eq:PCnsigma}) into the general condition
(\ref{eq:UC-IEn}), we thus arrive at the following   
{\it improved estimate} of the unitarity bound,
\beq
\label{eq:UBus}
\dis
E^\star ~=~ v\[C_0\, 2^{4n-2}\pi^{2(n-1)}
          (n-1)!\,(n-2)!
\(2N_c\)^{\d-2}
\(\f{v}{\,m_f\,}\)^{2(2-\d)}\]^{\f{1}{2(n-2+\d)}}
,
\eeq
where the process-dependent constant $C_0$ will be evaluated
by an exact calculation (cf. Sec.\,4-5).  
For the current estimate it is reasonable to assume
\,$C_0 = \O(1)\,$\, or, in a more prescise form,
\beq
\label{eq:C0}
C_0^{\f{1}{\,2(n-2+\d)\,}} ~\sim~ 1\,,
\eeq
which becomes increasingly more accurate as $n$ gets larger.
In fact, Eq.\,(\ref{eq:UBus}) may be regarded as an exact
bound (with the constant $C_0$ to be derived from the precise
calculation), and
thus the formula (\ref{eq:C0}) may actually represent 
{\it all our approximations} made in the current estimate
of the unitarity bound:
{\bf (i).} ignore the dimensionless coefficient \,$c_0^{~}(\theta_j)$\,
in the amplitude-square (\ref{eq:T2c}) including its possible
angular dependence [that may lead to an $\O(1)$ correction
to the $n$-body phase space integration (\ref{eq:PSnexact})];
{\bf (ii).} ignore the symmetry factor \,$1/\varrho$\,
arising from possible identical particles in a given final state.

The crucial observation  from Eq.\,(\ref{eq:UBus})
is to note that the $n$-dependent dimensionless factor
$~\dis\[(n-1)!\,(n-2)!\]^{1\over{2(n-2+\d)}}~$
does not approach unity  as $n$ becomes large, 
rather, it increases almost linearly  with $n$. This
can be understood from Stirling's formula\footnote{The
relative error of this formula is less than 
$4\%$ for \,$n\geqq2$\,.},
$~n!\simeq n^ne^{-n}\sqrt{2\pi n}\,$,\, from which
we can estimate, for large $n$,
\beq
\label{eq:stirling}
\dis\[(n-1)!\,(n-2)!\]^{\f{1}{2(n-2+\d)}} 
~\simeq~   \(n!\)^{\f{1}{n}} ~\simeq~ 
            \f{n}{e} ~>~ \f{n}{3}\,,~~~~~~
            (\,{\rm for}~n\gg1\,)\,,
\eeq
where the constant $\,e=2.7182818\cdots$.
It is this feature that makes our bound
(\ref{eq:UBus}) fully consistent with the kinematic condition
(\ref{eq:KCx})! 
For the processes \,$f\bar{f},\nL\nL \to nV_L^a$\, ($\d=1$) with light
quarks (leptons) or neutrinos, i.e., $v^2/m^{2}_f\gg 1$, we 
observe that there is a {\it competition} in the unitarity bound
(\ref{eq:UBus}) between the quickly decreasing factor 
$~\dis\({v}/{m^{~}_f}\)^{\f{1}{n-1}}~$ and
the almost linearly increasing factor 
$~\dis\[(n-1)!\,(n-2)!\]^{1\over{2(n-2+\d)}}~$ 
when $n$ becomes large.
As a consequence we expect the strongest unitarity bound to
occur at a value \,$n=n_s > 2$\, and, after $n$ exceeds
\,$n_s$\,,\, the factor $~\dis\[(n-1)!\,(n-2)!\]^{1\over{2(n-2+\d)}}~$ 
will become dominant and eventually force the bound
(\ref{eq:UBus}) to grow almost linearly  with  \,$n$\,.\,
Only for the heaviest SM fermion, 
the top quark (where \,$v^2/m_t^{2} \simeq 2 \not\gg 1$), 
can the value of \,$n_s$\, be equal (or very close) to $2$.  
On the other hand, for the process
\,$V_L^{a_1}V_L^{a_2}\to nV_L^a$\, ($\d=2$),
when $n$ increases, 
the only possible {\it decreasing} factor on the
right-hand side of Eq.\,(\ref{eq:UBus}) 
is ~$\dis\(\sqrt{C_0}/(2\pi)\)^{\f{1}{n}}$\,,\,
provided \,$\sqrt{C_0} > 2\pi$\,.\,  
But our current estimate under the approximation
(\ref{eq:C0}) implies the setting \,$C_0\sim 1$\,.\,
So it seems  that we will have \,$n_s=2$\, for the
EWSB scale which corrsponds to 
the customary $\,2\to 2$\, unitarity limit
discussed in Sec.\,2.3.2 [cf. Eq.\,(\ref{eq:EWSB-UB22})].
Since the current estimate does not precisely fix the value of
$\,C_0$,\, a reliable determination of \,$C_0$\, and thus 
\,$n_s$\, should be given by a quantitative calculation, 
which will indeed support \,$n_s=2$\, (cf. Sec.\,6).

Considering the region \,$n\gg 1$\, and thus 
\,$\dis\O(1)^{\f{1}{2(n-2+\d)}}\sim \O(1)^{\f{1}{2n}}
   \sim 1$\,,\,
we can further derive a complete asymptotic formula for
Eq.\,(\ref{eq:UBus}),
\beq
\ba{ll}
\label{eq:UBasymt}
\dis
E^\star 
& \dis\approx~ v\(\f{v}{m_f^{~}}\)^{\f{2-\d}{n}}
                 \f{4\pi}{e}n ~, ~~~~~~~
(\,{\rm for}~ n\gg 1\,)
\\[6mm]
& \dis\approx~ \f{\,4\pi v\,}{e} 
  \[n + (2-\d)\ln\(\f{v}{m^{~}_f}\) + \O\(\f{(2-\d)^2}{n}\)\].
\ea
\eeq
From this, we deduce that for each asymptotic value of $n$, the
leading difference between the unitarity bounds of two types of
fermions with masses $m_{f_1}^{~}$ and  $m_{f_2}^{~}$ is
given by\footnote{For the analysis in the current section, 
$f$ denotes either a quark,  a charged lepton, or a Majorana neutrino.}
\beq
\label{eq:UBf-dif}
E^\star_{f_1} - E^\star_{f_2} ~\approx~ \dis
\f{\,4\pi v\,}{e} 
\[\ln\(\f{m^{~}_{f_2}}{m^{~}_{f_1}}\) +\O\(\f{1}{n}\)\],
\eeq
which is independent of the value of $n$. 
Similarly, the difference
between the unitarity limit $E_f^\star$ for the fermion $f$
and the limit $E^\star_W$ for the scale of EWSB behaves as
\beq
\label{eq:UBW-dif}
E^\star_{f} - E^\star_W ~\approx~ \dis
\f{\,4\pi v\,}{e} \[\ln\(\f{v}{m^{~}_{f}}\) +\O\(\f{1}{n}\)\],
\eeq
and is again independent of \,$n$\, in its asymptotic region.

Finally, in Fig.\,2,
we numerically plot out the bound (\ref{eq:UBus})
under the approximation (\ref{eq:C0}), for
a number of typical processes including 
\,$V_L^{a_1}V_L^{a_2},\,t\bar{t},\,b\bar{b},\,
   \tau^-\tau^+,\,e^-e^+,\,\nL\nL
   \to nV_L^a$\,,\, where we have input the following central   
values for the SM fermion masses\,\cite{PDG},\footnote{Recently 
both D0 and CDF announced their updated 
analyses of the Run-I data for top quark mass 
measurements\,\cite{CDF-D0mt}.
The resulting world average for the top quark pole-mass is
$\,m_t=178.0\pm 4.3\,$GeV~\cite{mt-world}.}
\beq
\label{eq:mf-input}
\ba{rcl}
(m_t,\,m_b,\,m_c,\,m_s,\,m_u,\,m_d) &=&
   (178.0,\,4.85,\,1.65,\,0.105,\,0.003,\,0.006)\,{\rm GeV},
\\[2.5mm]
(m_\tau,\,m_\mu,\,m_e) &=& (1.777,\,0.1057,\,0.000511)\,{\rm GeV}.
\ea
\eeq
For leptons and heavy quarks $(c,b,t)$ the pole-masses are
used, while for light quarks $(u,d,s)$ the current-quark masses 
under $\ov{\rm MS}$ \cite{PDG}
are chosen as the light quark pole-masses
are hard to directly extract. 
Our $2\to n$ analysis shows that for light fermions 
the minimum $\,n=n_s\,$ is sizable
and the unitarity limit is insensitive to the fermion masses,
so a small difference between the pole-mass and  $\ov{\rm MS}$ mass 
for light fermions does not cause a visible effect in Fig.\,2. 
The absolute scale for neutrino masses will be further determined
by various experiments.
As mentioned earlier, the recent 
WMAP data\,\cite{WMAP}, in conjunction with the 
2dF Galaxy Redshift Survey\,\cite{2dF} put a 95\%\,C.L.
upper limit 
\,$\,\dis\sum_j m_{\nu j} \leqq 1.01$\,eV \cite{WMAP-2dF}.
From the WMAP, SDSS and Lyman-$\alpha$ Forest data, a stronger
bound\,\cite{Hannest},  
\,$\,\dis\sum_j m_{\nu j} \leqq 0.65$\,eV
(95\%\,C.L.), may be derived.
Neutrino oscillations measure the
difference of squared mass-eigenvalues and 
the atmospheric data require the larger mass gap
bounded by\,\cite{atm,nu-rev},
\beq
\dis
0.04\,{\rm eV} ~\leqq~ \Delta_{\rm atm}^{\f{1}{2}} ~\leqq~ 
0.06\,{\rm eV},
~~~~~~(99\%\,{\rm C.L.})\,,
\eeq
where \,$\Delta_{\rm atm} = |m^{2}_{\nu_{1,2}}-m_{\nu_3}^2|$\,.\,
This requires the neutrino mass scale to be bounded from below,
i.e.,
~$m_\nu \gtrsim \Delta_{\rm atm}^{1\over 2}\sim 0.05$\,eV,\,
and when the neutrino mass spectrum exhibits a hierarchical structure,
the quantity \,$\Delta_{\rm atm}^{1\over2}$\, actually sets the mass
scale for one or two of the active neutrinos\,\cite{nu-rev,HDN}.
The upcoming laboratory experiments on neutrinoless double 
$\beta$-decay\,\cite{0nu2Ba,0nu2Bb} are indispensable to further pin down
the neutrino mass scale. Without losing generality, we choose
a sample neutrino mass value 
\,$m_\nu \simeq 0.05$\,eV\, for the analyses
in Fig.\,2 (and in the Table\,1 below).

%%%%%%%%%%%%%%%%%%%%%%%  Fig.2  %%%%%%%%%%%%%%%%%%%%%%%%%%%
\begin{figure} [h]
\label{fig:Fig2}
\begin{center}
%\hspace*{-3mm}
\vspace*{-2mm}
\includegraphics[width=16.5cm,height=12.5cm]{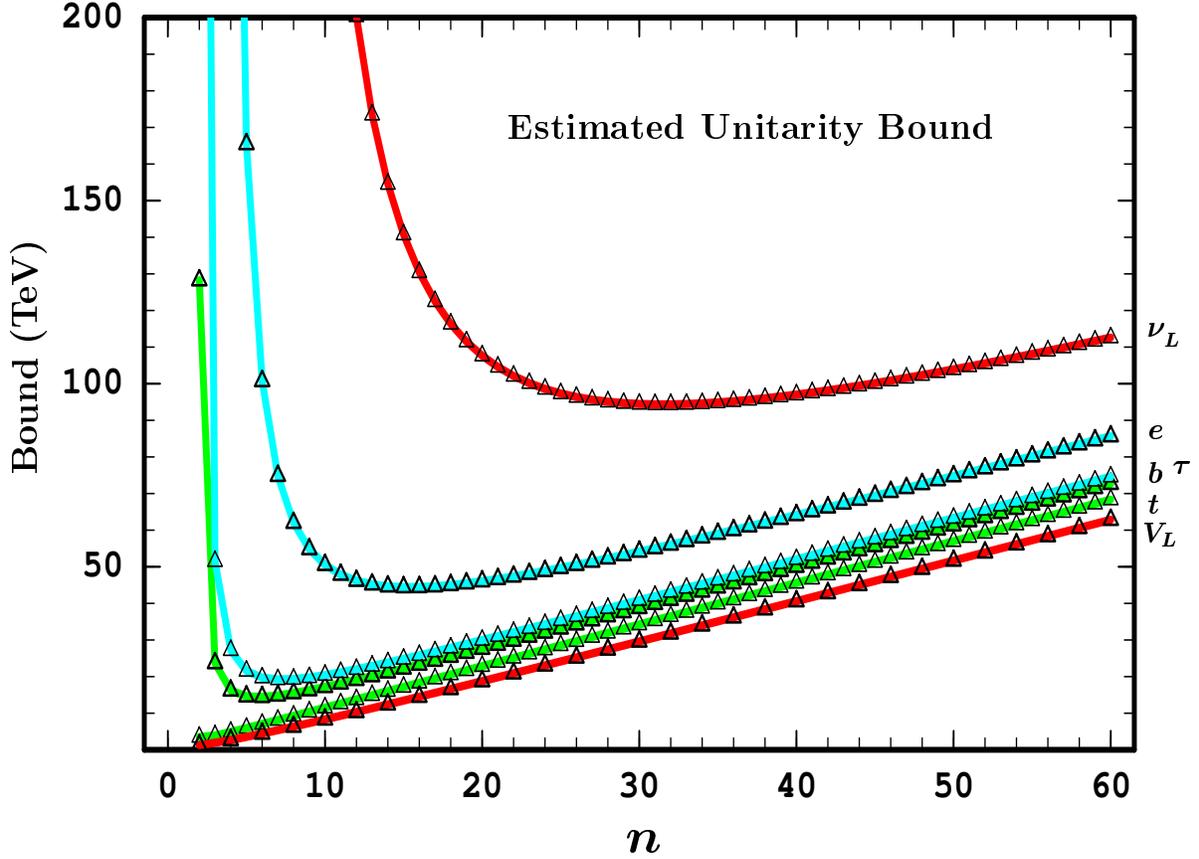} 
\vspace*{-10.5mm}
\caption{Realistic estimate of the unitarity bound $E^\star$,
based upon Eq.\,(\ref{eq:UBus}) with the approximation (\ref{eq:C0}),
for the scattering processes 
\,$V_L^{a_1}V_L^{a_2},\,t\bar{t},\,b\bar{b},\,
   \tau^-\tau^+,\,e^-e^+,\,\nL\nL
   \to nV_L^a$\, (curves from bottom to top),
as a function of \,$n\,(\geqq 2)$\,.\,
Only those points corresponding to integer $n$ have physical
meaning, and on the lowest curve $n$ is restricted to the 
even integers. 
}
\vspace*{-4mm}
\end{center}
\end{figure}

Fig.\,2 shows that the 
scattering processes \,$f\bar{f},\nL\nL \to nV_L^a$\, 
indeed put new unitarity limits on the scale of fermion
mass generation,  {\it independent of the scale $v$,} and in
agreement with the original motivation of 
Appelquist and Chanowitz\,\cite{AC}.  However, strikingly,
our analysis further reveals that for the light fermions including
neutrinos, leptons and quarks,
%(except possibly the top quark),
the strongest bounds generally occur at a value \,$n=n_s > 2$\,.\,
The current improved estimates of 
$\,\(n_s,\,E^{\star}_{\min} \)\,$
are summarized in Table\,1.
Note that these bounds easily satisfy the kinematic constraint 
Eq.\,(\ref{eq:KC}).

%\vspace*{-2mm}
Another feature of Fig.\,2 is that  the curves for the 
quarks and charged leptons
become almost parallel  in the asymptotic region
$\,n\gtrsim 15$\,,\, and the upper curve (for neutrinos) 
also becomes parallel when 
\,$n\gtrsim 35$\,.\,
This fact can be  understood from the asymptotic
formula, 
Eqs.\,(\ref{eq:UBf-dif}) and (\ref{eq:UBW-dif}).
Fig.\,2 further shows that for the scattering
\,$t\bar{t} \to nV_L^a$\, 
the unitarity bounds are very close for \,$n=2,3,4$,\, and it is
easy to check that the minimum of the bound can shift to
\,$n_s=4$\, if we just average over the spin and color for
the  \,$t\bar{t}$\, pair instead of composing a
spin-0 and color-singlet initial state.
The precise calculations in Sections 4,  5, and 6 below
will quantitatively improve the above estimated numbers but, as we
will see, all the qualitative features in Fig.\,2 and
Table\,1  remain.
Finally, it is very useful to compare these estimated bounds 
(Table\,1)
with the customary \,$2\to 2$\, limits (\ref{eq:EWSB-UB22}), 
(\ref{eq:ACx}) and (\ref{eq:nuB22})
[cf. Sec.\,2.3.2] whose predictions are summarized in Table\,2.
Comparing Table\,1 and 2, we see 
that for the scales of EWSB and top mass generation,
our estimates in Table\,1 agree with the customary limits 
in Table\,2;  however, for all the light fermions and 
for the Majorana neutrinos, our bounds are substantially stronger 
than the AC\,\cite{AC} and MNW\,\cite{scott,scott-PRL} 
limits by {\it many orders of magnitude.}

\begin{table}[H]
\label{Tab:Tab1}
\caption{  
The estimated {\it lowest} unitarity bound 
\,$E^\star_{\rm min}$\, 
[derived from the Eqs.\,(\ref{eq:UBus}) and (\ref{eq:C0})]
for the scattering process $\,\xi^{~}_1\xi^{~}_2\to nV_L^a$\, 
and the corresponding number of final state particles 
\,$n=n^{~}_s$\,,\, 
where \,$\xi^{~}_{1,2} = V_L,\,f,\,\nL$\, or their anti-particles.
}
\vspace*{-1mm}
\begin{center}
\begin{tabular}{c||c|cccccc|ccc|c}
\hline\hline
&&&&&&&&&&&\\[-2.5mm]
~$\xi_1^{~}\xi_2^{~}$~      
& ~$V_L^{a_1}V_L^{a_2}$~     & ~$t\ov{t}$  &   
$b\ov{b}$     & $c\ov{c}$     &  $s\ov{s}$ &
$d\bar{d}$    & $u\ov{u}$    ~&~ $\tau^-\tau^+$  & 
$\mu^-\mu^+$  & $e^-e^+$     & ~$\nL\nL$~~        
\\ [1.5mm] 
\hline\hline
&&&&&&&&&&&\\[-2.5mm]
$n_s$ & $2$ & ~$2$ & $6$ & $7$ & $10$ & $12$ & $13$ ~& $7$ & $10$ & 
       $15$ & ~$32$~~   
\\[1.5mm]
\hline
&&&&&&&&&&&\\[-2.5mm]
~$E^\star_{\rm min}$\,(TeV)~ 
& $1.2$ & ~$3.5$ & $14$ & $18$ & $26$ & $35$ & $37$ ~& $19$ & $28$ & 
       $44$ & ~$95$~~   
\\[1.5mm]
\hline\hline
\end{tabular}
\end{center}
\end{table}

\vspace*{-3mm}
\begin{table}[H]
\label{Tab:Tab2}
\caption{  
Predictions of the customary unitarity limits for the
scattering process   
$\,\xi^{~}_1\xi^{~}_2\to V_L^{a_1}V_L^{a_2}$\,,\, 
based upon Eqs.\,(\ref{eq:EWSB-UB22}), 
(\ref{eq:ACx}) and (\ref{eq:nuB22x}). 
}
\vspace*{-6mm}
\begin{center}
\begin{tabular}{c||c|cccccc|ccc|c}
\hline\hline
&&&&&&&&&&&\\[-2.5mm]
   $\xi_1^{~}\xi_2^{~}$      
&  $V_L^{a_1}V_L^{a_2}$       &   $t\ov{t}$  &   
$b\ov{b}$     & $c\ov{c}$     &   $s\ov{s}$ &
$d\bar{d}$    & $u\ov{u}$     &   $\tau^-\tau^+$  & 
$\mu^-\mu^+$  & $e^-e^+$      &   $\nL\nL$     
\\ [1.5mm] 
\hline
&&&&&&&&&&&\\[-2.5mm]
$E^\star_{2\to 2}$\,(TeV)
& $1.2$  & $3.5$         & $128$ & $377$ & $6\!\!\times\!\!10^3$ 
& $10^5$ & $2\!\!\times\!\!10^5$ & $606$ & $10^4$ 
& $2\!\!\times\!\! 10^6$ & $1.1\!\!\times\!\!10^{13}$   
\\[1.5mm]
\hline\hline
\end{tabular}
\end{center}
\end{table}

\newpage
%\vspace*{7mm}
\section{\hspace*{-6mm}.$\!$
Scales of Mass Generation for Quarks and Leptons}
\vspace*{4mm}

In this section we will attempt to improve on our 
new limits on the scales of fermion mass generation
by computing the matrix elements more precisely.
A parallel treatment of Majorana neutrinos will be 
given in Sec.\,5.
In Sec.\,4.1, we first 
systematically derive the precise unitarity limits for
various  \,$f\bar f \to n\pi^a\,(nV_L^a)$\, scatterings with quarks/leptons
in the initial state,  which result in optimal
upper bounds on the scales of the fermion mass generation. 
Then, we compare these quantitative limits with our estimates 
given in the previous section (Table\,1 and Fig.\,2).
In Sec.\,4.2 we include EWSB effects among the $\pi^a$
[cf. Fig.\,1(c-d)].
We also comment on fermion mass generation with dynamical EWSB, 
where the $V_L^a$ and
its would-be Goldstone boson $\pi^a$ are composite fields.

\vspace*{4mm}
\subsection{\hspace*{-5mm}.$\!$
Unitarity Bound on the Mass Generation Scale from
${\mbox {\boldmath  $f\bar{f} \to n\pi^a$}}$
}
\vspace*{2mm}

From the nonlinear fermion mass term (\ref{eq:f-mass}),
we deduce the following fermion-Goldstone interaction
Lagrangian,
\beq
\vspace*{-2mm}
\label{eq:Lf-int}
\dis
{\cal L}_f^{\rm int} ~=~
-\sum_{n=1}^\infty \f{i^n}{v^n n!}
\left\{
\ba{ll}
\left|\vec{\pi}\right|^n\(m^{~}_f\bar{f}f+
                          m^{~}_{f'}\ov{f'}f'\),
& ~~(n={\rm even})\,,
\\[6mm]
\left|\vec{\pi}\right|^{n-1}\[
\pi^0\(m^{~}_f\ov{f}\gamma_5^{~}f
      -m^{~}_{f'}\ov{f'}\gamma_5^{~}f'\)
\right.
&
\\[3mm]
\left. \hspace*{8mm}
~~+\sqrt{2}\pi^-
\ov{f'}\( m_f^{~}P_L - m^{~}_{f'}P_R\)f + {\rm H.c.}
\],
& ~~(n={\rm odd})\,,
\ea
\right.
\eeq
where $~|\vec{\pi}|=\[2\pi^+\pi^-+\pi^0\pi^0\]^{1\over 2}\,$\,
and $\,P_{L(R)}=\f{1}{2}(1\mp\gamma^5)$\,.\,
This leads to three types of $2\to n$ processes,
containing the leading contact diagrams shown in
Fig.\,1(a),
\beq
\label{eq:abc}
\ba{lll}
({\mathrm a}). & \dis
f\ov{f},\,f'\ov{f'} ~\to~ 
\(\pi^+\)^{\l} \(\pi^-\)^{\l} \(\pi^0\)^{n-2\l} \,,
& ~~(n={\rm even})\,,
\\[2mm]
({\mathrm b}). & \dis
f\ov{f},\,f'\ov{f'} ~\to~ 
\(\pi^+\)^{\l} \(\pi^-\)^{\l} \(\pi^0\)^{n-2\l} \,,
& ~~(n={\rm odd})\,,
\\[2mm]
({\mathrm c}). & \dis
f\ov{f'} ~\to~ 
\(\pi^+\)^{\l+1} \(\pi^-\)^{\l} \(\pi^0\)^{n-2\l-1} \,,
& ~~(n={\rm odd})\,,
\ea
\eeq
which may be generically denoted as
~$f^a\ov{f^b}\to 
\(\pi^+\)^{k} \(\pi^-\)^{\l} \(\pi^0\)^{n-k-\l}
$\,,\, where $\,f^a,f^b\in \(f,\,f'\)$ as appropriate.
The helicity amplitudes 
$~\TT\[f^a_{\pm}\ov{f^b}_{\pm};\,n,k,\l\]$~
are given by
\beq
\label{eq:Amp-all}
\vspace*{-5mm}
\ba{lll}
\hspace*{-3mm}
n={\rm even}\!:~\, & \!\!\!
\TT\!\[f_{\pm}\ov{f}_{\pm}^{~};\, n,\l,\l\]\! =
\dis\mp i^{n}{\cal C}_1\f{\,m_f^{~}\sqrt{s}\,}{v^n}  \,,
&~ \TT\!\[f'_{\pm}\ov{f}_{\!\pm}'; \,n,\l,\l\]\! =
\dis\mp i^{n}{\cal C}_1\f{\,m_{f'}^{~}\sqrt{s}\,}{v^n} \,;
\\[3mm]
\hspace*{-3mm}
n={\rm odd}\!:~\,  &\!\!\!
\TT\!\[f_{\pm}\ov{f}_{\pm}^{~};\, n,\l,\l\]\! =
\dis -i^{n}{\cal C}_2\f{\,m_f^{~}\sqrt{s}\,}{v^n} \,,
&~ \TT\!\[f'_{\pm}\ov{f}_{\!\pm}';\, n,\l,\l\]\! =
\dis +i^{n}{\cal C}_2\f{\,m_{f'}^{~}\sqrt{s}\,}{v^n} \,,
\\[4mm]
& \!\!\!
\TT\!\[f_{\pm}\ov{f}_{\!\pm}';\, n,\l+1,\l\]\! =
\dis -i^{n}{\cal C}_3
\!\(\!\!\!\ba{l} m_f^{~} \\ m_{f'}^{~}\ea\!\!\!\)\!
\f{\sqrt{s}\,}{\,v^n} \,,
&~\, \TT\!\[f'_{\pm}\ov{f}_{\pm};\, n,\l,\l+1\]\! =
\dis -i^{n}{\cal C}_3
\!\(\!\!\!\ba{l} m_{f'}^{~} 
               \\ m_{f}^{~}\ea\!\!\!\)\!
\f{\sqrt{s}\,}{\,v^n} \,;
%\\[4mm]
\ea
\eeq
where the coefficients $\,\(\C_1,\C_2,\C_3\)$\, are defined as
\beq
\ba{lcl}
\C_1 & =& 
\dis\f{1}{\,n!\,}\,C^\ell_{n\over2} 2^\ell \(\ell!\)^2\(n-2\ell\)! \,,
\\[5mm]
\C_2 & =& 
\dis\f{1}{\,n!\,}\,C^\ell_{{n-1}\over2} 2^\ell \(\ell!\)^2\(n-2\ell\)! \,,
\\[5mm]
\C_3 & =&  
\dis\f{\,\sqrt{2}~}{n!}C^\ell_{\f{n-1}{2}} 2^\ell\, 
       \ell!\,\(\ell+1\)!\,\(n-2\ell-1\)! \,,
\ea
\eeq
and $~C_n^\ell \equiv \dis\f{n!}{\,\ell!\,(n-\ell)!\,}\,$.\,
Now we define a color-singlet initial state with a proper helicity
combination,
\beq
\label{eq:in}
\dis
|{\rm in}\rangle ~=~
\left\{
\ba{l}
\dis\f{1}{\,\sqrt{2N_c}\,}
\sum_{\beta =1}^{N_c}
\[\left|f^{a\beta}_+ \bar{f}^{a\beta}_+ \right\rangle  
 \mp
  \left|f^{a\beta}_- \bar{f}^{a\beta}_- \right\rangle  
\] ,
\\[5mm]
\dis\f{1}{\,\sqrt{N_c}\,}
\sum_{\beta =1}^{N_c}
\left|f^{a\beta}_+ \bar{f}^{b\beta}_+ \right\rangle  ,
~~~~~(a\neq b)\,,
\ea
\right.
\eeq
where on the right-hand side
\,$N_c=3\,(1)$\, for quarks (leptons), and for the first definition
the sign $-(+)$ between the two helicity states corresponds to 
the \,$n={\rm even}\,({\rm odd})$\, case for an initial state
$|{\rm in}\rangle $ of spin-0(1).
With these we compute the cross section for
$~|{\rm in}\rangle \to \(\pi^+\)^{k} \(\pi^-\)^{\ell} 
  \(\pi^0\)^{n-k-\ell}$~
as
\beq
\label{eq:ffnsigma}
\ba{rl}
\sigma[{\rm in};\,n,k,\l]  &=~ \dis
\f{C_{0j}\(2^x N_c\)
}{~2^{4(n-1)}\pi^{2n-3}\,(n-1)!(n-2)!~}
\f{1}{\,s\,}
\(\f{\,\sqrt{s}\,}{v}\)^{2(n-1)}
\(\f{\,m^{~}_{\widehat{f}}\,}{v}\)^{2}
\,,
\\[5mm]
C_{0j} &\equiv ~ \dis\f{\,\C_j^2\,}{\varrho} ~=~ 
           \f{\C_j^2}{~k!\,\l!\,(n-k-\l)!~} \,,
\ea
\eeq
where \,$\widehat{f} \in (f,\,f')$,\, and \,$x=1\,(0)$\,
for the first (second) definition of the initial state
in Eq.\,(\ref{eq:in}).\,
For the three processes in Eq.\,(\ref{eq:abc}), 
the maximal values of $\,C_{0j}$\, can be derived as
\beq
\label{eq:C0j-max}
\dis
C_{01}^{\max} =\f{~ 2^n\(\f{n}{2}!\)^2~}{\(n!\)^2} \,,
~~~~
C_{02}^{\max} =\f{1}{\,n!\,}\,,
~~~~
C_{03}^{\max} =\f{~ 2^n\(\f{n-1}{2}\)!\(\f{n+1}{2}\)!~}
                 {(n!)^2} \,,
\eeq
corresponding to 
\,$k=\ell=\dis\f{n}{2}$,\,  $k=\ell =0$,\, and
\,$k=\ell+1=\dis\f{n+1}{2}$,\,
respectively.
Thus, for each given type of fermion, 
using the general unitarity condition (\ref{eq:UC-IEn}),
we arrive at the bound $~\sqrt{s}\leqq E_f^\star$~
with
\beqa
\label{eq:UBf-final}
E^\star_f &=& v\[\(\f{v}{m_{\widehat{f}}^{~}}\)^2
               \f{4\pi}{\,(2^xN_c)\,\R_j^{\max}\,}
             \]^{\f{1}{2(n-1)}}  .
\eeqa
$~\R_j^{\max}$~ is given by
\beq
\label{eq:Rmaxj}
\dis
\R_j^{\max} ~=~
\f{C_{0j}^{\max}}{~2^{4(n-1)}\pi^{2n-3}(n-1)!\,(n-2)!~} \,, 
\eeq
or, using (\ref{eq:C0j-max}),
\beq
\ba{lcll}
\R_1^{\max} &=&
\dis\f{\(\f{n}{2}!\)^2
}{~2^{3n-4}\pi^{2n-3}\,(n!)^2\,(n-1)!\,(n-2)!~} \,,
&~~~~~(n={\rm even})\,,
\\[5mm]
\R_2^{\max} &=&
\dis\f{1}{~2^{4(n-1)}\pi^{2n-3}\,n!\,(n-1)!\,(n-2)!~} \,,
&~~~~~(n={\rm odd})\,,
\\[6mm]
\R_3^{\max} &=&
\dis\f{\(\f{n-1}{2}\)!\(\f{n+1}{2}\)!}
{~2^{3n-4}\pi^{2n-3}\,(n!)^2(n-1)!\,(n-2)!~} \,,
&~~~~~(n={\rm odd})\,.
\ea
\eeq
where $\,\R_{1,2,3}^{\max}$\,
have $\,x=1,\,1,\,0\,$  and
correspond to the processes (a), (b) and (c) in 
Eq.\,(\ref{eq:abc}), respectively.

%\newpage
%\vspace{0mm}
\begin{table}[H]
\label{Tab:Tab3}
\caption{  
The precise minimum unitarity bound $E^\star_{\rm min}$ 
[derived from Eq.\,(\ref{eq:UBf-final})]
for the scattering process 
$\,\xi^{~}_1\xi^{~}_2\to n\pi^a\,(nV_L^a)$\, 
and the corresponding number of final state particles 
\,$n=n^{~}_s$\,,\, 
where \,$\xi^{~}_{1,2} = f$\,or\,$\bar f$\,.
}
\vspace*{-1mm}
\begin{center}
\begin{tabular}{c||cccccc|ccc}
\hline\hline
&&&&&&&&&\\[-2.5mm]
~$\xi_1^{~}\xi_2^{~}$~      
& ~$t\ov{t}$  &   
$b\ov{b}$     & $c\ov{c}$     &  $s\ov{s}$ &
$d\bar{d}$    & $u\ov{u}$    ~&~ $\tau^-\tau^+$  & 
$\mu^-\mu^+$  & $e^-e^+$~~        
\\ [1.5mm] 
\hline\hline
&&&&&&&&&\\[-2.5mm]
$n_s$ & ~$2$ & $4$ & $6$ & $8$ & $10$ & $10$ ~& $6$ & $8$ & $12$~~   
\\[1.5mm]
\hline
&&&&&&&&&\\[-2.5mm]
~$E^\star_{\rm min}$\,(TeV)~ 
& ~$3.49$ & $23.4$ & $30.8$ & $52.1$ & $77.4$ & $83.6$~
&  $33.9$ & $56.3$ & $107$~~   
\\[1.5mm]
\hline\hline
\end{tabular}
\end{center}
\vspace*{-2mm}
\end{table}

Finally, using the bound (\ref{eq:UBf-final}), we derive the 
lowest bound for each scattering and the corresponding number
of final state particles \,$n=n_s$\,,\, as shown in Table\,3.
This is to be compared with our estimated
bounds in Table\,1  and the customary limits in Table\,2.
We see that the precise bounds agree well with our estimate for the
case of a heavy top quark, but significantly improve 
the estimates in Table\,1
for all other light fermions; the limits in Table\,3
become about a factor of \,$1.5-2$\, higher.
The shape of each bound as a function of $\,n\,$
is shown in Fig.\,3 for all SM quarks and leptons,  
where  $\,n_s$\, corresponds to the minimum of each curve.
The values of $n_s$ are visibly smaller than
those in Table\,1 and Fig.\,2.
Once $\,n>n_s\,$, the curves also
grow much faster than Fig.\,2, but still exhibit very
similar slopes for \,$n\gtrsim 15$\,.\,
This is because for the large \,$n$\, the power factor
\,$\(v/m_f\)^{\f{1}{n-1}}$\, is essentually unity 
and the dimensionless phase-space
factor [cf. Eqs.\,(\ref{eq:stirling}) and (\ref{eq:Rmaxj})]
becomes dominant.  
Using the Stirling formula, we find that the three functions
$\,\R_{1,2,3}^{\max}\,$ in Eq.\,(\ref{eq:Rmaxj})
exhibit the same asymptotic behavior
\beq
\label{eq:Rj-asympt}
\dis\R_{1,2,3}^{\max}
~\,\approx~\, \dis 
\[\f{e^{\f{3}{2}}_{~}}{~4\pi n^{\f{3}{2}}_{~}~}\]^{2n} \,,
~~~~~~(\,{\rm for}~ n\gg 1\,)\,,
\eeq
so that the final unitarity limit (\ref{eq:UBf-final}) behaves
as
\beqa
\label{eq:UBQL-asympt}
%\vspace*{-8mm}
%\ba{ll}
\dis
\(E_f^\star\)_{\rm asym}  &
\approx& \dis v\(\f{v}{m^{~}_{\widehat{f}}}\)^{\f{1}{n-1}}
\[\f{~4\pi~}{e_{~}^{3\over2}}\,n^{3\over2}\],
~~~~~~~~(\,{\rm for}~ n\gg 1\,)
\\[2mm]
& 
\dis =&
\f{~4\pi v~}{e_{~}^{3\over2}}\[
n_{~}^{3\over 2} + n_{~}^{1\over 2} \ln\(\f{v}{m^{~}_{\widehat{f}}}\)
+\f{1}{2}n_{~}^{-\f{1}{2}}\ln^2\(\f{v}{m^{~}_{\widehat{f}}}\)
+\O\(n_{~}^{-\f{3}{2}}\)
\],
\label{eq:UBQL-asympt2}
%\ea
\eeqa
where we approximate 
$\,\dis\O(1)^{\f{1}{2(n-1)}} \approx 1\,$ for \,$n\gg 1$\,.\,
We see that unlike Eq.\,(\ref{eq:UBasymt}), the leading term
in the above bound has $\,n^{3\over2}$\, (rather than
$\,n^1$) power dependence. This is why the curves in Fig.\,3
grow much faster than those in Fig.\,2 in the asymptotic 
region of \,$n$\,.\,
Furthermore, we note that the second term in 
the expansion (\ref{eq:UBQL-asympt2}) depends on both 
\,$n_{~}^{\f{1}{2}}$\, and 
\,$m^{~}_{\widehat{f}}$\,,\, contrary to (\ref{eq:UBasymt}).
But, the third term in the expansion of (\ref{eq:UBQL-asympt2}) 
is only suppressed by \,$n_{~}^{-\f{1}{2}}$\, (not $\dis n^{-1}$)
and thus can be sizable due to the squared logarithm.
For instance, comparing the limits for $e^-e^+$ and $t\bar{t}$
scatterings and using Eq.\,(\ref{eq:UBQL-asympt}),
we obtain, numerically,
\beq
\dis
\(E^\star_e - E^\star_t\)_{\rm asym} ~\approx~
\left\{
\ba{ll}
61.1\,{\rm TeV} \,,~~~ & (n=15)\,,
\\[4mm]
63.4\,{\rm TeV} \,,~~~ & (n=30)\,,
\ea
\right.
\eeq
which  explains why in Fig.\,3 the 
curves are almost parallel
for the displayed asymptotic region $\,15\lesssim n \leqq 30$.\,
Finally, the charm quark has a slightly stronger bound than
the tau lepton despite \,$\,m_c<m_\tau$\, (cf. Fig.\,3 and Table\,3)
because the color factor \,$N_c=3$\,
in (\ref{eq:UBf-final}) helps to
lower the bound for charm.

%\newpage
%%%%%%%%%%%%%%%%%%%%%%%%   Fig.3  %%%%%%%%%%%%%%%%%%%%%%%%%%%%%
\begin{figure}[H]
\label{fig:Fig3}
\vspace*{-5mm}
\begin{center}
\hspace*{-5.5mm}
\includegraphics[width=17cm,height=12.5cm]{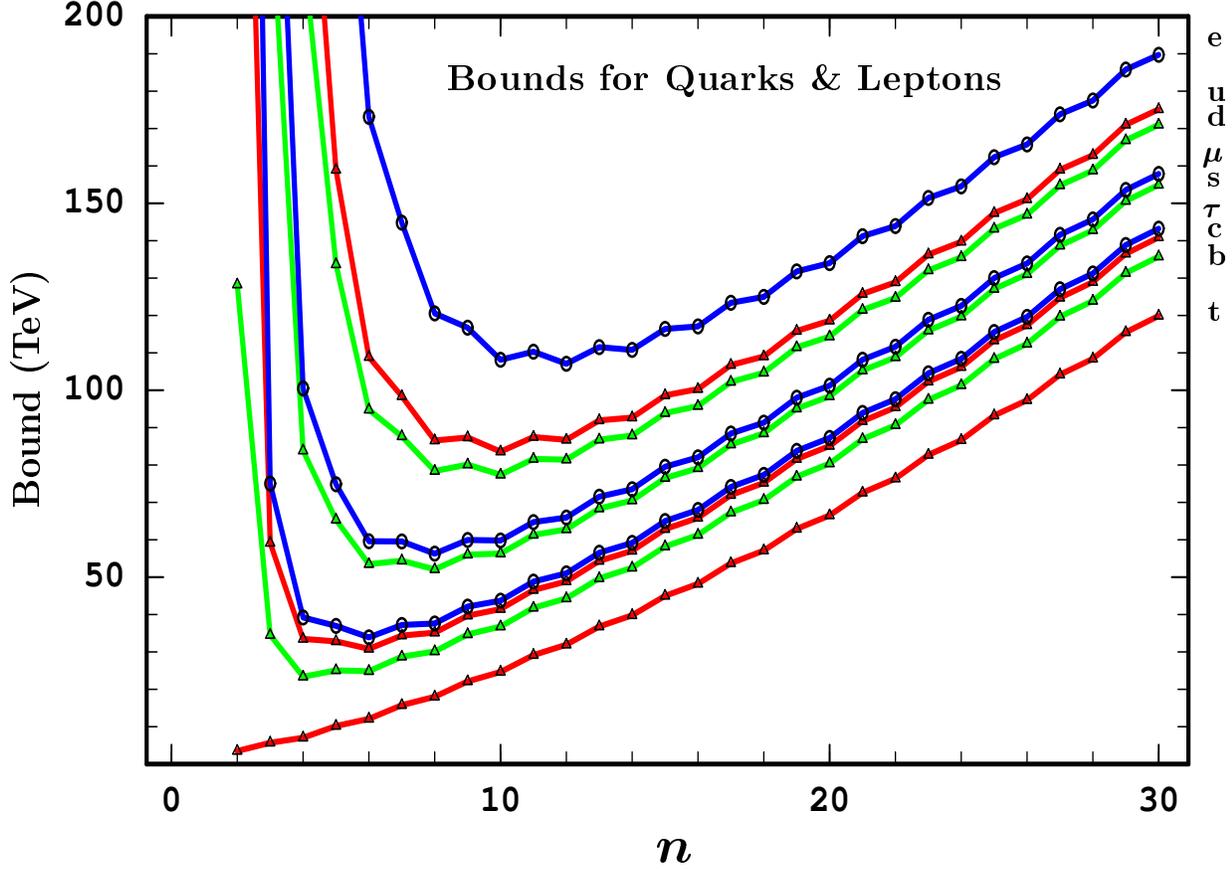} 
%\hspace*{5mm} 
\vspace*{-11mm}
\caption{Precise unitarity bound $E^\star_f$
based upon Eq.\,(\ref{eq:UBf-final}), for the scattering
processes 
$\,t\bar{t},\,b\bar{b},\,c\bar{c},\,\tau^-\tau^+,\,
   s\bar{s},\,\mu^-\mu^+,\,d\bar{d},\,u\bar{u},\,e^-e^+
\to n\pi^a~(nV_L^a)$\, (curves from bottom to top),
as a function of \,$n\,(\,\geqq 2\,)$\,.\, 
For the points with odd $n$ values, the best bounds are actually
given by $\,t_+\bar{b}_+,\,t_-\bar{b}_-,\,c_+\bar{s}_+,\,
\tau_-^+\nu_-,\,
   c_-\bar{s}_-,\,\mu_-^+\nu_-,\,u_-\bar{d}_-,\,
   u_+\bar{d}_+,\,e^+_-\nu_-
\to n\pi^a$\, (curves from bottom to top), respectively.
Only integer values of \,$n$\, have physical meaning.
The curves marked by circles represent bounds for leptons and
those marked by triangles are for quarks; also, for clarity, 
the up-type, down-type quarks and the leptons are plotted in 
three different colors.}
\end{center}
\end{figure}
%%%%%%%%%%%%%%%%%%%%%%%%%%%%%%%%%%%%%%%%%%%%%%%%%%%%%%%%%%%%%%%%%%%
%\vspace*{-3mm}

\vspace*{5mm}
\subsection{\hspace*{-5mm}.$\!$
Model-Dependent Effects from EWSB Sector}
\vspace*{3mm}

As shown in Fig.\,1(c,d),  for \,$n\geqq 3$\,
the tree-level scattering process
\,$f\bar{f}\to n\pi^a$\, 
can have additional contributions involving the 
EWSB sector, i.e., the Goldstone boson self-interaction
vertices and the exchange of the new quanta that unitarize
the EWSB sector (such as the SM Higgs boson).
These contributions depend on the details of the EWSB mechanism
and are thus highly model-dependent. In the following, we
will take the simplest EWSB sector of the SM as an explicit
example and examine how our universal estimates of the 
unitarity bounds [based upon the pure contact interactions
among fermions and Goldstone bosons in Fig.\,1(a)]
are affected by the EWSB contributions via 
Fig.\,1(c,d).  The diagrams in Fig.\,1(a)
and (c,d) are of the same $\O(E^1)$ by power counting, so that 
a contribution of type-(c,d) may affect the amplitude
by a factor of \,$\O(2-3)$\, and thus the total cross section
by a factor of \,$[\O(2-3)]^2 \leqq \O(10)$\,.\,
Thus we expect the unitarity bound of 
Sec.\,4.1 to be changed by at most 
a factor of \,$[\O(2-3)]^{\f{1}{(n-1)}}$\, ($n\geqq 3$)
since the $n$-body phase space integration contributes
an energy factor of \,$s^{n-2}$\, (cf. Sec.\,3.1-3.2).
Numerically the factor 
\,$[\O(2-3)]^{\f{1}{(n-1)}}$ is at most $1.7$\, (for $n=3$),
and becomes  close to unity as $n$ increases above $3$. 
This means that, fortunately, 
%except for the top-quark,
the effect of the EWSB contribution is small
for light fermions since their strongest unitarity bounds
occur at  significantly large $n$ values (cf. Fig.\,2-3).
The effect for top-quark case is interesting since the
analysis based on contact contribution alone shows that
the unitarity bound is strongest for $n_s=2$ and
only becomes slightly weaker for $n_s=3$. Hence, adding the
EWSB contribution associated with \,$t\bar{t}\to n\pi^a$\,
($n\geqq 3$) processes could push the location of the 
minimum  unitarity limit up to $n_s=3$. 
As explicitly shown below, this
is indeed the case, but the improvement in the unitarity
bound for all fermions (including the top-quark) 
is always less than a factor of 2, as expected.

For convenience  we first consider the SM with
its EWSB sector (Higgs sector) nonlinearly realized.
The Higgs-Goldstone fields of the SM can then be formulated 
as $2\times2$ matrix,
\beq
\label{eq:nonlinearH}
\Phi = (v+h^0)U\,,~~~~~ U=\exp[i\pi^a\tau^a/v]\,,
\eeq
where the physical Higgs field $h^0$ transforms as a singlet
under $SU(2)_L\otimes U(1)_Y$.
Correspondingly, the EWSB Lagrangian of the SM takes the form,
\beq
\label{eq:SM-EWSB-L}
{\cal L}_{\rm EWSB} ~=~ 
{\cal L}^{\rm kin}_\Phi - V(h) ~=~
\f{1}{4}{\rm Tr}[(D^\mu\Phi)(D_\mu\Phi)^\dag]
-\f{\lambda}{4}\[(v+h^0)^2-v^2\]^2 \,,
\eeq
where \,$\,D_\mu\Phi = (\partial_\mu h^0)U+(v+h^0)D_\mu U\,$
and \,$D_\mu U$\, is defined below Eq.\,(\ref{eq:V-mass}).
The kinematic term $\,{\cal L}^{\rm kin}_\Phi$\, can be
simplified as 
\beqa
{\cal L}^{\rm kin}_\Phi &=&
\f{1}{2}(\partial_\mu h^0)(\partial^\mu h^0)
+\f{1}{4}\(v^2+2vh^0+{h^0}^2\){\rm Tr}[(D^\mu U)(D_\mu U)^\dag]\,.
\eeqa
The part 
$\,\dis\f{v^2}{4}{\rm Tr}[(D^\mu U)(D_\mu U)^\dag]\,$
is the same as Eq.\,(\ref{eq:V-mass}) and contains the
following pure Goldstone Lagrangian,
\beq
\label{eq:L-GBn}
\ba{ll}
{\cal L}_{\rm GB} 
& \!\! =~ \dis\f{v^2}{4}{\rm Tr}[(\partial^\mu U)(\partial_\mu U)^\dag]
\\[3mm]
& \!\! =~
\dis\f{1}{2}\partial^\mu\overrightarrow{\pi}\cdot
                    \partial_\mu\overrightarrow{\pi}
+\!\sum^\infty_{n({\rm even})=4}
\dis\f{(-)^{n\over 2}2^{n-2}}{n!\,v^{n-2}}
\(\overrightarrow{\pi}\cdot
  \overrightarrow{\pi}\)_{~}^{\f{n}{2}-2}
\[\(\overrightarrow{\pi}\cdot
    \partial_\mu\overrightarrow{\pi}\)_{~}^2 -
\(\overrightarrow{\pi}\cdot
  \overrightarrow{\pi}\)
\(\partial^\mu\overrightarrow{\pi}\cdot
  \partial_\mu\overrightarrow{\pi}\)
\],
\ea
\eeq
where 
\beq
\label{eq:L-GBnx}
\ba{l}
\overrightarrow{\pi}\cdot\overrightarrow{\pi}
   =2\pi^+\pi^- \!+\pi^0\pi^0 \,,
\\[3mm]
\(\overrightarrow{\pi}\cdot\partial_\mu
    \overrightarrow{\pi}\)_{~}^2 -
\(\overrightarrow{\pi}\cdot
  \overrightarrow{\pi}\)
\(\partial^\mu\overrightarrow{\pi}\cdot
  \partial_\mu\overrightarrow{\pi}\)
\\[2mm]
\hspace*{12mm}
=\(\pi^+\partial_\mu\pi^-\!+\pi^-\partial_\mu\pi^+\!+
   \pi^0\partial_\mu\pi^0\)^2   -
   \(2\pi^+\pi^- \!+{\pi^0}^2\)
   \(2\partial^\mu\pi^+\partial_\mu\pi^-\!+
      \partial^\mu\pi^0\partial_\mu\pi^0\).
\ea
\eeq 
The next crucial step in the SM 
is to assume that the physical Higgs boson 
$h^0$, in addition to breaking the electroweak gauge symmetry, 
also couples to the fermions, 
generating all the fermion masses via the Yukawa interactions,
whose strength is given by ratio of the fermion masses $m_{f,f'}$ 
over the VEV $v$, 
\beq
\label{eq:YukawaC}
\dis
y_f^{~} ~=~ \sqrt{2}\,m_f^{~}/v\,,~~~~~~~~
y_{f'}^{~} ~=~ \sqrt{2}\,m_{f'}^{~}/v\,.
\eeq 
Thus the bare fermionic mass-term \,$- m_f\ov{f}f - m_{f'}\ov{f'}f'$\,
will be generated from the following gauge-invariant Yukawa Lagrangian
of the nonlinearly realized SM,
\beq
\label{eq:LY-NL}
{\cal L}^{\rm NL}_{\rm Y} ~=~ \dis
-m_f^{~}\(1+\f{h^0}{v}\)\ov{F_L}U\(\ba{c} 1 \\ 0 \ea\)f_R
-m_{f'}^{~}\(1+\f{h^0}{v}\)
\ov{F_L}U\(\ba{c} 0 \\ 1 \ea\)f'_R + {\rm H.c.} 
\eeq
The expanded interaction Lagrangian of (\ref{eq:LY-NL}) can be
directly obtained from Eq.\,(\ref{eq:Lf-int}) 
by adding an {\it overall factor} \,$\(1+{h^0}/{v}\)$\,.\,

The Feynman rules of this nonlinearly realized SM given by
Eqs.\,(\ref{eq:SM-EWSB-L})  and   (\ref{eq:LY-NL}) are very
different from the usual linearly realized SM. But, as they both
share the {\it same} unitary-gauge Lagrangian, they are actually
equivalent and thus both renormalizable. In particular, we note 
that the above nonlinearly realized Higgs boson 
couples to the Goldstone fields  with two derivatives,
similar to the Goldstone self-interaction vertices. This is
crucial for maintaining the unitarity of the Goldstone boson
scatterings \,$\pi^{a_1}\pi^{a_2}\to n\pi^a$\, and the
fermion-anti-fermion scatterings 
\,$f\bar{f}\to n\pi^a$\, in the nonlinear SM.
For instance, in the latter case, 
the energy power counting shows that
the contributions of the EWSB sector in Fig.\,1(c,d)
are of the same $\O(E^1)$ as that of the universal
contact contributions in Fig.\,1(a). Hence, the unitarity of the
SM will force all $\O(E^1)$ terms in Fig.\,1(a,c,d) and  Fig.\,1(e)
to at least cancel down to $\O(E^0)$, i.e.,
${\rm Fig}.\,1(a)+(c)+(d)+(e) = \O(E^0)$, or,
\beq
\label{eq:fig1sum}
{\rm Fig}.\,1(a)+(c)+(d) ~=~ -\, {\rm Fig}.\,1(e)+\O(E^0) 
                         ~=~ \O(E^1) ~.
\eeq

Now we will explicitly analyze the unitarity violation for the scattering 
\,$f\ov{f}, f\ov{f'}\to 3\pi^a$\, by including
the EWSB contributions of the SM [cf. Fig.\,1(c$+$d)], 
but {\it without assuming} the Yukawa interactions of 
the Higgs boson $h^0$ for fermion mass generation [cf. Fig.\,1(e)].
This means that we will only include leading contributions from 
Fig.\,1(a)$+$(c)$+$(d) at $\O(E^1)$ [cf. Eq.\,(\ref{eq:fig1sum})].
Then, we will explicitly show how the contributions from 
a given type of EWSB mechanism such as that in the SM [cf. Fig.\,1(c,d)]
could affect the final unitarity bound. We choose the
process \,$f\ov{f}, f\ov{f'}\to n\pi^a$\, with \,$n=3$\, not only
because \,$n=3$\, is the simplest nontrivial case where the
EWSB effect first shows up, but also because the 
\,$n=3$\, case is expected
to have {\it the largest EWSB effect} on the unitarity limit,
as explained earlier in this subsection.

%%%%%%%%%%%%%%%%%%%%%%%   Fig.4  %%%%%%%%%%%%%%%%%%%%%%%%%%%%%%%%
\vspace*{-5mm}
\begin{figure}[H]
\label{fig:ff-3pi-ppm}
\begin{center}
\hspace*{-10.5mm}
%\vspace*{-1cm}
\includegraphics[width=18cm,height=11.6cm]{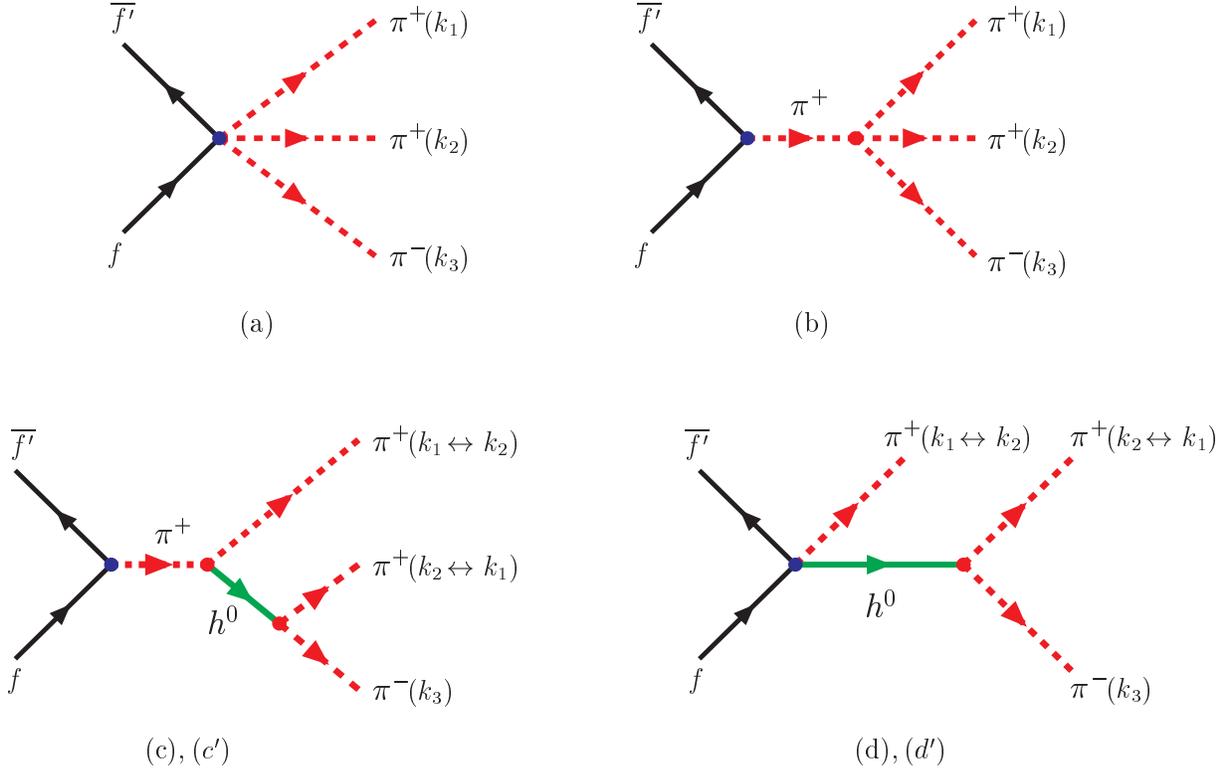}
\vspace*{-7mm}
\caption{Leading Feynman diagrams of 
\,$\O(E^1)$\, for the scattering 
\,$f\ov{f'}\to \pi^+\pi^+\pi^-$\, in the nonlinearly realized SM.
(a) Model-independent universal ``contact'' contribution;
(b) $s$-channel contribution due to the Goldstone boson
     self-interactions from the SM EWSB sector;
(c,c$'$) $s$-channel contributions due to Higgs-Goldstone boson 
     interactions from the SM EWSB sector;
(d,d$'$) Yukawa-type contributions due to the {\it assumption} 
     that the {\it same} SM Higgs boson $h^0$ is responsible for the
     fermion mass generation.
}
\vspace*{-3mm}
\end{center}
\end{figure}
%%%%%%%%%%%%%%%%%%%%%%%%%%%%%%%%%%%%%%%%%%%%%%%%%%%%%%%%%%%%%%%%%%%%

For the $n=3$ case, the best unitarity limit comes from the
scattering \,$f\ov{f'}\to \pi^+\pi^+\pi^-$\,.\, 
The relevant diagrams are explicitly shown in Fig.\,4.
We see that Fig.\,4(a) is the model-independent universal ``contact'' 
contribution (studied in Sec. 4.1);
Fig.\,4(b) is the $s$-channel contribution 
from the Goldstone self-interactions
     in the SM EWSB sector;
Fig.\,4(c,c$'$) are 
the $s$-channel contributions from Higgs-Goldstone interactions
     in the SM EWSB sector;
Fig.\,4(d,d$'$) are the 
Yukawa-type contributions due to the {\it assumption} that the 
{\it same} SM Higgs boson $h^0$ is responsible for 
fermion mass generation.
We then compute the helicity amplitudes for all
the individual contributions in Fig.\,4,
at the leading \,$\O(E^1)$\,,
\beq
\label{eq:Amp-ppm}
\ba{ll}
i\T\[f_\pm\ov{f'}_\pm\to \pi^+\pi^+\pi^-\](a) \hspace*{-2mm}
&=
\dis\f{\sq\,}{v^3}\!\(\!\!\!\ba{c} m_f^{~} \\[2mm] m_{f'}^{~} \ea\!\!\!\!\)
    \!\!\[-\f{2}{3}\]\!\sqrt{s} \,,
\\[6mm]
i\T\[f_\pm\ov{f'}_\pm\to \pi^+\pi^+\pi^-\](b) \hspace*{-2mm}
&=
\dis\f{\sq\,}{v^3}\!\(\!\!\!\ba{c} m_f^{~} \\[2mm] m_{f'}^{~} \ea\!\!\!\!\)
    \!\!\[-\f{1}{3}+\f{\,\widehat{s}_{12}\,}{s}\]\!\sqrt{s} \,,
\\[6mm]
i\T\[f_\pm\ov{f'}_\pm\to \pi^+\pi^+\pi^-\](c+c') \hspace*{-2mm}
&=
\dis\f{\sq\,}{v^3}\!\(\!\!\!\ba{c} m_f^{~} \\[2mm] m_{f'}^{~} \ea\!\!\!\!\)
    \!\!\[\(\!\f{\,\widehat{s}_{23}\,}{s}\!-\! 1\!\)\!
        \f{\widehat{s}_{23}}{\widehat{s}_{23}\!-\!m_H^2}
     +\(\!\f{\,\widehat{s}_{13}\,}{s}\!-\!1\!\)
        \f{\widehat{s}_{13}}{\widehat{s}_{13}\!-\!m_H^2}
\]\!\sqrt{s} \,,
\\[6mm]
i\T\[f_\pm\ov{f'}_\pm\to \pi^+\pi^+\pi^-\](d+d') \hspace*{-2mm}
&=
\dis\f{\sq\,}{v^3}\!\(\!\!\!\ba{c} m_f^{~} \\[2mm] m_{f'}^{~}\ea \!\!\!\!\)
    \!\!\[\f{\,\widehat{s}_{23}\,}{\widehat{s}_{23}\!-\!m_H^2}
         +\f{\,\widehat{s}_{13}\,}{\widehat{s}_{13}\!-\!m_H^2}
        \]\!\sqrt{s} \,,
\ea
\eeq
where \,$\widehat{s}_{ij}=(k_i+k_j)^2$\, and
\,$s=(k_1+k_2+k_3)^2 \simeq 
   \widehat{s}_{12}+\widehat{s}_{23}+\widehat{s}_{13}
\gg M_W^2$\,.\,  
Summing up all contributions in
(\ref{eq:Amp-ppm}), we arrive at
\beq
\label{eq:3pi-sum}
\ba{ll}
\hspace*{-4mm}
\T\[f_\pm\ov{f'}_\pm\to \pi^+\pi^+\pi^-\]_{\rm SM} 
&\!\! =~ (a)+(b)+(c+c')+(d+d')
\\[3mm]
&\!\! =~ -i\dis 
    \f{\sq\,}{v^3}\!\(\!\!\!\ba{c} m_f^{~} \\[2mm] m_{f'}^{~} \ea\!\!\!\!\)
    \!\!\[\f{\,\widehat{s}_{23}\,}{\widehat{s}_{23}\!-\!m_H^2}
         +\f{\,\widehat{s}_{13}\,}{\widehat{s}_{13}\!-\!m_H^2}
        \]\!\f{m_H^2}{\sqrt{s}\,} ~=~ \O\!\(\f{1}{\sqrt{s}\,}\),
\ea
\eeq
which means that the corresponding $2\to 3$
total cross section will behave as $\O(1/s)$ or smaller.
According to the
unitarity condition (\ref{eq:UC-IEn}), the scattering 
$\,f_\pm\ov{f'}_\pm\to \pi^+\pi^+\pi^-$\, is indeed unitary in the SM, 
as expected. 
If we {\it do not assume} the Higgs boson to be responsible for the
fermion mass generation, the diagrams in Fig.\,4(d)$+$(d$'$) must be
removed, and thus the remaining contributions give 
\beq
\label{eq:3pi-bare-mf}
\ba{ll}
\hspace*{-4mm}
\T\[f_\pm\ov{f'}_\pm\to \pi^+\pi^+\pi^-\] 
&\!\! =~ (a)+(b)+(c+c') 
\\[4mm]
&\!\! =~ i\dis 
    \f{\sq\,}{v^3}\!\(\!\!\!\ba{c} m_f^{~} \\[2mm] m_{f'}^{~} \ea\!\!\!\!\)
    \!\!\[\f{\,\widehat{s}_{23}\,}{\widehat{s}_{23}\!-\!m_H^2}
         +\f{\,\widehat{s}_{13}\,}{\widehat{s}_{13}\!-\!m_H^2}
        \]\!\sqrt{s} \,+\, \O\!\(\f{1}{\sqrt{s}}\)
\\[7mm]
&\!\! =~ -(d+d') \,.
\ea
\eeq
Finally, we evaluate the total cross section 
numerically using the leading order amplitude in
(\ref{eq:3pi-bare-mf}) and plot the results in Fig.\,5(a)-(d) 
in comparison with the imposed unitarity condition
(\ref{eq:UC-IEn}).

In this figure, we consider four types of initial helicity states,
$f_\lambda\ov{f'}_\lambda 
 \in ( t_+\ov{b}_+,\, u_+\ov{d}_+,\,
      \tau^+_-\ov{\nu}_-,\, e^+_-\ov{\nu}_-)$,\,
where the corresponding cross sections are proportional to the fermion 
masses \,$(m_t^2,\,m_u^2,\,m_\tau^2,\,m_e^2)$,\, respectively.
We have further included the Higgs width 
in Eq.\,(\ref{eq:3pi-bare-mf}) for  the actual calculations of Fig.\,5,
but we see that the effects of the Higgs mass and width are significant 
only in Fig.\,5(a)
where the unitarity violation scale occurs around 3\,TeV or so;
for all other light fermions such effects are almost invisible
because the relevant unitarity violation scales are substantially 
above the largest values of the Higgs boson mass and width.
We see that for all cases in Fig.\,5 
the inclusion of the EWSB contributions 
makes the cross sections larger for the whole energy range
and thus further enhances the unitarity constraints, 
but the corrections to the bounds are small, around 
$40-50\%$ or so, i.e., always less than about a factor of \,2\, 
for the \,$2\to 3$\, processes.\, 
As explained earlier, with the increasing number $n\,(>3)$ of
the final state particles, 
such corrections will become even
smaller, approaching unity like  
\,$[\O(2-3)]^{\f{1}{n-1}}$\,.\,
Therefore, we conclude that the estimated unitarity limits based
on model-independent, universal ``contact'' interactions are robust
and reliable up to a factor of \,$\sim\!2$\, or much less.

The Fig.\,5(a) has another very interesting feature, namely, the
inclusion of the EWSB effect pushes the unitarity limit for the 
scale of the top-quark mass generation down to the region of 
\beq
\label{eq:UB-tppm}   
E^\star ~=~ 2.8-3.3\,{\rm TeV}, ~~~~~
{\rm for}~~m_H^{~}=115-800\,{\rm GeV},
\eeq
which is below the best bound derived from the scattering 
\,$t\ov{t}\to 2\pi^a$\, in Table\,3. So, we see that by
including the EWSB contributions (from the minimal SM), 
the strongest unitarity limit for the scale of the top-quark 
mass generation occurs at \,$n=3$\,,\, 
rather than \,$n=2$\,.\,
Hence, the present unitarity limits from 
\,$2\to n$\, ($n\geqq 3$) scattering
gives  better upper limits on the scale of fermion mass
generation than the classic Appelquist-Chanowitz bounds
via $2\to 2$ scattering, for all known fermions 
including the top quark!

Finally, we note that in addition to Fig.\,4
there are three independent $2\to 3$ 
scattering processes,
$f\ov{f}\to \pi^0\pi^0\pi^0,~\pi^+\pi^-\pi^0$ and
$f\ov{f'}\to \pi^+\pi^0\pi^0$. 
We have verified the energy cancellation for each of
the above amplitudes at $\O(E^1)$ when
the complete SM diagrams are included. 
Furthermore, without assuming the Yukawa interaction of the
Higgs boson $h^0$ for fermion mass generation, we compute
the amplitudes including the  contributions
from the EWSB sector of the SM. We find, 
%derive the following leading amplitudes for these processes, 
at $\O(E^1)$,
\beq
\label{eq:3pi-others}
\ba{ll}
\hspace*{-8mm}
\T\[f_\pm\ov{f}_\pm\to \pi^0\pi^0\pi^0\] 
&\!\! =~\, i\dis 
    \f{\,m_f\,}{v^3}
      \!\[\f{\,\widehat{s}_{12}\,}{\,\widehat{s}_{12}\!-\!m_H^2\,}
         +\f{\,\widehat{s}_{23}\,}{\,\widehat{s}_{23}\!-\!m_H^2\,}
         +\f{\,\widehat{s}_{13}\,}{\,\widehat{s}_{13}\!-\!m_H^2\,}
        \]\!\sqrt{s} ~,
\\[7mm]
\hspace*{-8mm}
\T\[f_\pm\ov{f}_\pm\to \pi^+\pi^-\pi^0\] 
&\!\! =~\, i\dis 
    \f{\,m_f\,}{v^3}
       \!\[\f{\,\widehat{s}_{12}\,}{\,\widehat{s}_{12}\!-\!m_H^2\,}\]\!
       \sqrt{s} ~,
\\[7mm]
\hspace*{-8mm}
\T\[f_\pm\ov{f'}_\pm\to \pi^+\pi^0\pi^0\] 
&\!\! =~\, i\dis 
    \f{\sq\,}{v^3}\!\(\!\!\!\ba{c} m_f^{~} \\[2mm] m_{f'}^{~} \ea\!\!\!\!\)
    \!\[\f{\,\widehat{s}_{23}\,}{\,\widehat{s}_{23}\!-\!m_H^2\,}
        \]\!\sqrt{s} ~.
\ea
\eeq
Accordingly, we have computed their cross sections numerically and found the 
unitarity limits for these processes to be always 
weaker than the limits from $f\ov{f'}\to \pi^+\pi^+\pi^-$ .

\vspace*{2mm}
Before concluding this subsection, we
would like to remark that the current analysis only
makes the mild restriction that the longitudinal weak boson 
$V_L^a(=W_L^\pm,\,Z_L^0)$,
or its corresponding would-be Goldstone boson $\pi^a(=\pi^\pm,\,\pi^0)$,
behaves as a fundamental field below the unitarity violation scale.
This includes all supersymmetry
(SUSY) or non-SUSY models with fundamental Higgs doublets 
for the EWSB. There are dynamical models in which 
$V_L^a$ or $\pi^a$ is composite but
the compositeness scale is very high, around the GUT scale or above.
Examples of this kind of theory 
include the classic minimal top-condensate
model\,\cite{MTY,BHL} and its various viable supersymmetric 
extensions\,\cite{TOPC-Rev}, and more recently the dynamical EWSB
models via a neutrino condensate\,\cite{DEWSBnu}. 
In these cases the current analysis of
the scattering 
\,$f\ov{f}\to nV_L^a ~(n\pi^a)$\, will apply.
For the case where $V_L^a$ or $\pi^a$ is composite at a scale of
only a few TeV, we may need other types of processes based on
effective higher dimensional fermionic operators (such as
$\,f\ov{f}\to (f\ov{f})^k\,$ with $k\geqq 1$)\,\footnote{Recent
complementary studies\,\cite{4F-taumu} analyzed the {\it low energy} 
phenomenological constraints on 
the scale of new physics via generic four-fermion contact  
interactions involving the 
leptonic \,$\tau-\mu$\, flavor violation.}\, 
for a generic unitarity analysis of the scales of light 
fermion mass generation.

%%%%%%%%%%%%%%%%%%%%%%%  Fig.5  %%%%%%%%%%%%%%%%%%%%%%%%%%%%%%
\begin{figure}[H]
\label{fig:ff-3Vh}
\begin{center}
%\vspace*{-16mm}
%\vspace*{8cm}
\includegraphics[width=18.5cm,height=16cm]{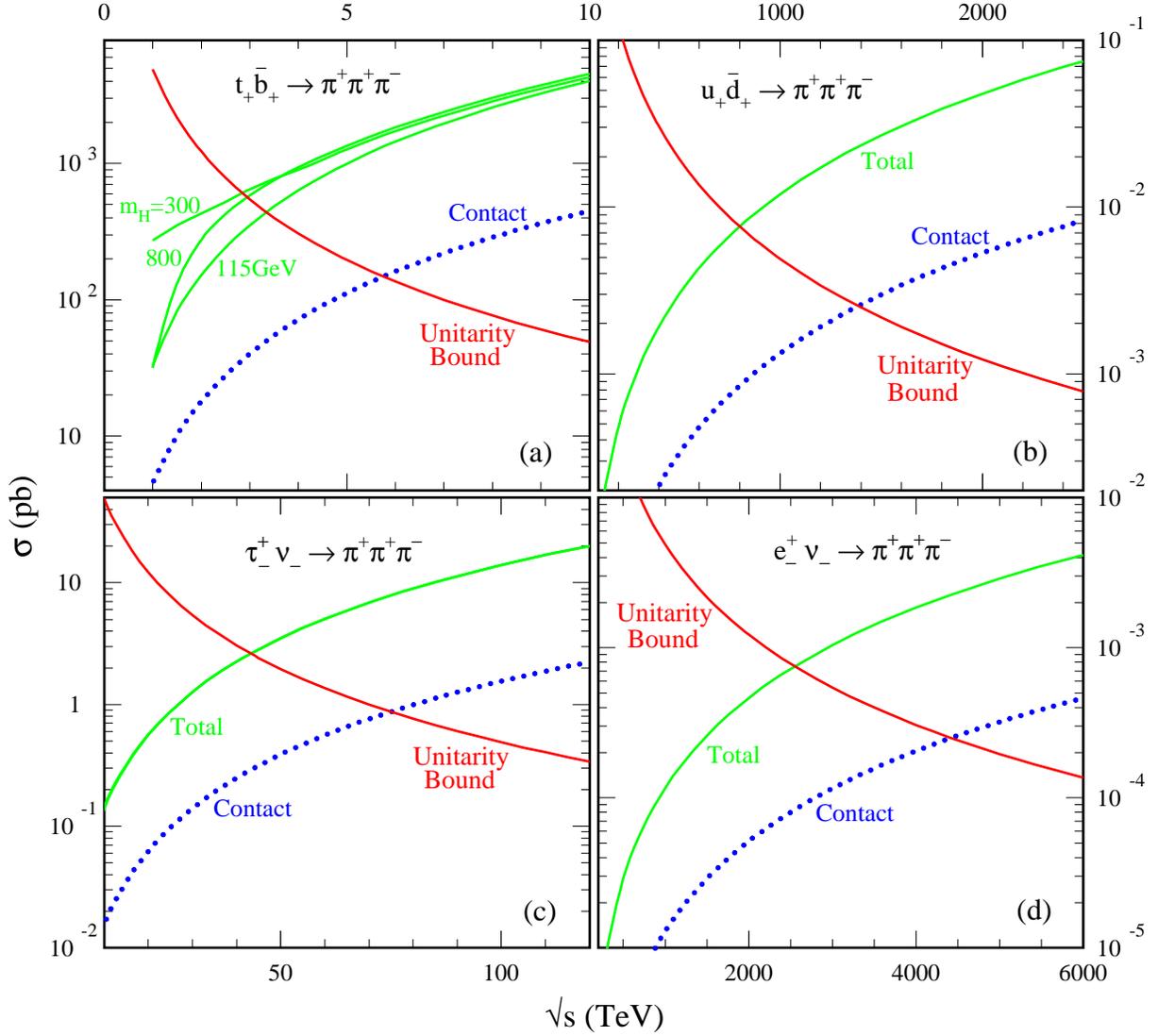}
\vspace*{-9mm}
\caption{The effect of the SM EWSB sector on the unitarity
bound for the scale of fermion mass generation.
(a). Same helicity scattering 
$t_+\bar{b}_+ \to \pi^+ \pi^+ \pi^-$:
the total cross sections including the contributions from 
the SM Higgs sector are shown in solid curves,
for $m_H^{~}=115,\,300,\,800$\,GeV, respectively.
The cross section  given by the universal
``contact'' contribution alone is shown as dotted curve. 
(b). Same as (a), but for the scattering
$u_+\bar{d}_+ \to \pi^+ \pi^+ \pi^-$.
(c). Same as (a), but for the scattering
$\tau^+_-\nu_- \to \pi^+ \pi^+ \pi^-$.
(d). Same as (a), but for the scattering
$e^+_-\nu_- \to \pi^+ \pi^+ \pi^-$.
In all cases, including the contribution of  EWSB
makes the cross section larger and thus 
tends to  enhance the unitarity limit, but the effect
is always less than a factor of two for the current 
\,$2\to 3$\,
scattering. Such effects will become even smaller for 
\,$2\to n$\, scattering as \,$n$\, increases above $3$\,.  
}
%\vspace*{-1cm}
\end{center}
\end{figure}
%%%%%%%%%%%%%%%%%%%%%%%%%%%%%%%%%%%%%%%%%%%%%%%%%%%%%%%%%%%%%%%

\newpage
%\vspace*{5mm}
\section{\hspace*{-6mm}.$\!$
Scale of Mass Generation for Majorana Neutrinos}
\vspace*{4mm}

Parallel to Sec.\,4, this section is devoted to a
quantitative analysis of the scale of mass generation
for Majorana neutrinos. 
In Sec.\,5.1, we will derive the 
interactions of Majorana neutrinos with weak gauge
bosons and Goldstone bosons. Then we systematically 
analyze the scattering processes 
$\,\nL\nL\to n\pi^a\,(nV_L^a)\,$ and deduce new unitarity
limits on the scale of neutrino mass generation.
In Sec.\,5.2, we discuss the interpretation and implication
of our new bounds for the underlying mechanism
of neutrino mass generation. Three types of mass mechanisms
will be addressed, including the 
seesaw mechanism\,\cite{nu-seesaw}, 
the radiative mechanism\,\cite{nu-rad}
and the proposal for generating neutrino
masses from supersymmetry breaking\,\cite{nu-susy}.

\vspace*{5mm}
\subsection{\hspace*{-5mm}.$\!$
Majorana Masses and Neutrino-Neutrino Scattering}
\vspace*{3mm}

Given the observed particle spectrum of the SM, 
neutrinos can only have the Majorana mass term,
$~
%{\cal L}_\nu^{\rm m}  =
\dis -\f{1}{\,2\,}{m_\nu^{ij}}\,\nu_{Li}^T \widehat{C} \nu_{Lj} 
       + {\rm H.c.}\,,\,
$
which violates lepton number by two units. 
With the nonlinearly realized Goldstone field $U$ 
[cf. (\ref{eq:U})],  such a dimension-3 bare mass term 
can be made gauge-invariant but is necessarily nonrenormalizable. 
As we showed earlier in Eq.\,(\ref{eq:nu-mass}), this
uniquely takes the form of the Weinberg operator\,\cite{weinberg5} 
with the nonlinear field defined as
\,$\Phi =\dis\f{v}{\sqrt{2}}\ov{\Phi}$\, with
\,$\ov{\Phi} = U\(0,\,1\)^T $,\, i.e.,
\beq
\label{eq:nu-mass2}
{\cal L}_\nu ~=~ \dis
-\f{1}{2}m_\nu^{ij}{L^{\alpha}_i}^T \widehat{C}L^{\beta}_j
 \ov{\Phi}^{\alpha'}\ov{\Phi}^{\beta'}
\epsilon^{\alpha\alpha'} \epsilon^{\beta\beta'} + {\rm H.c.}
\,,
\eeq
where 
~$m_\nu^{ij} = \dis {\cal C}_{ij}\f{v^2}{\cut}$~
is the neutrino Majorana mass derived in
Eq.\,(\ref{eq:mij-nu}).
The neutrino Majorana mass term
allows us to introduce, for convenience, a (real) Majorana 
neutrino field \,$\X\,(\,= \X^c\,)$\, by defining
\beqa   
\label{eq:X-def}
\chi &=& \f{1}{\sqrt{2}\,}\(\,\nuL+\nu_L^c\,\) \,, 
\eeqa
where
$\,\nuL = \sqrt{2}{P_L}\X\,$ and
$\,\nu_L^c \equiv (\nuL)^c = \sqrt{2}{P_R}\,\X\,$
with \,$P_{L,R}=\f{1}{2}(1\mp\gamma^5)$\,.\,
Using a properly normalized notation for the lepton doublet
$\,L_j=\(\f{1}{\sqrt 2\,}\nu^{~}_{Lj}\,,\,\l_{Lj} \)^T\,$
and the operator (\ref{eq:nu-mass2}),  
we arrive at the following kinetic and mass terms for the
Majorana neutrino field $\X$\,,
\beqa
\label{eq:X-Kin}
{\cal L}^{\rm kin}_{\X} &=&
\f{1}{\,2\,}\,\ov{\X}_j \(i\! \not\!\partial -m_{\nu j}^{~}\)\X_j^{~}
\,,
\eeqa
where $m_{\nu j}^{~}$ is the $j$th real mass eigenvalue derived from
diagonalizing the symmetric $3\times 3$ mass matrix $\{m_\nu^{ij}\}$\,.\,
The gauge interactions of $\X$, 
which  can be deduced from the SM Lagrangian, are
\beqa
\label{eq:X-gauge}
%&& \nonumber
%\\[-17mm]
{\cal L}_{\X}^{\rm G} &=&
-\f{g}{\,4c_w\,}\ov{\X}_j^{~}\gamma^\mu\gamma^5\X_j^{~} Z_\mu^0 -
\f{g}{\,2\sqrt{2}\,} \mathbb{V}_{ij}
\ov{\l_{Li}}\gamma^\mu\gamma^5\X_j^{~} W^-_\mu
+{\rm H.c.}\,,
\eeqa
where \,$c_w=\cos\theta_w$\, with $\theta_w$  the
Weinberg angle, and $\mathbb{V}_{ij}$ is 
the Maki-Nakagawa-Sakata-Pontecorvo (MNSP) mixing\,\cite{MNS} 
from diagonalizing the lepton and neutrino mass matrices.
In this expression, we have used the fact that 
the Majorana field $\X_j^{~}$ has vanishing vector coupling.
Despite the difference between the gauge couplings of the Majorana  
neutrinos and the massless SM-neutrinos, their contributions to
the existing physical observables, such as the invisible decay width
of $Z^0$ etc, differ only by a tiny amount of  
\,$\O(m_{\nu}^2/M_Z^2)$\,.\,  
The real test of the Majorana nature of neutrinos has to come
from the next generation of neutrinoless double-beta decay
experiments.

Comparing the operator (\ref{eq:nu-mass2})
with the nonlinear Dirac mass term (\ref{eq:f-mass}),
we see that (\ref{eq:nu-mass2}) also appears as
dimension-3 but contains two nonlinear $U$ matrices and violates
lepton number by two units. Hence, the Feynman rules for the
neutrino-Goldstone interactions are different from
those in Eq.\,(\ref{eq:f-mass}) though the relevant vertices
always have a contact structure with 2 fermions and $n$ 
Goldstone fields $\pi^{a_1}$-$\pi^{a_2}$-$\cdots$-$\pi^{a_n}$
($n\geqq 1$).\,
It is clear that this does not affect our general power counting
for the amplitude $\,\T [\nL\nL\to n\pi^a]$\,
which takes the same form as $\,\T[f\bar{f}\to n\pi^a]$\,
[cf. Eq.\,(\ref{eq:Torder})]. This is why our  estimates in 
Fig.\,2 and Table\,1  apply to \,$\nL\nL\to n\pi^a$\,
scattering as well.

To carry out the quantitative analysis, we deduce from
(\ref{eq:nu-mass2}) the following neutrino-Goldstone interaction
Lagrangian, in the neutrino mass-eigenbasis,\footnote
{The extension to an initial state like 
$\ell^+_i\X_j$ is straightforward.}
\beq
\label{eq:Lnu-int}
{\cal L}^{\rm Int}_{\nu\nu\pi^n} ~=~ \dis
-\f{1}{2}m^{~}_{\nu j}\ov{\X}_j^{~}
 \[\mathbb{A} (\pi)+i\gamma_5^{~}\mathbb{B} (\pi)
 \]\X_j^{~}   \,,
\eeq
where $\mathbb{A} (\pi)$ and $\mathbb{B} (\pi)$ are given by
\beq
\label{eq:Lnu-AB}
\ba{lcl}
\dis 
\mathbb{A} (\pi) \equiv & \dis\RE\(U_{22}\)^2 - 1 & =
\dis\sum_{n({\rm even})=2}^\infty \f{i^n}{\,v^n\,}
\[K_{1,n}2^{\f{n}{2}}_{~} \(\pi^+\pi^-\)^{\f{n}{2}}
  +\sum_{\l=0}^{\f{n}{2}-1}K_{4,n\l}
         \(\pi^+\pi^-\)^{\l}\(\pi^0\)^{n-2\l}
\],
\\[7mm]
\dis
\mathbb{B} (\pi) \equiv & \dis -\IM\(U_{22}\)^2  & =
\dis\sum_{n({\rm odd})=1}^\infty \f{i^{n-1}}{\,v^n\,}
K_{2,n}\sum_{\l=0}^{\f{n-1}{2}}
C^\l_{\f{n-1}{2}} 2^\l \(\pi^+\pi^-\)^{\l}\(\pi^0\)^{n-2\l} ,
\ea
\eeq
and
\vspace*{-4mm}
\beq
\label{eq:Kj}
\ba{lll}
\dis
K_{1,n} &=~\, \dis\sum_{k=0}^{\f{n}{2}}\f{1}{~(2k)!(n-2k)!~} 
~=~\f{\,2^{n-1}\,}{n!} \,,
& ~~~(n={\rm even})\,,\\[5mm]
K_{2,n} &=~\, \dis\sum_{k=0}^{\f{n-1}{2}}
           \f{n+1}{~(2k)!(n-2k-1)!(2k+1)(n-2k)~} 
~=~\f{\,2^{n}\,}{n!} \,,
& ~~~(n={\rm odd})\,, \\[5mm]
K_{3,n} &=~\, \dis\sum_{k=0}^{\f{n}{2}-1}
           \f{1}{~(2k)!(n-2k-2)!(2k+1)(n-2k-1)~}
~=~\f{\,2^{n-1}\,}{n!}  \,,
& ~~~(n={\rm even})\,,
%\\[6mm]
\ea
\eeq
$$
\ba{lll}
\dis
K_{4,n\ell} &=~\, \dis
2^\l \[K_{1,n}C^\ell_{\f{n}{2}} +
       K_{3,n}C^\ell_{\f{n}{2}-1} \]  
~=~\f{\,2^{n+\ell -1}\,}{n!}
\[C^\ell_{\f{n}{2}} + C^\ell_{\f{n}{2}-1} \],
%~~~~~~
%C_n^\ell ~=~ \dis\f{n!}{\,\ell!\,(n-\ell)!\,} \,,  
& ~~~(n={\rm even})\,,
\ea
%\eeq
$$
\noindent
with the notation 
$~C_n^{\ell} \,\equiv\, {n!}/[\,\ell!\,(n-\ell )!\,] \,$.\,
Consider the generic scattering process
$\,|\X^{~}_{j\pm}\ov{\X}^{~}_{j\pm}\rangle \to 
   \(\pi^+\)^\ell\(\pi^-\)^{\ell}\(\pi^0\)^{n-2\ell}\,$
$(\ell=0,1,2,3,\cdots)$
for three cases:
(a) $n({\rm even})=2\ell$\,;\,   
(b) $n({\rm even}) > 2\ell$\,;\,
(c) $n({\rm odd})\geqq 3$\,.\,
We compute the helicity amplitudes
$\,\T\[\X^{~}_{j\pm}\ov{\X}^{~}_{j\pm} ;
 \,\ell,\ell,n-2\ell\]$,\,
from the leading contact diagrams [cf. Fig.\,1(a)],
\beq
\label{eq:Amp-nu}
\ba{rll}        
%\hspace*{-3mm}
n={\rm even:}~~~ &  \dis
   \T \[\X^{~}_{j\pm}\ov{\X}^{~}_{j\pm} ;\,\f{n}{2},\f{n}{2},0\] 
  & =~\dis\pm i^{n+1}\C^\nu_{1} 
      \f{~m_{\nu j}^{~}\sqrt{s}~}{v^n} \,,
\\[5mm]   
             & \dis
   \T \[\X^{~}_{j\pm}\ov{\X}^{~}_{j\pm} ;\,\l,\l,n-2\l \] 
  & =~\dis\pm i^{n+1}\C^\nu_{2} 
      \f{~m_{\nu j}^{~}\sqrt{s}~}{v^n} \,;
\\[5mm]
n={\rm odd:}~~~ &  \dis
   \T \[\X^{~}_{j\pm}\ov{\X}^{~}_{j\pm} ;\,\l,\l,n-2\l \] 
  & =~\dis + i^{n+1}\C^\nu_{3} 
      \f{~m_{\nu j}^{~}\sqrt{s}~}{v^n} \,;
\ea
\eeq
where the dimensionless coefficients $\,\C_{1,2,3}^\nu$\, are
given by
\beq
\ba{lll}
\C^\nu_1 &=&  2^{{n}\over{2}}_{~}\(\f{n}{2}!\)^2K_{1,n} \,,
\\[3.5mm]
\C^\nu_2 &=& \dis 2^{\ell}\(\l!\)^2(n-2\l)!
                 \[K_{1,n}C^\l_{\f{n}{2}}+K_{3,n}C^\l_{\f{n}{2}-1}\] ,
\\[3.5mm]
\C^\nu_3 &=& \dis 2^{\ell}\(\l!\)^2(n-2\l)!C^\l_{\f{n-1}{2}}K_{2,n} \,.
\ea
\eeq
We further define the normalized combination for the
initial state of Majorana neutrinos with spin-0
(or spin-1),
\beq
|{\rm in}\rangle ~=~ \dis
\f{1}{\sqrt{2}\,}\[|\X^{~}_{j+}\ov{\X}^{~}_{j+}\rangle \mp
          |\X^{~}_{j-}\ov{\X}^{~}_{j-}\rangle\] ,
\eeq
corresponding to the case of \,$n=$\,even(odd),\,
and find the corresponding $S$-matrix element
\beq
\ba{ll}
\T[{\rm in};\l,\l,n-2\l] & =~
\dis\f{1}{\sqrt{2}\,}\(\,\T[\X^{~}_{j+}\ov{\X}^{~}_{j+};\l,\l,n-2\l] 
            \mp\T[\X^{~}_{j-}\ov{\X}^{~}_{j-};\l,\l,n-2\l]\,\)
\\[6mm]
& =~ \sqrt{2}~\T[\X^{~}_{j\pm}\ov{\X}^{~}_{j\pm};\l,\l,n-2\l] \,.
\ea
\eeq
This means that the cross section for the initial 
state \,$|{\rm in}\rangle$\, is enhanced by
a factor $\sqrt{2}$\,.
Using this we derive the $2\to n$ inelastic cross section as
\beq
\label{eq:nu-sigma}
\ba{rl}
\sigma[{\rm in};\,\l,\l,n-2\l]  &=~ \dis
\f{2C^\nu_{0r}
}{~2^{4(n-1)}\pi^{2n-3}\,(n-1)!(n-2)!~}
\f{1}{\,s\,}
\(\f{\,\sqrt{s}\,}{v}\)^{2(n-1)}
\(\f{~m^{~}_{\nu r}\,}{v}\)^{2}
\,,
\\[5mm]
C_{0r}^{\nu} &=~ \dis\f{\,(\C_r^\nu )^2\,}{\varrho} ~=~
           \f{(\C_r^\nu )^2}{(\l!)^2\,(n-2\l)!~} \,,
~~~~~(r=1,2,3)\,.
\ea
\eeq
Finally, with the condition (\ref{eq:UC-IEn}) and
\,$\varrho_{e}^{~} =2!$\,,\, 
we derive the inelastic unitarity limit, 
\,$\sqrt{s} \leqq E^\star_\nu$\,,\, with
\beqa
\label{eq:UBnu-final}
E^\star_\nu &=& v\[\f{v}{m_{\nu r}^{~}}
               \sqrt{\f{4\pi}{\,\R_{\nu r}^{\max}\,}}
               \,\]^{\f{1}{n-1}}  ,
\eeqa
where $~\R_{\nu r}^{\max}$~ is given by   
\beq
\label{eq:Rmax-nu}
\ba{lcll}
\R_{\nu 1}^{\max} &=&
\dis\f{\(\f{n}{2}!\)^2 (K_{1,n})^2
}{~2^{3n-4}\pi^{2n-3}\,(n-1)!\,(n-2)!~} \,,
&~~~~~(n={\rm even},~\ell=\f{n}{2})\,,
\\[4mm]
\R_{\nu 2}^{\max} &=&
\dis\f{~n\(K_{1,n}+K_{3,n}\)^2~}
{~2^{4(n-1)}\pi^{2n-3}\,(n-2)!~} \,,
&~~~~~(n={\rm even},~\ell < \f{n}{2})\,,
\\[4mm]
\R_{\nu 3}^{\max} &=&
\dis\f{~n\,(K_{2,n})^2~}{~2^{4(n-1)}\pi^{2n-3}\,(n-2)!~} \,,
&~~~~~(n={\rm odd})\,,
\ea            
\eeq
and for $n=$\,even,\, the best limits are from
$\,\(\R_{\nu 1}^{\max},\,\R_{\nu 2}^{\max}\)_{\max} $.\,
Numerically, we find that 
the bounds given by 
\,$\R_{\nu 1}^{\max}$\, and \,$\R_{\nu 2}^{\max}$\,
differ very little, always less than 2\%.\,

%%%%%%%%%%%%%%%%%%%%%  Fig.6  %%%%%%%%%%%%%%%%%%%%%%%
\begin{figure}[H]
\label{fig:Fig5}
\vspace*{-2mm}
\begin{center}
%\hspace*{7mm}
\includegraphics[width=16.5cm,height=12.5cm]{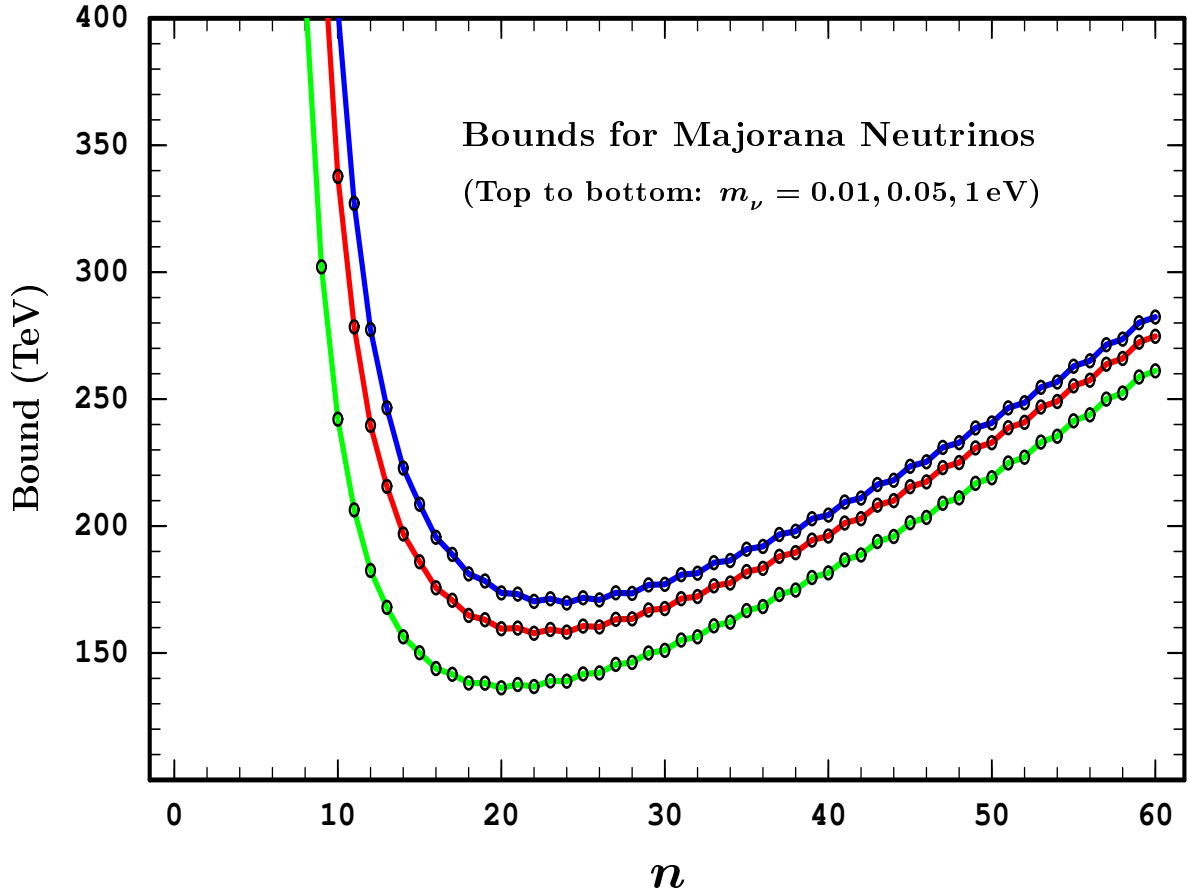}   
\vspace*{-8mm}
\caption{Scales of mass generation for Majorana neutrinos:
the precise unitarity bounds \,$E^\star_\nu$\,
from Eq.\,(\ref{eq:UBnu-final}) are plotted as a function
of \,$n\,(\geqq 2)$\,,\,  where
\,$m_{\nu j}^{~} =0.01,\, 0.05,\,1$\,eV\,
(curves from top to bottom).
Only integer values of $n$ have a physical meaning.
The minimum point of each curve is given in 
Eq.\,(\ref{eq:nuB-min}).
}
\end{center}
\end{figure}
%%%%%%%%%%%%%%%%%%%%%%%%%%%%%%%%%%%%%%%%%%%%%%%%%%%%%%%%%%%%%%%%

With the formulas (\ref{eq:nu-sigma})-(\ref{eq:Rmax-nu}),
we plot the numerical results of the unitarity limits
in Fig.\,6 for three typical values of the neutrino masses
$m_{\nu j}$.
We find that the best unitarity bound in each case to be
\beq
\label{eq:nuB-min}
E^\star_{\min} = 136,\,158,\,170\,{\rm TeV},~~~~~~
{\rm located~at}~~n=n_s = 20,\,22,\,24,
\eeq
for
\,$m_{\nu j}^{~} = 1,\, 0.05,\, 0.01$\,eV,\, respectively.
Under the same input \,$m_\nu = 0.05$\,eV,\,
this agrees to our early generic  
{\it estimate} (Table\,1) within about a factor of $1.7$.\,
Comparing Eq.\,(\ref{eq:nuB-min}) with Table\,3, it is striking
to note that even though the neutrino mass is about ten million times 
smaller than the electron mass ($m^{~}_\nu/m^{~}_e \sim 10^{-7}$),
the unitarity violation scale associated with its mass generation
is only about a factor of
\,$1.3$-$1.6$\, higher than that for the electron!\footnote{If we
only look at the customary limit from $2\to 2$ scattering
(cf. Sec.\,2.3.2C and Table\,2), we would naively expect the
mass-generation scale for Majorana neutrinos to be as high
as the GUT scale ($\sim\! 10^{16}$\,GeV).  
The reason that the scale is
substantially lower will be discussed in the following subsection.}\,
For neutrinos such an unitarity violation occurs in the
multiple $nV_L^a$ ($n\pi^a$) production with the number
\,$n=n_s = 20-24$,\, while for the electron it appears
at \,$n=n_s = 11$.\,  This leads to an exciting observation
that the mass-generation mechanisms for both the electron
and neutrinos have to reveal themselves at or below the scale of
$\sim\!10^2$\,TeV.  The scales for the mass generations of
other leptons and light quarks are much lower, as shown in our
Table\,2 of Section\,4.
This not only provides an important guideline for model-building, 
but also indicates where to seek definitive  
{\it experimental tests,}  such as the relevant 
astrophyical processes\,\cite{HeDicus2}.

\vspace*{5mm}
\subsection{\hspace*{-5mm}.$\!$
Unitarity Violation vs.~Mechanisms for Majorana Mass Generation 
}
\vspace*{3mm}

The unitarity violation scale (\ref{eq:nuB22x})
derived from the customary $2\to 2$ scattering
indicates that the scale of mass generation for Majorana neutrinos
could be as high as the GUT scale. At first sight, this
sounds quite natural if we note that the traditional 
seesaw mechanism\,\cite{seesaw} invokes a heavy right-handed
neutrino whose mass is intrinsically tied to the
GUT scale. But, knowing there are other low scale
mechanisms for neutrino mass 
generation\,\cite{nu-rad,nu-susy,2Loopnu,Ma,nu-susy2},
we then wonder how the unitarity violation
limit would just fit into the seesaw mechanism rather than other
mechanisms especially since the analysis of unitarity
violation is universal, independent of a particular 
mechanism. Furthermore, now we have shown in Sec.\,5.1 that
the unitarity violation
scale associated with  nonzero neutrino mass can be
as low as $\O(100)$\,TeV once we open up the 
inelastic multiple gauge boson channels for  neutrino-neutrino
scattering. 
What is the physics interpretation and implication of our new bounds 
for the underlying mechanism of neutrino mass generation?
We address this issue 
by analyzing three types of neutrino mass generation mechanisms, 
including the seesaw mechanism\,\cite{nu-seesaw},
the radiative mechanism\,\cite{nu-rad} 
and neutrino masses induced from SUSY breaking\,\cite{nu-susy}.

\vspace*{4mm}
\noindent 
{\bf 5.2.1.
Unitarity Violation vs.$\!$ Seesaw and Radiative Mechanisms} 
\vspace*{2mm}

In our unitarity analysis of Majorana neutrinos (cf. Sec.\,5.1), 
we have strictly followed the same logic 
as  for Dirac fermions or weak gauge bosons to
define the scale \,$\cut_\nu$\, for generating
the Majorana mass \,$m_\nu^{~}$,\,  i.e., the scale 
$\cut_\nu$ is the minimal energy at which the 
{\it bare Majorana mass term,}
\beq
\label{eq:bare-mnu}
{\cal L}_\nu^{\rm m} ~=~
\dis -\f{1}{2}m_\nu^{ij}\,\nu_{Li}^T\widehat{C} \nu_{Lj}^{~} 
       + {\rm H.c.} \,,
\eeq
has to be replaced by a renormalizable interaction [cf. the
general definition given above Eq.\,(\ref{eq:cutx-UB})].
A crucial difference between the current study and
Refs.\,\cite{scott,scott-PRL} for the scale of Majorana neutrino
mass generation is that the neutrinos are not assumed
to couple to the same Higgs doublet which breaks
the electroweak gauge symmetry for their mass generation.
We put in the {\it bare mass terms} by hand for any given type of 
fermions and derive the corresponding unitarity violation scale
as the most general and model-independent limit on their mass generation.
For instance, if we replace the nonlinear field $\Phi$ in
Eq.\,(\ref{eq:nu-mass}) or (\ref{eq:nu-mass2}) by the usual
SM Higgs doublet $H$, we recover the original Weinberg dimension-5
operator\,\cite{weinberg5} which is {\it not} a pure mass term even
in the unitary gauge and implies the assumption that the SM Higgs 
doublet actually participates in  neutrino mass generation. 
To see this more clearly consider, for instance, 
the traditional seesaw mechanism with a heavy singlet 
Majorana neutrino $\nu_R^{~}$ \cite{nu-seesaw}, which gives
the renormalizable Lagrangian\footnote{By definition, 
$H$ has hypercharge $\f{1}{2}$.}
\beq
\label{eq:Lseesaw}
{\cal L}_{\rm seesaw}~=~ \dis
-y^\nu_j\ov{L}_j\,\ep{H^\ast}\nu_R^{~} 
-\f{1}{2}M_R\nu_R^T\widehat{C}\nu_R^{~} + {\rm H.c.} \,,
\eeq
and leads to the dimension-5 Weinberg operator 
after integrating out the heavy $\nu_R^{~}$. 
Because of the assumption that the SM Higgs
doublet $H$ generates the Dirac mass term 
$\, -m_{Dj}^{~}\ov{\nu_L^{~}}_j\nu_R^{~}$\,,\,  
the resulting
dimension-5 Weinberg operator invokes the Higgs $H$
for Majorana neutrino mass generation. 
But, by our general proceedure, we  put in this
Dirac mass term by hand and replace (\ref{eq:Lseesaw})
by
\beq
\label{eq:Lseesaw-mass}
{\cal L}'_{\rm seesaw}~=~ \dis
 -m_{Dj}^{~}\,\ov{\nu_{L}^{~}}_j\nu_R^{~}
-\f{1}{2}M_R\,\nu_R^T\widehat{C}\nu_R^{~} + {\rm H.c.} \,,
\eeq
which, after integrating
out the heavy $\,\nu_R^{~}$,\, generates the
{\it bare} Majorana mass term of Eq.\,(\ref{eq:bare-mnu}),
and can be further made 
gauge-invariant as in Eq.\,(\ref{eq:nu-mass2})
with the nonlinear field \,$\ov{\Phi}$.\,
Hence, studying the unitarity violation of
the scattering $\,\nL\nL\to n\pi^a\,$ based on a
dimension-5 Weinberg operator [by integrating out
$\nu_R^{~}$ from (\ref{eq:Lseesaw})] 
{\it only probes  the scale of the singlet 
mass $M_R$} which obeys the 
decoupling theorem\,\cite{DCT}\footnote{
As expected, the resulting unitarity violation scale
from the high energy 
$\f{1}{2}\[|\nLp\nLp\rangle -|\nLm\nLm\rangle\]
 \to \f{1}{2}|Z_LZ_L+hh\rangle$ 
scattering\,\cite{scott} is, 
$\,\dis\f{4\pi v^2}{m_\nu^{~}}
 \sim 4\pi M_R \sim (10^{15}-10^{17})$\,GeV
$\sim \cut_{\rm GUT}$, [cf. Eq.\,(\ref{eq:nuB22})],
which is \,$12-14$\, orders of magnitude higher than the
weak scale.}.\,
But the seesaw neutrino mass is really a {\it combined outcome
of both bare mass terms in Eq.\,(\ref{eq:Lseesaw-mass})}
which corresponds to the operator (\ref{eq:nu-mass2}) 
after integrating out  $\nu_R^{~}$.  Unlike the singlet 
Majorana mass term for $\nu_R^{~}$, the non-singlet Dirac mass term
in Eq.\,(\ref{eq:Lseesaw-mass}) does not respect
the decoupling theorem\,\cite{DCT}. 
Hence, studying the unitarity violation of
the scattering $\,\nu_L^{~}\nu_L^{~}\to n\pi^a\,$ based on 
(\ref{eq:nu-mass2})  probes 
the scale at which the {\it full} seesaw mechanism
(\ref{eq:Lseesaw-mass}) has to set in.
As will be clear below, even though the singlet neutrino
$\nu_R^{~}$ can be still as heavy as $\cut_{\rm GUT}$, 
the inelastic $2\to n$ unitarity requires the
seesaw mechanism to {\it start} at a scale not much
above $\O({\rm TeV})$\,!

%%%%%%%%%%%%%%%%%%%%%%  Fig.7  %%%%%%%%%%%%%%%%%%%%%%%%%%%%%
\begin{figure}[H]
\label{fig:Fig4}
\vspace*{-4mm}
\begin{center}
\includegraphics[width=16cm,height=6cm]{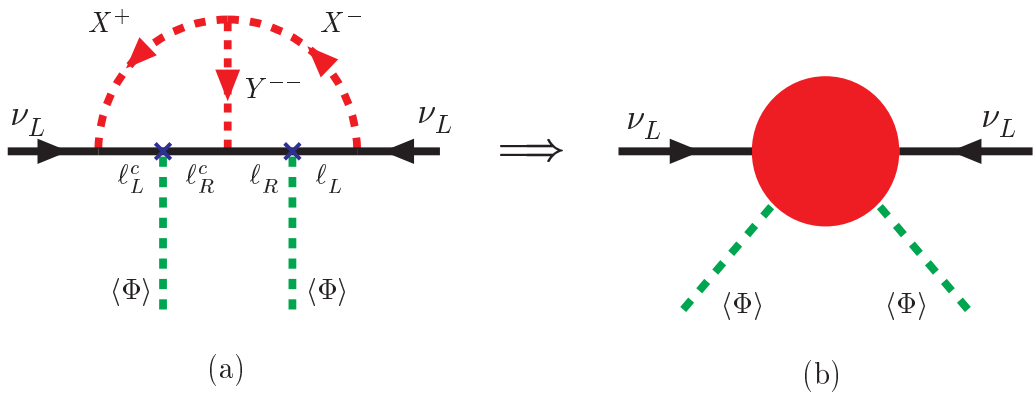} 
\vspace*{-2mm}
\caption{Majorana neutrino masses radiatively generated 
at a low scale of \,$\O({\rm TeV}) \lll \cut_{\rm GUT}$\,.\,
~~~
(a) An example of two-loop radiative neutrino mass generation
with two extra charged singlet scalars $(X^-,\,Y^{--})$. 
Here the inserted lepton Dirac masses are given by 
Eq.\,(\ref{eq:f-mass}).
(b) The  Majorana neutrino mass contained in the
effective mass operator (\ref{eq:nu-mass2}), after integrating 
out all heavy scalars  $(X^-,\,Y^{--})$ from (a). 
}
\vspace*{-5mm}
\end{center}
\end{figure}
%%%%%%%%%%%%%%%%%%%%%%%%%%%%%%%%%%%%%%%%%%%%%%%%%%%%%%%%%%%%

We further note
that the most general operator (\ref{eq:nu-mass2})
need not necessarily arise from
a seesaw mechanism as in Eq.\,(\ref{eq:Lseesaw-mass}).
It was demonstrated that even without introducing a heavy
$\,\nu_R^{~}$,\,  the desired Majorana mass term
(\ref{eq:bare-mnu}) can be generated from a
radiative mechanism at the {\it TeV scale} \cite{nu-rad}.
An explicit example of a two-loop mechanism\,\cite{2Loopnu} for
radiative Majorana neutrino mass generation is illustrated
in Fig.\,7(a). 
After integrating out the two extra charged singlet scalars
$(X^-,\,Y^{--})$,  
this generates an effective Majorana mass
term for neutrinos in Fig.\,7(b) 
which is just contained in the operator (\ref{eq:nu-mass2}). 
It is very interesting to note that in Fig.\,7(a) 
this two-loop mechanism also connects the Majorana neutrino masses to the
lepton Dirac masses given by Eq.\,(\ref{eq:f-mass}).
The step from Fig.\,7(a) to 7(b) of integrating out heavy
singlet scalars $(X^-,\,Y^{--})$ is fully {\it parallel to}
the derivation of Majorana neutrino masses from the seesaw Lagrangian
(\ref{eq:Lseesaw-mass}) by integrating out the singlet $\nu_R^{~}$\,!
If we consider the special case that 
the lepton masses are generated by the SM Higgs
doublet $H$, we can replace the two nonlinear Goldstone fields $\Phi$
in Fig.\,7(a) by $H$ and the corresponding effective neutrino
mass operator in Fig.\,7(b) is just the original dimension-5
Weinberg operator. Again this is fully parallel to the step of deriving
the seesaw neutrino masses from the {\it renormalizable}
Lagrangian (\ref{eq:Lseesaw}) by integrating out $\nu_R^{~}$,
which results in the dimension-5 Weinberg operator.

So, once we  put in all fermion
mass-terms by hand (without making extra model-dependent assumptions
about their origins), the Majorana neutrino masses generated either
from the seesaw mechanism or from the radiative mechanism can be formulated
as the generic effective mass operator  (\ref{eq:nu-mass2}).
However, in the seesaw mechanism the new singlet fermion \,$\nu_R^{~}$\,
is known to be as heavy as the GUT scale\,\cite{nu-seesaw} 
while the new singlet scalars
such as \,$(X^-,\,Y^{--})$\, 
in the radiative mechanism can naturally have 
masses of $\O({\rm TeV})$ \cite{nu-rad}.   
Then it is intriguing to ask:
how does the unitarity violation given by inelastic neutrino scattering  
$\,\nL\nL\to n\pi^a$ [based on (\ref{eq:nu-mass2})]
constrain the {\it new physics scale} for neutrino
mass generation? We know that {\it this limit must be universal, 
independent of whether the underlying dynamics is 
a seesaw mechanism or a radiative mechanism.}
Starting with the general mass operator  (\ref{eq:nu-mass2}),
we thus expect that the optimal bound on the unitarity violation scale
should be neither simply fixed by the mass $M_R\,(\sim \cut_{\rm GUT})$
of $\nu_R^{~}$ in the seesaw mechanism nor by the masses
$M_{X,Y}\,(\sim {\rm TeV})$ of   $(X^-,\,Y^{--})$ in the radiative
mechanism.  This is because neither $M_R$ nor $M_{X,Y}$ 
are the whole story in each mechanism: there has to be additional
Dirac mass terms [either the first term in Eq.\,(\ref{eq:Lseesaw-mass})
or the lepton mass insertions in Fig.\,7(a)] which
do not respect the usual decoupling theorem\,\cite{DCT}.
This non-decoupling feature means
either a weak-doublet Higgs or a nonlinear field $U$ has
to be invoked to make the Dirac mass term invariant under the SM gauge
group, where in the first case the Dirac mass is proportional to
the Higgs Yukawa coupling (forbidding the decoupling of the 
corresponding Higgs scalar), and in the second case the nonlinear
$U$ field links the nonrenormalizable Goldstone-fermion coupling to the
Dirac mass.  In general, such a nondecoupling occurs 
when we build {\it any} fermion mass term involving
the left-handed $SU(2)_L$ doublet fermion together with 
(i) a weak singlet fermion (giving a Dirac mass), or, 
(ii) itself (giving a Majorana mass).
Indeed, it is this common non-decoupling feature that 
forces the unitarity violation scale of Majorana neutrinos 
to be around $\O(100)$\,TeV, substantially below the GUT
scale but not much above the scale for electron mass generation
despite the fact of $~m_\nu^{~}/m_e^{~} \sim 10^{-7}\,$.

For the above explicit models of seesaw and radiative neutrino
mass generation, this means that our new limits constrain the 
new physics scale of the leptonic Higgs Yukawa interaction, 
which must invoke extra new fields 
(such as right-handed neutrinos or Zee-scalars or triplet-Higgs)
as needed for ensuring renormalizability and
generating lepton number violation 
(although our new bounds do not directly constrain
the masses of these new fields themselves 
which obey the decoupling theorem).

%\newpage
\vspace*{6mm}
\noindent 
{\bf 5.2.2. Mass Generation Scale for Majorana Neutrinos 
            in Supersymmetric Theories}
\vspace*{3mm}

Supersymmetry (SUSY) is attractive
for stabilizing the weak scale, but does not explain 
the huge hierarchies in the fermion mass-spectrum
or the observed flavor mixings. In fact, it necessarily
extends the SM with extra three-family superpartners and thus 
{\it adds} new puzzles to the ``flavor problem''\,\cite{susy-f}.
Furthermore, similar to the SM, 
the minimal supersymmetric SM (MSSM)\,\cite{SUSY} 
also does not provide nonzero neutrino masses. 
Therefore, it is important to extend MSSM appropriately
to accommodate the massive neutrinos.

\vspace*{5mm}
\noindent 
{\bf 5.2.2A.}  
{\tt Scale of Mass Generation for Majorana Neutrinos via MSSM Seesaw
}
\vspace*{3mm}

A minimal extension of the MSSM, for instance,
is to introduce a heavy right-handed 
singlet neutrino $\nu_R^{~}$ to 
form a seesaw mechanism\,\cite{nu-seesaw}.
The MSSM superpotential can thus be generalized to
\beq
\label{eq:W-MSSM}
\ba{lcl}
W & =& \dis W_{\rm MSSM} +\Delta W\\[3mm]
  & =& \dis \[H_d L {\bf y_e^{~}} \ov{e} + H_d Q{\bf y_d^{~}}\ov{d}
                              - H_u Q{\bf y_u^{~}}\ov{u}
              -\mu H_d H_u\]
      +\[-H_u L {\bf y^{~}_\nu} \ov{n} -\ov{n}M_R^{~}\ov{n}\,\],
\ea  
\eeq
where \,$\ov{n}$\, is a chiral superfield\footnote{The singlet 
field \,$\nb$\, can expand in flavor space, e.g., one for each 
family, and the mass $M_R^{~}$ then becomes a $3\times 3$ matrix.
In fact, the leptogenesis mechanism\,\cite{LEPG} for explaining 
the cosmological baryon asymmetry of the universe 
necessarily requires at least two right-handed
neutrinos to ensure the existence of the 
needed {\tt CP}-violation phase from the seesaw 
sector (called minimal neutrino seesaw)\,\cite{2nuLEPG}.} 
containing a right-handed singlet neutrino  \,$\ov{\nu}_R^{~}$\,
and its scalar-partner \,$\widetilde{\ov{\nu}}_R^{~}$\,,\, 
while the other chiral superfields have the same meaning as in MSSM. 
Integrating out the heavy $\ov{n}$ field, we obtain a generic
(model-independent) dimension-5 effective operator for the MSSM
Lagrangian, 
\beqa
\label{eq:dim5-Hu}
\Delta {\cal L}_5 &=&\dis
-\f{\,m_\nu^{ij}\,}{v_u^2}{L^{\alpha}_i}^T \widehat{C}L^{\beta}_j
 H_u^{\alpha'}H_u^{\beta'}
\epsilon^{\alpha\alpha'} \epsilon^{\beta\beta'} \,+\, {\rm H.c.}
\,,
\eeqa
where 
$\,\dis H_u = 
   \(\phi_u^+,\,\f{v_u+\phi_u^0+ip_u^0}{\sqrt{2}}\)^T\,$,\,
\,$v_u = v\sin\beta \,$,\,
and $\,\tan\beta = \lan H_u\ran /\lan H_d\ran\,$.\,
This generates new interactions of the 
Majorana neutrinos with the Goldstone bosons and Higgs
particles,
\beqa
\label{eq:dim5-HH-nu}
\Delta {\cal L}_5^{\nu\nu} &=&\dis
-\f{\,m_\nu^{ij}\,}{2v_u^2}\({\nu_{Li}}^T\, 
    \widehat{C}\,\nu_{Lj}\)\(H_{u2}\)^2
\,+\, {\rm H.c.}
\,.
\eeqa
This Lagrangian becomes, in the mass eigenbasis,
\beqa
\label{eq:dim5-HH-XX}
\Delta {\cal L}_5^{\nu\nu} &=&\dis
-\f{\,\,m_{\nu j}^{~}\,}{2}\ov{\chi}_j^{~}\chi_j^{~}
\,-\,\f{m_{\nu j}^{~}}{~v^2\sin^2\!\beta~}
\ov{\chi}_j^{~}\[\,{\mathbb C}(H_u) - i\gamma_5^{~}
                   {\mathbb D}(H_u)\,\]\!\chi_j^{~} \,,
\eeqa
where
\beq
\label{eq:dim5-HHXX-CD}
\ba{ll}
{\mathbb C}(H_u) & =  \dis
v\sb\,\phi^0_u + \f{1}{2}\[{\phi_u^0}^2-{p_u^0}^2\]  \,,
\\[4mm]
&= \dis
v\ssb\(\cca h^0\!+\!\ssa H^0\) + \f{1}{2}
\[\ccaa {h^0}^2 \!+\! \ssaa {H^0}^2 \!+\! s_{2\alpha}^{~}\,h^0H^0
\!-\!\ssbb {\pi^0}^2 \!-\!\ccbb {A^0}^2 
\!-\! s_{2\beta}^{~}\,\pi^0A^0
\]  ,
\\[5mm]  
{\mathbb D}(H_u) & =  \dis
v\sb\,p^0_u + \phi_u^0\,p_u^0  
\\[4mm]
& = \dis
v\ssb\(\ssb \pi^0\!+\!\ccb A^0\) + \[
\cca\ssb \,\pi^0h^0+\cca\ccb\, h^0A^0 +
\ssa\ssb\,\pi^0H^0 +\ssa\cca   H^0A^0
\],
\ea
\eeq
and
$\,(\ssa ,\,\cca )\equiv (\sin\alpha,\,\cos\alpha)\,$,\,
$\,(\ssb ,\,\ccb )\equiv (\sin\beta,\, \cos\beta)\,$\,
and
$\,(s_{2\alpha}^{~},\,s_{2\beta}^{~})\equiv 
(\sin 2\alpha ,\,\sin 2\beta)\,$.\,
The fields \,$(\pi^0,\,\pi^\pm)$\, and 
$\,(h^0,\,A^0,\,H^0,\,H^\pm)\,$ are would-be Goldstone bosons
and physical Higgs scalars, respectively.
Here \,$\alpha$\, denotes the usual mixing angle in the neutral
Higgs sector of the MSSM\,\cite{2HDM-alpha}.  
As an important feature of the linearly realized MSSM Higgs
sector with fundamental Higgs bosons, we see that
the new interactions in 
(\ref{eq:dim5-HH-XX}) contain at most two Goldstone/Higgs 
bosons, involving the neutral (pseudo-)scalars only.
This means that the leading amplitudes of 
high energy Majorana neutrino scattering come
from the contact diagrams with a two-body
final state, i.e., $\,n=2\,$ [similar to Fig.\,1(a)], 
which are of order  $\,E^1\,$.\,  
Hence, the best unitarity limit is expected to be
of  $\,O(v^2/m_{\nu j}^{~})\,$, similar to the classic 
bound derived for the SM case\,\cite{scott}.     
Also, with (\ref{eq:dim5-HH-XX})
we immediately deduce the scattering amplitudes
\beq
\label{eq:MSSM-nunuWW}
\ba{lcl}
 \T\[\chi_{j\pm}^{~}\ov{\chi}_{j\pm}^{~}\to W^+_LW^-_L\] &=&
-\T\[\chi_{j\pm}^{~}\ov{\chi}_{j\pm}^{~}\to \pi^+\pi^-\] 
~=~ 0 + \O(E^0) \,,
\\[4mm]
\T\[\chi_{j\pm}^{~}\ov{\chi}_{j\pm}^{~}\to H^+H^-\] 
&=& 0 + \O(E^0) \,,
\ea
\eeq
whose leading order behavior only starts at $\,\O(E^0)\,$,\,
and is due to the exchange of $t$ channel charged leptons.
Comparing the first amplitudes in (\ref{eq:MSSM-nunuWW})
and (\ref{eq:nunu-VV}), we note that the reason that
\,$\T [\chi_{j\pm}^{~}\ov{\chi}_{j\pm}^{~}\to W^+_LW^-_L ]$\,
vanishes at $\,\O(E^1)\,$ in the present case is due to the 
cancellation from the $s$-channel Higgs contribution which is 
absent in (\ref{eq:nunu-VV}) where we did not assume,
{\it a priori,}
the Higgs Yukawa interactions for the neutrino mass generation.
In (\ref{eq:MSSM-nunuWW}),
the absence of a residual $\O(E^1)$ contribution to the
$W^+_LW^-_L$ channel becomes evident by inspecting 
the corresponding Goldstone amplitude 
$\T [\chi_{j\pm}^{~}\ov{\chi}_{j\pm}^{~}\to \pi^+\pi^- ]$
and the interaction Lagrangian (\ref{eq:dim5-HH-XX}).

For the current analysis of the Majorana neutrino scattering
into Higgs/Goldstone boson pairs in the MSSM, it is
clear that the relevant scattering energy is at the GUT scale.
So it is safe to ignore all the masses of the final state
Higgs bosons, and  accordingly the mixing angle \,$\alpha$\,
in the neutral Higgs mass matrix.
Under this massless limit, we have 
$\,\phi_u^0\simeq h^0\,$.\,
Also, we will consider the parameter region with
moderate to large 
\,$\tan\beta \sim 10-50$\, in the discussion below, 
which implies \,$\ssb \geqq 0.995\simeq 1$\, and 
\,$\ccb \leqq 10^{-1}\ll 1$\,\,
so that $\,p_u^0\simeq \pi^0\,$.\,
Thus, for estimating the optimal unitarity limit,
it suffices to set \,$\ccb \approx 0$\,.\,
With these we see that the interactions in
(\ref{eq:dim5-HH-XX}) are greatly simplified and reduce to the
case of one Higgs doublet,
\beqa
\label{eq:dim5-HH-XX-SM}
\Delta {\cal L}_{5{\rm int}}^{\nu\nu} & \simeq & \dis
-\,\f{\,m_{\nu j}^{~}\,}{~v^2~}\,
\ov{\chi}_j^{~}\[\,\(vh^0+\f{1}{2}\({h^0}^2-{\pi^0}^2\)\)
- i\gamma_5^{~}\(v\pi^0 + \pi^0h^0\)
\,\]\!\chi_j^{~} \,.
\eeqa
%
%for $\,\ssa \approx 0\,$ and
%    $\,\ssb \approx 1\,$.\,
From this we compute the leading amplitudes,
\beq
\ba{lcl}
\dis
 \T\[\chi_{j\pm}^{~}\ov{\chi}_{j\pm}^{~}\to \f{1}{\sqrt{2}}Z_LZ_L\] 
&=& \dis
-\T\[\chi_{j\pm}^{~}\ov{\chi}_{j\pm}^{~}\to \f{1}{\sqrt{2}}\pi^0\pi^0\] 
~\simeq~ \dis\pm\f{\,\sqrt{2}m_{\nu j}^{~}\,}{v^2}\sqrt{s} \,,
\\[5mm]
\dis
\T\[\chi_{j\pm}^{~}\ov{\chi}_{j\pm}^{~}\to \f{1}{\sqrt{2}}h^0h^0\] 
&\simeq&  \dis\pm\f{\,\sqrt{2}m_{\nu j}^{~}\,}{v^2}\sqrt{s} \,,
\\[5mm]
\dis
\T\[\chi_{j\pm}^{~}\ov{\chi}_{j\pm}^{~}\to \f{1}{\sqrt{2}}Z_Lh^0\] 
&=& \dis
-i\,\T\[\chi_{j\pm}^{~}\ov{\chi}_{j\pm}^{~}\to \pi^0 h^0\] 
~\simeq~ \dis +\f{\,2m_{\nu j}^{~}\,}{v^2}\sqrt{s} \,.
\ea
\eeq
To derive the optimal unitarity limit, it is desirable to consider the
scattering in the spin-singlet channel with a normalized final state,
$\,\dis\f{1}{\sqrt{2}}
\[|\chi_{j+}^{~}\ov{\chi}_{j+}^{~}\ran -
  |\chi_{j-}^{~}\ov{\chi}_{j-}^{~}\ran \]
\to \f{1}{2}\[|Z_LZ_L\ran + |h^0h^0\ran\]
\,$.\,
Evaluating the corresponding $s$-wave unitarity limit
gives,
\beqa
\label{eq:UB-MSSM} 
E^{\star}_{\nu} &<& \dis\f{\,4\pi v^2\,}{m_{\nu j}} \,, 
\eeqa
which is at the GUT scale, in agreement 
with the bound mentioned below Eq.\,(\ref{eq:nuB22}).

It is no surprise that 
the MSSM unitarity limit (\ref{eq:UB-MSSM}) on the scale of
Majorana neutrino mass generation from the Weinberg
operator (\ref{eq:dim5-Hu}) is as high as the GUT scale. 
This is because the nonrenormalizability of the dimension-5 
operator (\ref{eq:dim5-Hu}) arises solely from integrating
out the heavy right-handed singlet neutrino  $\nu_R^{~}$
{\it which obeys the
decoupling theorem.}   
The other part of the neutrino seesaw sector is 
the Dirac mass term $m_D^{~}$, 
which is already generated from the {\it renormalizable} Yukawa
interactions with the fundamental Higgs doublet $H_u$.
As a result, the MSSM Higgs doublet $H_u$ does participate
in the neutrino mass generation, and therefore {\it the minimal scale
of the mass generation for Majorana neutrinos in the MSSM is
already known and is partly set 
by the Higgs masses of $\,\O(v)\,$, 
which are extremely low and right at the weak scale!}   
This is why including a fundamental Higgs doublet in
the dimension-5 operator (\ref{eq:dim5-Hu}) causes  
unitarity violations at the GUT scale 
which could be saturated only
by the heavy right-handed singlet neutrino $\nu_R^{~}$\,.\,
In the next subsection we will provide an interesting counter 
example showing that invoking the SUSY breaking sector for the
neutrino mass generation would even allow the right-handed
neutrino mass $M_R$ to be as low as the weak scale
and thereby evade the bound (\ref{eq:UB-MSSM}).

\vspace*{5mm}
\noindent 
{\bf 5.2.2B.}  
{\tt
Unitarity Violation vs.$\!$ SUSY Breaking Induced Neutrino Masses} 
\vspace*{3mm}

In addition to the SUSY flavor problem mentioned earlier,
the MSSM further suffers a conceptual $\mu$-problem, namely, the
supersymmetric $\mu$ parameter in (\ref{eq:W-MSSM}) is expected
to be at the Planck scale ($\MP$) 
which contradicts the original motivation
of stabilizing the Higgs mass at the weak scale. 
A resolution of this is to invoke a global symmetry
$\mathbb{G}$ which forbids the vector-like mass term $H_dH_u$
and is broken only by the SUSY breaking sector.
In fact, both the supersymmetric $\mu$ and $M_R$ terms 
in Eq.\,(\ref{eq:W-MSSM}) are bi-linear and invariant under the
SM gauge group, so we wonder why the same suppression mechanism 
would not simultaneously keep both of them 
around the weak scale.
This motivates the recent proposal of generating small neutrino 
masses from SUSY breaking\,\cite{nu-susy}. 
For the current discussion, 
we will analyze a variation of the model
in \cite{nu-susy} and effectively generate both a dynamical
$\mu$-term and a seesaw mechanism at the weak scale.
Consider that the SUSY breaking comes from a hidden sector at 
an intermediate scale \,$\MI\approx \sqrt{v\MP}$\, via fields
$\B$ with 
\,$\lan\B\ran_F = F \approx \MII \approx v\MP$\,
and
\,$\lan\B\ran_A =\sqrt{F} \approx \MI$\,.\, 
The SUSY breaking
is communicated to the SM via the gravitational interaction so
that the effective theory is cut off at the scale 
\,$\cut \approx \MP$\,.\, 
Then, take a subset of fields 
\,$X,Y \in \B$\,,\, which, together with
the field $\ov{n}$ in the visible sector, are charged under
\,$\mathbb{G}$\,.\,
Thus, the $\mu$-term itself is forbidden by $\mathbb{G}$, but 
can be dynamically generated from the SUSY-breaking operator 
\beq
\label{eq:XYHH}
\f{1}{\,\cut\,}\[X^2 H_dH_u\]_F^{~}\,,
\eeq
leading to the dynamical $\mu$-parameter,
\beq
\mu ~\approx~ \dis\f{\,\lan X\ran_A^2\,}{\cut}
    ~\approx~ \f{\MII}{\cut} 
    ~\approx~ \f{\MII}{\MP} ~\approx~ v \,.
\eeq
Similarly, there are additional operators associated
with $\ov{n}$, which contribute the following Lagrangian terms,
\beq
\label{eq:sMaj}
\Delta {\cal L} ~=~
\dis -\f{c^{~}_{1}}{\cut}[ Y^\dag \ov{n}\,\ov{n} ]_D^{~}
     -\f{c^{~}_{2}}{\cut}\[YH_uL\ov{n}\]_F^{~} 
     \,+\, {\rm H.c.} 
\eeq
The operators (\ref{eq:XYHH}) and (\ref{eq:sMaj}) 
can be justified from the $R$-symmetry 
under which $(\,X,\,Y,\,\nb )$ have charges 
$\(1,\,\f{4}{3},\,\f{2}{3}\)$ and
$L$ and $H_u$ are $R$-singlets.
In (\ref{eq:sMaj}) 
we could have additional operators involving the
$X$ fields but they are suppressed 
by higher powers of \,$1/\cut$.\, 
A proper superpotential in the hidden sector will ensure 
that $(X,\,Y)$ develop VEVs for their $F$ and $A$ components,
i.e.,
$\,{\lan X\ran}_F = {\lan Y\ran}_F
   =F \approx v\MP 
   \approx (10^{10}\,{\rm GeV})^2$ \,  and 
$\,{\lan X\ran}_A = {\lan Y\ran}_A =\sqrt{F}$\,.\,
Now, the first term in (\ref{eq:sMaj}) generates
a weak-scale Majorana mass for $\ov{n}$, 
$M_R^{~} \approx F/\cut\approx F/\MP \approx v$\,,\, 
while its second
term gives the Dirac mass term
\,$m_D^{~} \approx v\sqrt{F}/\cut 
         \approx v\sqrt{v\MP}/\MP
         \approx \sqrt{v^3/\MP} 
         = \O({\rm keV})$\,.\, 
Hence, a non-canonical seesaw mechanim occurs at  
tree-level,
\beq
\ba{lcl}
{\cal M}_\nu \approx
\left\lgroup
\ba{cc}
& \\[-2.5mm]
0                      & \dis\f{\,v\sqrt{F}}{\cut}
\\[5mm]
\dis\f{\,v\sqrt{F}}{\cut} & \dis\f{\,F\,}{\cut} 
\\[-2.5mm]
&
\ea 
\right\rgroup ,
&~~\Longrightarrow~~ & 
{\cal M}_\nu^{\rm diag} =
\left\lgroup
\ba{cc}
& \\[-2.5mm]
\dis\O\(\f{\,v^2}{\cut}\) & 0 
\\[6mm]
0 & \O\(v\)
\\[-2.5mm]
&
\ea 
\right\rgroup ,
\ea
\eeq
which predicts a right-handed singlet neutrino
$\nu_R^{~}$ at the weak-scale
in addition to the conventional seesaw
Majorana masses of \,$\O\!\(\dis\f{v^2}{\cut}\)$\, 
for the active neutrinos. 
The light seesaw mass is retained in a nontrivial way: 
the global symmetry $\mathbb{G}$ properly suppresses
both the Dirac mass $m_D^{~}$ and Majorana mass $M_R$ 
simultaneously.
Now, we can re-analyze the unitarity violation 
limit from Majorana neutrino scattering 
\,$\ov{\X}\X\to V_L^aV_L^a$\,  in the effective theory 
with $\nb$ ``integrated out''. Since the effective
Lagrangian remains the same as Eq.\,(\ref{eq:dim5-Hu})
in Sec.\,5.2.2A,
we find the unitarity violation scale to be still as
high as the GUT scale, 
despite the fact that the right-handed $\nu_R^{~}$
mass is actually at the weak scale. Indeed, such a
SUSY induced neutrino mass mechanism has a 
``screening'' feature that {\it hides} the real neutrino
mass generation scale to well below that revealed from
the unitarity violation of high energy neutrino-neutrino
scattering\footnote{This ``seesaw screening'' 
occurs at tree level rather than loop level,
unlike Veltman's screening theorem\,\cite{scr} 
for the SM Higgs sector.}.
However, since the $\nu_R^{~}$ is so light, 
the most effective way to probe the scale of neutrino mass generation 
in this scenario is by  the {\it direct production} of
$\nu_R^{~}$ and its scalar-partner $\widetilde{\nu}_R^{~}$,
which can be the LSP (lightest supersymmetric particle),
and has unconventional predictions for cold dark 
matter\footnote{Ref.\,\cite{rLPG} also analyzed a model similar
to \cite{nu-susy}
for TeV-scale resonant leptogenesis.} 
and for the collider signatures of 
Higgs bosons\,\cite{nu-susy}.

\newpage
%\vspace*{6mm}
\section{\hspace*{-6mm}.$\!$  %$V_LV_L\to nV_L$: 
On the Scale of Electroweak Symmetry Breaking}
\vspace*{4mm}

The weak gauge bosons $V_L^a$ $(=W^\pm,\,Z^0)$ have the
bare mass term,
$M_W^2W^{+\mu}W^-_\mu+\f{1}{2}M_Z^2Z^\mu Z_\mu$,
which can be made gauge-invariant by the
dimension-2 nonlinear operator of the Goldstone kinetic term
in Eq.\,(\ref{eq:V-mass}) and is necessarily nonrenormalizable. 
The unitarity violation in the high energy scattering
$V_L^{a_1}V_L^{a_2}\to nV_L^a$ $(n\geqq 2)$
can be conveniently analyzed in the
corresponding Goldstone boson scattering
$\pi^{a_1}\pi^{a_2}\to n\pi^a$ according to the 
equivalence theorem (ET).
The power counting method in Sec.\,2.2 shows that the amplitude for this
$2 \to n$ scattering
behaves as $\O\!\(\dis\f{E^2}{v^n}\)$, where the $E$-power dependence
is independent of the number of external lines 
[cf. Eq.\,(\ref{eq:Torder})].

%%%%%%%%%%%%%%%%%%%%%%%%% Fig.8 %%%%%%%%%%%%%%%%%%%%%%%%%%%%%%%%%%%%
\begin{figure}[h]
\label{fig:VV-4V}
\begin{center}
\vspace*{-10mm}
\hspace*{-4mm}
\includegraphics[width=17cm,height=7.6cm]{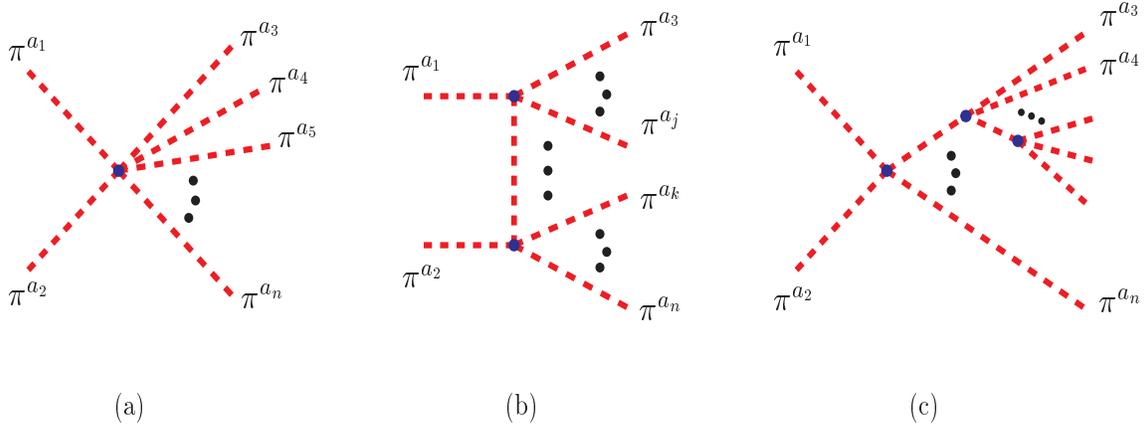} 
\vspace*{-14mm}
\caption{Illustration of the relevant Feynman diagrams for the
scattering amplitudes of
\,$\pi^{a_1}\pi^{a_2}\to n\pi^a$\, ($n\geqq 4$)\, 
which are of \,$\O(E^2/v^n)$\,.\, 
(a) The ``contact'' diagrams;
(b) the ``$t(u)$-channel'' type diagrams;
(c) the ``jet-like'' diagrams. 
}
\end{center}
\end{figure}
%%%%%%%%%%%%%%%%%%%%%%%%%%%%%%%%%%%%%%%%%%%%%%%%%%%%%%%%%%%%%%%%%%%%%
%\vspace*{-4mm}

For a quantitative analysis, we derive the pure Goldstone
Lagrangian from  Eq.\,(\ref{eq:V-mass}) and arrive at
\beq
\label{eq:L-npi}
\ba{l}
{\cal L}_\pi ~=~
\dis
\f{1}{2}\partial^\mu\overrightarrow{\pi}\cdot
                    \partial_\mu\overrightarrow{\pi}
+\!\sum^\infty_{n({\rm even})=4}
\dis\f{(-)^{n\over 2}2^{n-2}}{n!\,v^{n-2}}
\(\overrightarrow{\pi}\cdot
  \overrightarrow{\pi}\)_{~}^{\f{n}{2}-2}
\[\(\overrightarrow{\pi}\cdot\partial_\mu
    \overrightarrow{\pi}\)_{~}^2 -
\(\overrightarrow{\pi}\cdot
  \overrightarrow{\pi}\)
\(\partial^\mu\overrightarrow{\pi}\cdot
  \partial_\mu\overrightarrow{\pi}\)
\],
\ea
\eeq
where 
\beq
\label{eq:L-npiX}
\ba{l}
\overrightarrow{\pi}\cdot\overrightarrow{\pi}
   =2\pi^+\pi^- \!+\pi^0\pi^0 \,,
\\[3mm]
\(\overrightarrow{\pi}\cdot\partial_\mu 
    \overrightarrow{\pi}\)_{~}^2 -
\(\overrightarrow{\pi}\cdot
  \overrightarrow{\pi}\)
\(\partial^\mu\overrightarrow{\pi}\cdot
  \partial_\mu\overrightarrow{\pi}\)
\\[2mm]
\hspace*{12mm}
=\(\pi^+\partial_\mu\pi^-\!+\pi^-\partial_\mu\pi^+\!+
   \pi^0\partial_\mu\pi^0\)^2  -
   \(2\pi^+\pi^- \!+{\pi^0}^2\)
   \(2\,\partial^\mu\pi^+\partial_\mu\pi^-\!+
      \partial^\mu\pi^0\partial_\mu\pi^0\).
\ea
\eeq
The cancellation of $(\pi^0)^4$ terms in the second equation of
(\ref{eq:L-npiX}) explicitly shows 
that all vertices of type $\(\pi^0\)^n$
($n\geqq 4$) in (\ref{eq:L-npi})
are absent, implying that the scattering process
$\,\pi^0\pi^0\to n\pi^0$\, ($n\geqq 2$) 
is forbidden at the tree-level.
Note that in Eq.\,(\ref{eq:L-npi})
each interaction vertex contains an even number
of Goldstone fields\footnote{This is why in our early
estimate of Fig.\,2 only the even $n$ values are marked for 
the unitarity bound on the scattering
$V_L^{a_1}V_L^{a_2}\to nV_L^a$\,. 
}.\, 
In contrast to the $2\to 2$ scattering discussed earlier
in Sec.\,2.3.2A, the scattering
$\pi^{a_1}\pi^{a_2}\to n\pi^a$ ($n\geqq 4$) contains additional
non-contact  Feynman diagrams as shown in Fig.\,8(b)-(c),
which all have the same leading $E^2$ power dependence. 
This greatly complicates the calculation of the cross sections so
that a numerical evaluation becomes necessary.
Other contributions with 
internal gauge boson exchanges are at most of $\O(g^2/v^{n-2})$
so that they do not threaten  unitarity and thus are not shown
in Fig.\,8.

To study the $2\to 4$ scattering, the relevant Goldstone
interaction Lagrangian is,
\beq
\label{eq:L-6pi}
\ba{l}
{\cal L}_\pi^{\rm Int}[n\leqq 6] ~=~
\dis
\dis
\[\f{1}{\,6v^2\,}-\f{1}{\,45v^4\,}
\(\overrightarrow{\pi}\cdot\overrightarrow{\pi}\)
\]
\[\(\overrightarrow{\pi}\cdot\partial_\mu
    \overrightarrow{\pi}\)_{~}^2 -
\(\overrightarrow{\pi}\cdot
  \overrightarrow{\pi}\)
\(\partial^\mu\overrightarrow{\pi}\cdot
  \partial_\mu\overrightarrow{\pi}\)
\] .
\ea
\eeq
From (\ref{eq:L-6pi}), we consider all possible $2\to 4$ scatterings,
classified into four types according to their
initial states:

\beq
\label{eq:2to4all}
\ba{lcll}
\pi^0\pi^0 &\to& \pi^0\pi^0\pi^0\pi^0 \, &~~
({\rm absent~at~tree~level})\,,      
\\
&\to& \pi^0\pi^0\pi^+\pi^- \,,  &         
\\
	   &\to& \pi^+\pi^+\pi^-\pi^- \, & ~~(\surd \,)\,;
\\[2mm]
\pi^+\pi^- &\to& \pi^0\pi^0\pi^0\pi^0 \,, &
\\
           &\to& \pi^+\pi^-\pi^0\pi^0 \, & ~~(\surd \,)\,,
\\
	   &\to& \pi^+\pi^+\pi^-\pi^- \,; &
\\[2mm]
\pi^\pm\pi^0 &\to& \pi^\pm\pi^0\pi^0\pi^0 \,, &
\\
	   &\to& \pi^\pm\pi^+\pi^-\pi^0 \,  & ~~(\surd \,)\,;
\\[2mm]
\pi^\pm\pi^\pm &\to& \pi^\pm\pi^\pm\pi^0\pi^0 \,, &
\\
	  &\to& \pi^\pm\pi^\pm\pi^+\pi^- \,  & ~~(\surd \,)\,;
\ea
\eeq
where among each type of processes, the one which has the largest cross 
section is marked by $\surd$ and accordingly, shown in 
Fig.\,9(a). In our numerical calculation, 
each incoming/outgoing
Goldstone boson field $\pi^a$ is put on the mass-shell
of its corresponding longitudinal weak boson $V^a_L$.
In Fig.\,9(a), we also depict the 
corresponding unitarity bound based on Eq.\,(\ref{eq:UC-IEn}).
We see that among all processes the channel
$\,\pi^0\pi^0 \to \pi^+\pi^+\pi^-\pi^- \,$ gives the
largest cross section and thus the best unitarity bound,
\beq
E^\star ~=~ 3.9\,{\rm TeV}\,. 
\eeq
In this channel, we observe that
the contribution from the contact graph is about a factor
of $1/(2-3)$ of that from the non-contact
graphs, and their interference is negative so that the
total cross section is smaller than the contact contribution
by about a factor of \,$2-3$\,.\, But, from Fig.\,9(b)
we see that the inclusion of all non-contact graphs only weakens
the unitarity bound slightly, from $E^\star = 3.6$\,TeV
to $E^\star = 3.9$\,TeV. This shows that using contact graphs
alone can provide a good estimate of the unitarity bound even
when the non-contact contributions have the 
same energy-power dependence. This is due to the fact that
an $\O(1)$ correction to the cross section of the 
$\pi^{a_1}\pi^{a_2}\to n\pi^a$ scattering
only affects the unitarity bound by a factor of
$\,[\O(1)]^{\f{1}{2n}}$ [cf. Eq.(\ref{eq:UBus})], 
which approaches unity as \,$n$\, 
becomes much larger than \,$2$\,.\,
Finally, we may expect to derive a
stronger bound  by considering the isospin
singlet channel, $I=0$\,.\, For the initial state, 
the \,$I=0$\, combination is given by %\,\cite{LQT,MVW},
$%\label{eq:I0-VVin}
\,|0_{\rm i}\rangle = \f{1}{\sqrt{3}\,}\dis\sum_a
|\pi^a\pi^a\rangle \,$.\,
For the $4\pi^a$ final state, using the
Clebsch-Gordan algebra we derive the 
following \,$I=0$\, combination,
\beq
\label{eq:I0-4Vf}
|0_{\rm f}\rangle ~\,=~\, \f{1}{\,\sqrt{120\,}\,}
 \[4|\pi^+\pi^+\pi^-\pi^-\rangle + 
   4|\pi^+\pi^-\pi^0\pi^0\rangle +
    |\pi^0\pi^0\pi^0\pi^0\rangle\] \,.
\eeq
Thus the scattering amplitude 
$\T[0]$ for the \,$I=0$\, channel is given by
\beq
\label{eq:T24-0}
\ba{ll}
\T[0] ~=~ \dis \langle 0_{\rm f}|\T|0_{\rm i}\rangle

      ~=\!\!\! & \dis\f{1}{\,6\sqrt{10}\,}
\(\,
8\T[+-,++--]+8\T[+-,+-00]+2\T[+-,0000] \right.
\\[3mm]
& ~~~~~~~~~~
\left. +4\T[00,++--]+4\T[00,+-00]+ \T[00,0000]
\,\) \,,
\ea
\eeq
where \,$\T[00,0000]=0$\, at tree-level.
The corresponding cross section of the \,$I=0$\, channel is
given by, for \,$s \gg M_{W,Z}^2$\,,
\beq
\sigma [0] ~=~ \dis\f{1}{\,2s\,}\int_{{\rm PS}_4} |\T[0]|^2 \,.
\eeq
A rough estimate would be to simply take the first five individual
amplitudes in Eq.\,(\ref{eq:T24-0}) to be equal,
say to the one with the largest value, 
$\pi^0\pi^0\to\pi^+\pi^+\pi^-\pi^-$. 
Then, we would have
\,$\T[0]\sim \f{13}{\,3\sqrt{10}\,} \T[00,++--] \simeq
1.37\,\T[00,++--]$\,,\, which implies
\beq
\ba{ll}
\sigma [0] & 
\sim\,\, \dis\f{1}{\,2s\,}\int_{{\rm PS}_4} 
\left| 1.37\,\T[00,++--]\right|^2 
\\[5mm]
           & =\,\, \dis 1.37^2\times 2!^2\times \sigma[00,++--] 
              \,\simeq\, 7.5\,\sigma[00,++--] \,,
\ea
\eeq
where the factor \,$2!^2$\, is due to the conversion of the phase space
integration into the channel
\,$\pi^0\pi^0\to\pi^+\pi^+\pi^-\pi^-$\, 
which contains two pairs of identical
particle in the final state.
Hence, the above estimated enhancement factor $7.5$ indicates a
reduction of the unitarity bound of 
\,$\pi^0\pi^0\to\pi^+\pi^+\pi^-\pi^-$\,
channel by factor of
\,$(1/7.5)^{\f{1}{2n}} = (1/7.5)^{1\over 8} \simeq 0.78$\,.\,
We see that the
estimated unitarity bound for the singlet $I=0$ channel is
about \,$3.9\times 0.78\simeq 3.0$\,TeV\, 
which, as expected, is only slightly
stronger than that of the best individual channel 
\,$\pi^0\pi^0\to\pi^+\pi^+\pi^-\pi^-$\, 
shown in Fig.\,9(a).
Thus, a further precise numerical calculation of the cross section
\,$\sigma[0]$\, is expected not to change the conclusion.

%%%%%%%%%%%%%%%%%%%%%%%%%  Fig.9  %%%%%%%%%%%%%%%%%%%%%%%%%%%
\begin{figure}[H]
\label{fig:4V-CS}
\begin{center}
\vspace*{-11mm}
\includegraphics[width=17cm,height=16.8cm]{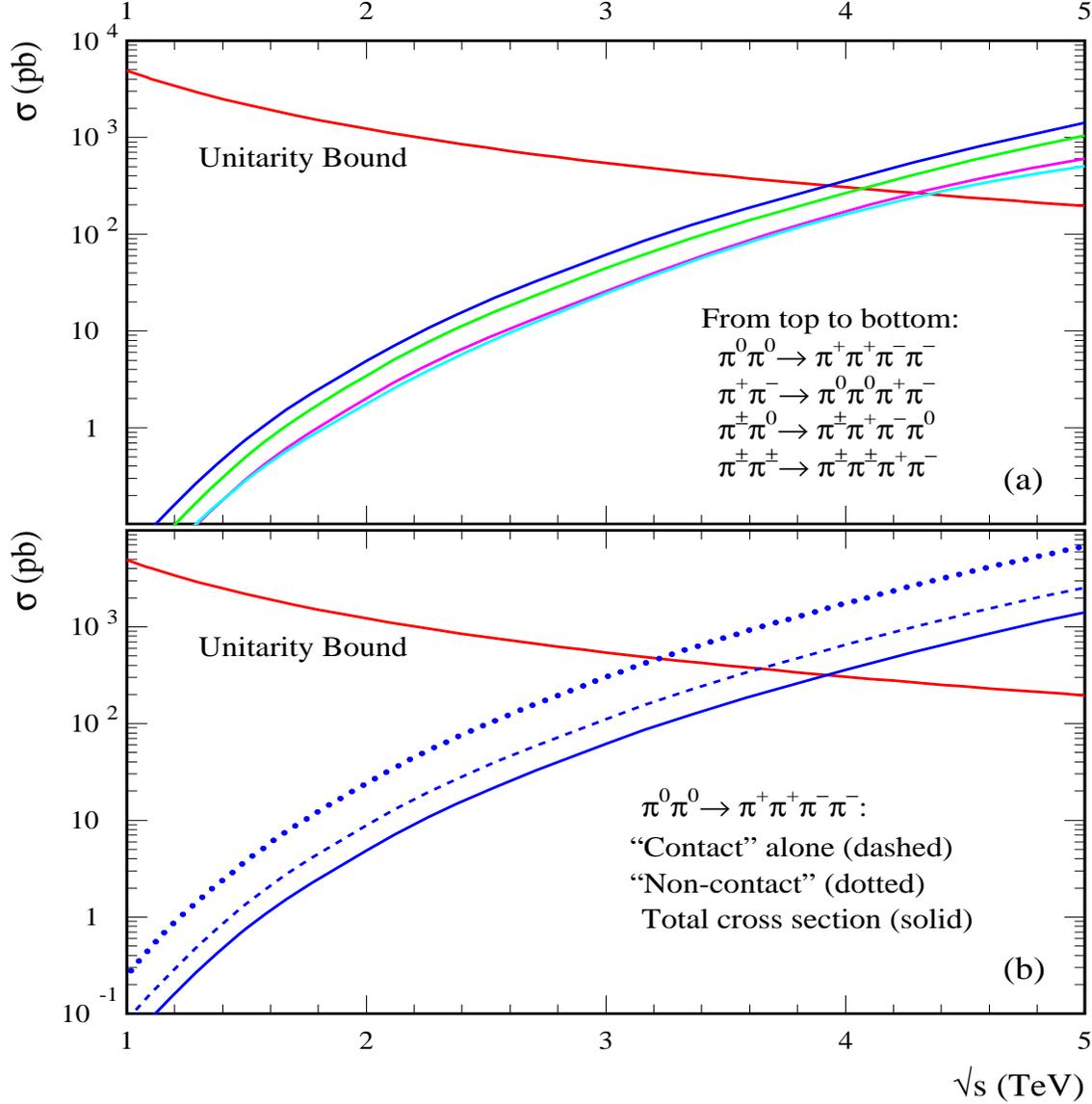} 
\vspace*{-13mm}
\caption{(a) Total cross sections of
the scattering processes 
\,$\pi^0\pi^0\to\pi^+\pi^+\pi^-\pi^-$,\,
\,$\pi^+\pi^-\to\pi^0\pi^0\pi^+\pi^-$,\,
\,$\pi^\pm\pi^0\to\pi^\pm\pi^+\pi^-\pi^0$,\, and\,
$\pi^\pm\pi^\pm\to\pi^\pm\pi^\pm\pi^+\pi^-$.\,
(b) For \,$\pi^0\pi^0\to\pi^0\pi^0\pi^+\pi^-$,\, we show 
the individual contributions from the contact graph alone
(dashed curve) and from the non-contact graphs
(dotted curve). The total cross section is depicted
by the solid curve. 
}
\vspace*{-4mm}
\end{center}
\end{figure}
%%%%%%%%%%%%%%%%%%%%%%%%%%%%%%%%%%%%%%%%%%%%%%%%%%%%%%%%%%%%%%%%%

In summary, Fig.\,9(a), together with
the above estimates, shows that the unitarity bound from 
$2\to 4$ scattering is about $3.0-3.9$\,TeV, significantly
weaker than that from the $2\to 2$ scattering [which is about
$1.2$\,TeV, cf. Eq.\,(\ref{eq:EWSB-UB22})].
This further supports our earlier estimate for 
$V_L^{a_1}V_L^{a_2}\to nV_L^{a}$ ($n\geqq 2$) in Fig.\,2 (Sec.\,3.2),
i.e., the strongest unitarity bound for the EWSB scale 
occurs at \,$n=2$,\, in contrast to the case for the
scales of fermion mass generation (cf. Sec.\,4-5).

%\newpage
\vspace*{6mm}
\section{\hspace*{-6mm}.$\!$
Conclusions}
\vspace*{4mm}

The requirement from unitarity that 
the partial wave amplitudes for $2\to 2$ scattering,
$V_L^{a_1} V_L^{a_2} \to V_L^{a_3} V_L^{a_4}$ and 
$f\bar{f},\,\nL\nL \to V_L^{a_1}V_L^{a_2}$,
be bounded for all energy has imposed nontrivial upper limits on the
energy scales at which the new physics 
responsible for the mass-generations
of the weak gauge bosons ($V^a=W^\pm,Z^0$) 
and fermions ($f,\nL$) must become evident.
The limit on the mass-generation scale for $V_L^a$ is about 1\,TeV
[cf. Eq.\,(1.1)] while the limits for the fermions are proportional to\, 
$v^2/m_{f,\nu}^{~}$\, 
[cf. Eqs.\,(\ref{eq:UBf22}) and (\ref{eq:nuB22x})] 
which is substantially above 1\,TeV  except for
the top-quark (cf. Table\,2), indicating that 
the scales of mass generation for all the light SM fermions might not be
directly accessible by  current or foreseeable 
high energy accelerators.
More recently it was pointed out\,\cite{scott}    
that the scattering \,$f\bar{f}\to n V_L^a$\,,\,  
with \,$n>2$\,,\,
could impose much stronger unitarity bounds for  
the light fermions when $n$ becomes
sufficiently large; these bounds 
are so severe that they approach the weak scale 
\,$v=(\sqrt{2}G_F)^{-\f{1}{2}}\simeq 246$\,GeV\, as $n$ becomes large 
[cf. Eq.\,(\ref{eq:UBfscot})]. 
This leads to the conclusion that the scattering
\,$f\bar{f}\to n V_L^a$\,  
does not reveal an independent new scale for  
fermion mass generation\,\cite{scott}.  
We further observe that such a bound would unavoidably contradict 
the kinematics [cf. Eq.\,(\ref{eq:KC})].
We have resolved this contradiction  
by treating the $n$-body phase space exactly.
The present analysis shows that {\it the scattering 
\,$f\bar{f}\to n V_L^a$\, ($n\geqq 2$) 
does reveal a separate scale for 
fermion mass generation.}  
Our bounds from  \,$2\to n$\, scattering with \,$n>2$\, 
establish a {\it new scale of mass generation for
all light fermions (including Majorana neutrinos)} 
and are {\it substantially stronger} 
than the conventional $2\to 2$ limits.

In the SM, the arbitrary Yukawa interactions of fermions with 
a single Higgs boson are assumed to be the
mechanism for the mass generation of all fermions; 
this is, however, not experimentally verified so far.  
Even if a SM-like light Higgs boson is
found to generate the electroweak symmetry breaking (EWSB) 
in forthcoming collider experiments, 
it could still be completely irrelevant to fermion mass generation 
because such a Higgs scalar 
is {\it not} required to couple to the SM fermions via Yukawa interactions 
by any known fundamental principle. This makes probing the scale of  
fermion mass generation an {\it independent} task.  
The crucial point is to note
that all the known fermions must join the SM gauge interactions 
and must also be chiral rather than
vector-like under the electroweak gauge group. 
Therefore, a bare mass term
for a SM fermion can be made gauge-invariant 
only in the nonlinear realization of
the electroweak gauge symmetry 
[cf. Eqs.\,(\ref{eq:f-mass}) and (\ref{eq:nu-mass})],
so it is necessarily nonrenormalizable and
thus will cause unitarity violation   
in high energy fermion-antifermion scattering.
The energy where such a unitarity violation happens provides a
{\it model independent upper bound} on the scale at which
the new physics responsible for fermion mass generation must be manifest.
The gauge-invariant, nonlinear formulation of the bare fermion mass terms 
also allows us to efficiently compute 
the high energy scattering amplitude of
\,$f\bar{f} \to n V_L^a$\, from that with
the corresponding final-state Goldstone bosons,
$f\bar{f} \to n \pi^a$,
by using the equivalence theorem. 
This is extremely valuable when the 
final state involves multiple longitudinal weak bosons.

For quarks and charged leptons, the scale of  unitarity violation 
is shown, as a function of
the number of final state weak bosons, $n$, in Fig.\,3.
These results depend on two opposing factors, one of which 
causes the cross section of \,$2 \to n$\, scattering to increase  
as $n$ becomes large, similar to Eq.\,(\ref{eq:UBfscot}), and
one of which causes the cross section to decrease with $n$  
until it eventually overcomes the first factor completely.   
Strikingly, the {\it competition} between these two opposing factors 
makes the strongest energy bounds
occur at a proper value of \,$n>2$\, (except, perhaps, for the top quark).
The values of $n$ (called $n_s$) which give the strongest upper bounds 
are summarized in Table\,3.
These new limits depend on the fermion mass and 
range from about \,3.5\,TeV\, ($n_s\,=\,2$)\,
for top quarks to \,84\,TeV\, ($n_s\,=\,10$)\, for up quarks, 
and from \,34\,TeV\, ($n_s\,=\,6$)\,
for tau leptons to \,107\,TeV\, ($n_s\,=\,12$)\, 
for electrons.\footnote{These are the energies which,
for instance, could be within the reach of a 
Very Large Hadron Collider (VLHC)\,\cite{VLHC}.}\,
Hence,  while the masses of the
involved fermions decrease by a factor up to 
\,$\sim\! 3\,\times\,10^5$,\, 
the bounds weaken by only a factor of \,$\sim\! 30$\,
in contrast to the customary
\,$2\to2$\, limits which are weakened 
by a factor of \,$\sim\! 6\times 10^5$\, (cf. Table\,2).
These unitarity bounds could be affected somewhat 
by model-dependent effects from 
the EWSB mechanism through self-interactions among the 
$V_L^a$'s  ($\pi^a$'s),
but such changes are shown to be small, 
to decrease rapidly with the increase of $n$, 
and furthermore, to be in the direction of lowering 
the above limits for a SM or SM-like EWSB sector.
Such effects indicate that 
even for the {\it top quark} the best bound    
becomes stronger than the customary \,$2\to 2$\, limit 
and occurs at \,$n_s=3$\, rather than 
\,$n_s=2$\, [cf. Eq.\,(\ref{eq:UB-tppm})].

The bounds for the Majorana neutrinos 
depend on the neutrino mass \,$m_{\nu}^{~}$\,.\,
With the typical values of $m_{\nu}^{~}$
between \mbox{1\,eV} and \mbox{0.01\,eV} 
(as suggested by the current experiments
on neutrino oscillations, 
neutrinoless double-beta decays and astrophysics
observations),   the strongest bound varies from 136\,TeV to 170\,TeV 
with \,$n=n_s\,=\,20\,-\,24$.\,
Thus the scale of  neutrino mass generation 
is only a factor of \,$1.3-1.6$\, higher
than that for the electron despite the tiny mass ratio of 
$m_\nu^{~}/m_e^{~}\simeq 2\times (10^{-6}-10^{-8})$.\,
These bounds are universal, independent of the details of the 
mechanism for the neutrino mass generation,
even though the underlying dynamics may invoke mass parameters of a very
different scale such as the right-handed singlet neutrino mass 
$M_R$ in the traditional seesaw mechanism 
(where $M_R\sim M_{\rm GUT}$) 
or in the SUSY-breaking induced seesaw mechanism
(where $M_R\sim \O({\rm TeV})$),
or the charged scalar masses $M_{X,Y}$ in the radiative mechanism
(where $M_{X,Y}\sim \O({\rm TeV})$). 
For all these models, our universal new bounds essentially constrain 
the new physics scale of the Higgs-fermion Yukawa interaction, 
which must invoke extra new fields 
(such as right-handed neutrinos or charged Zee-scalars or triplet-Higgs) 
as needed for ensuring renormalizability and
generating lepton number violation 
(although our new bounds do not directly constrain
the masses $M_R$ and $M_{X,Y}$ of these new fields 
as they obey the decoupling theorem\,\cite{DCT}).
Thus, for all known fermions including Majorana neutrinos, 
the new physics of mass generation
must manifest itself at a scale below about $170$\,TeV.

We see that these bounds are very insensitive to the variation
of the fermion masses and when the mass value decreases from
that of the bottom quark ($\sim\!4$\,GeV) down to that of the Majorana
neutrinos ($\sim\! 1-0.01$\,eV), by about \,$9-11$\, orders of magnitude,
the corresponding unitarity limits
are only weakened by about a factor of \,$7$\, or less.
This strong non-decoupling feature for the
scale of new physics associated with light fermion mass generation is 
essentially due to the {\it chiral structure} of 
the fermion bare mass terms, i.e.,
the fact that all the left-handed SM fermions are weak-doublets but their 
right-handed chiral partners are weak-singlets 
or, possibly absent
for Majorana neutrinos with radiative mass generation.
It is this feature that
makes the coupling strength of fermions to multiple Goldstones
(or effectively, multiple longitudinal gauge bosons)
{\it proportional to the fermion mass}, 
so the decoupling theorem\,\cite{DCT} no longer applies.  
A systematic discussion of this is given in Sec.\,5.2.1.

%%%%%%%%%%%%%%%%%%%%%%%%%%%%%%%%%%%%%%%%%%%%%%%%%%%%%%%%%%%%%%%%
\begin{table}[t]
%\vspace*{-4mm}
\label{Tab:TabF}
\caption{  
Summary of the {\it strongest} unitarity bound 
\,$E^{\star\min}_{2\to n}$\, (in TeV) 
for each scattering $\,\xi^{~}_1\xi^{~}_2\to nV_L^a$\, 
(\,$\xi^{~}_{1,2} = V_L,\,f,\,\bar{f},\,\nL$)
and the corresponding number of final state particles 
\,$n=n^{~}_s$\,,\, 
in comparison with the classic $2\to 2$ limit
\,$E^{\star}_{2\to 2}$\, (in TeV). 
For Majorana neutrinos, \,$m_{\nu}^{~}=0.05$\,eV is chosen.
}
\vspace*{3mm}
\hspace*{-3mm}
%\begin{center}
\begin{tabular}{c||c|cccccc|ccc|c}
\hline\hline
&&&&&&&&&&&\\[-2.5mm]
$\xi_1^{~}\xi_2^{~}$
& $V_L^{a_1}V_L^{a_2}$          &  $t\ov{t}$  &   
$b\ov{b}$       & $c\ov{c}$     &  $s\ov{s}$  &
$d\bar{d}$      & $u\ov{u}$     &  $\tau^-\tau^+$  & 
$\mu^-\mu^+$    & $e^-e^+$      &  $\nL\nL$        
\\ [1.5mm]  
\hline\hline
&&&&&&&&&&&\\[-2.5mm]
$n_s^{~}$ 
& $2$  &  $2$  & $4$ & $6$ & $8$ & $10$ & $10$ & $6$ & $8$ 
& $12$ &  $22$   
\\[1.5mm]
\hline
&&&&&&&&&&&\\[-2.5mm]
$E^{\star(\min)}_{2\to n}$
& $1.2$  & $3.49$ & $23.4$ & $30.8$ 
& $52.1$ & $77.4$ & $83.6$
& $33.9$ & $56.3$ & $107$  & $158$    
\\[1.5mm]
\hline\hline
&&&&&&&&&&&\\[-2.5mm]
$E^{\star}_{2\to 2}$
& $1.2$ & $3.49$ & $128$ & $377$ & $6\!\times\! 10^3$  & $10^5$ 
& $2\!\times\! 10^5$     & $606$ & $10^4$ & $2\!\times\! 10^6$ 
& $1.1\!\times\! 10^{13}$   
\\[1.5mm]
\hline\hline
\end{tabular}
%\end{center}
\vspace*{2mm}
\end{table}
%%%%%%%%%%%%%%%%%%%%%%%%%%%%%%%%%%%%%%%%%%%%%%%%%%%%%%%%%%%%%%%%%%%%%%

The summary Table\,4 shows that 
the established {\it new scales of mass
generation} for all fermions (including Majorana neutrinos)
fall in a narrow range, $3 \sim 160$\,TeV.
This not only provides an important guideline for model-building,
but also motivate and quantify where to look for definitive  
{\it experimental tests,}
such as the proposed VLHC experiment\,\cite{VLHC} and 
the relevant astrophysical processes\,\cite{HeDicus2}, etc.

Finally, for pure gauge boson scattering, 
$V_L^{a_1} V_L^{a_2} \to n V_L^a$ ($\pi^{a_1} \pi^{a_2} \to n \pi^a$)
with $n\geqq 4$,
the quantitative calculations are 
much more difficult due to nontrivial
angular dependences in the leading energy terms 
of the $S$-matrix elements, 
but a comparison of \,$2 \to 4$\, 
scattering with \,$2 \to 2$\, scattering indicates
that the rapidly decreasing phase space factor 
dominates and the best bound remains
the customary \,$2 \to 2$\, bound of \,$\sim\! 1$\,TeV. 
This picture is further 
supported by our general estimates for the $2\to n$ 
gauge boson scattering with an 
arbitrary value of $n$
as shown in Fig.\,2 and Table\,1, 
where our similar estimates for the upper bounds 
on the scales of mass generation
of the quarks, leptons and Majorana neutrinos are
consistent with the quantitative calculations 
within a factor 2 or so (cf. Sec.\,4-5).
Also, the current study of \,$f\fbar \to nV_L^a\,(n\pi^a)$\,
scattering assumes $V_L^a$ ($\pi^a$) 
to be a local field or remain local up to the limit 
\,$E^{\star}_{f}$\,.\,
If $V_L^a$ ($\pi^a$) becomes composite much below 
$E^{\star}_{f}$, the pure fermion process
$\,f\fbar \to (f\fbar)^n\,$
may be useful for a model-independent analysis.

In conclusion, 
Table\,4 summarizes all the optimal $2\to n$ unitarity bounds 
on the scales of mass-generation, as compared to the classic 
$2\to 2$ limits. 
A short summary of these key results is given in \cite{DH-PRL}.

\vspace*{5mm}\noindent
%\section*{Acknowledgments}
{\large\bf Acknowledgments}\\[2mm]
%\vspace{-3mm}
We thank S. Willenbrock for valuable discussions and a
careful reading of the manuscript. We also thank 
R. S. Chivukula, A. de Gouv\^{e}a, E. Ma, M. E. Peskin 
and W. W. Repko for discussing the unitarity issue, 
and U. Varadarajan and N. Weiner 
for discussing aspects of SUSY breaking
induced neutrino mass.
HJH thanks the SLAC Theory Group for hospitality
during the completion of this work.
This research was supported by U.S. Department of Energy under
grant No.~DE-FG03-93ER40757.

\newpage
%\vspace*{8mm}
\appendix

\noindent
{\bf {\Large Appendices}}
%\noindent

%\newpage
%\vspace{4mm}
\section{\hspace*{-6mm}.$\!$
Derivation of $\mathbf{n}$-Body Phase Space}
\vspace*{2mm}

For completeness, we summarize a derivation of the
$n$-body phase space formula (\ref{eq:PSnexact}) which 
is used in our analyses in Sec.\,3-5.
The phase space integral for $n$ particles is
\beq
\label{eqA:In}
\dis
{\cal I}_n=\int\f{d^3k_1\cdots d^3k_n}{2E_1\cdots 2E_n} \,
\d^{(4)}\!\(P-k_1-\cdots -k_n\) \,,
\eeq
where \,$P=p_1+p_2$\,.\, 
For the current purpose, it is safe to ignore the masses of
the final state particles, i.e., \,$m_j^2/E_j^2 \ll 1$\, 
($j=1,2,\cdots,n$)\,.\, Then we obtain the final formula,
\beqa
\label{eqA:Io}
{\cal I}_n &=& \(\f{\pi}{2}\)^{n-1}
\f{s^{n-2}}{\,(n-1)!(n-2)!\,} \,,
\eeqa
where $\,s=P^2\,$.

To derive Eq.(\ref{eqA:Io}), we carry out the following steps:
\\
{\bf (i).} Replace the delta function in (\ref{eqA:In})
with
\beq
\label{eqA:Qn2}
\dis
\int d^4 Q_{n-2}~\d^{(4)}\!\(Q_{n-2}-k_{n-1}-k_{n}\)
         \d^{(4)}\!\(P-k_1-\cdots -k_{n-2}-Q_{n-2}\) ,
\eeq
and multiply by
\beq
\label{eqA:sn2}
\dis
1 ~=~ \int ds_{n-2}~\d\!\(s_{n-2}-Q^2_{n-2}\)  .
\eeq

\noindent
{\bf (ii).} Use the first delta function
in (\ref{eqA:Qn2}) to do the integration over $k_{n-1}$
and $k_{n}$ in their center of mass frame, i.e.,
\beq
\dis
\int\f{\,d^3k_{n-1}\,d^3k_n\,}{2E_{n-1}\,2E_n}~
\d^{(4)}\!\(Q_{n-2}-k_{n-1}-k_n\) ~=~
\f{\pi}{2}  \,.
\eeq

\noindent
{\bf (iii).} Repeat step-{\bf (i)} by
replacing the second delta function in (\ref{eqA:Qn2}) by
\beq
\label{eqA:Qn3}
\dis
\int d^4Q_{n-3}~\d^{(4)}\!\(Q_{n-3}-k_{n-2}-Q_{n-2}\)
      \d^{(4)}\!\(P-k_1-\cdots -k_{n-3}-Q_{n-3}\)
\eeq
and multiplying by
\beq
\dis
1 ~=~ \int ds_{n-3}~\d\!\(s_{n-3}-Q^2_{n-3}\)  \,.
\eeq

\noindent
{\bf (iv).} Repeat step-{\bf (ii)} by
performing the integration over $k_{n-2}$ and $Q_{n-2}$
in their center of mass frame using the delta function in
(\ref{eqA:sn2}) and the first delta function in 
(\ref{eqA:Qn3}),
\beq
\dis
\int\f{\,d^3k_{n-2}\,}{2E_{n-2}}d^4Q_{n-2}~
\d^{(4)}\!\(Q_{n-3}-k_{n-2}-Q_{n-2}\) 
\d\!\(s_{n-2}-Q_{n-2}^2\)         
~=~ 
\f{\pi}{2}\f{\,s_{n-3}-s_{n-2}\,}{s_{n-3}}  \,.
\eeq

\noindent
{\bf (v).} Continue to repeat the steps {\bf (iii)} 
and  {\bf (iv)}.

The result of steps (i)-(v) is the integral
\beq
\dis{\cal I}_n ~=~    \(\f{\pi}{2}\)^{n-1}\!\!
\int_0^s \!\!         ds_1\f{s  \!-\!s_1}{s}
\!\int_0^{s_1} \!\!     ds_2\f{s_1\!-\!s_2}{s_1}
\!\int_0^{s_2} \!\!     ds_3\f{s_2\!-\!s_3}{s_2} \cdots
\!\int_0^{s_{n-3}} \!\! ds_{n-2}\f{s_{n-3}\!-\!s_{n-2}}{s_{n-3}}\,, 
\eeq
which can be directly evaluated to give (\ref{eqA:Io}).

As an alternate derivation, if we assume the validity
of (\ref{eqA:Io}) for an arbitrary \,$n\,(\geqq 2)$\,,\, 
then we deduce
\beq
\ba{ll}
\dis
{\cal I}_{n+1} &=~ \dis \(\f{\pi}{2}\)^n\int^s_0 ds_{n+1}
\f{\,s-s_{n+1}\,}{s}\f{\(s_{n+1}^{~}\)^{n-2}}{~(n-1)!\,(n-2)!~} 
\\[4mm]
&=~ \dis
\(\f{\pi}{2}\)^n
\f{s^{n-1}}{~n!\,(n-1)!~} \,,
\ea
\eeq
which proves Eq.\,(\ref{eqA:Io}) by induction.

\vspace*{4mm}
\section{\hspace*{-6mm}.$\!$
Refined Unitarity Conditions}
\vspace*{2mm}

For completeness we improve the unitarity conditions
in Sec.\,2.3.1 by keeping the masses of initial/final 
state particles.  
We first recompute the two-body phase space in Eq.\,(\ref{eq:SE22}) 
without approximation and obtain
\beqa
\label{eq:SE22x}
\dis\int_{{\rm PS}_2}
\left|\TT_{\rm el}[2\to 2]\right|^2
~=~ \dis\f{\,32\pi\eta_{\rm f}^{\rm el}\,}{\varrho_{\rm e}^{~}}
  \sum_{j} (2j+1)|a^{\rm el}_j|^2 \,,
\eeqa
where $\,\eta_{\rm f}^{\rm el}=\eta(s,m_3^2,m_4^2)\,$  with
$\,(m_3,\,m_4)\,$ the masses of final state particles,
and 
\beq
\label{eq:eta}
\eta(x,\,y,\,z) ~\equiv~ \dis\f{1}{x}
    \[x^2+y^2+z^2-2(xy+yz+zx)\]^{\f{1}{2}} \,.
\eeq
For $\,s\gg m_i^2,m_j^2$\,,\,
$\,\eta(s,\,m_i^2,\,m_j^2)\simeq 1$\, holds well. 
In the special case $\,m_3^{~}=m_4^{~}\equiv m\,$,\,
Eq.\,(\ref{eq:eta}) gives 
$~\eta(s,\,m^2,\,m^2)=\[1-4 m^2/s\]^{1/2}\,$.\,
For the elastic scattering $m_1+m_2 \to m_3+m_4$, we have
$\,\eta_{\rm in}^{~}  \equiv  
   \eta(s,\,m_1^2,\,m_2^2) = \eta(s,\,m_3^2,\,m_4^2)
   \equiv \eta_{\rm f}^{\rm el}\,$.\,
Thus we can derive a refined form of (\ref{eq:TUC2}),
\beq
\label{eq:TUC2x}
\dis
\sum_{j} (2j+1) \f{1}{\varrho_{\rm e}^{~}}
\[ \f{\varrho_{\rm e}^2}{4} - \( {\RE}\,\ah^{\rm el}_j\)^2
  -\( {\IM}\,\ah^{\rm el}_j - \f{\varrho_{\rm e}^{~}}{2}\)^2
\]
\,=~ \f{\eta_{\rm in}}{\,32\pi}\sum_n
\int_{{\rm PS}_n}\left|\TT_{\rm inel}[2\to n]\right|^2
~\geqq 0  \,,
\eeq
where
$\,\ah_j^{\rm el} \equiv \eta_{\rm in}^{~}a^{\rm el}_j
=\sqrt{\eta_{\rm in}^{~}\eta_{\rm f}^{\rm el}\,}
  a^{\rm el}_j
\,$
and 
$\,\eta_{\rm in}^{~} = \eta_{\rm f}^{\rm el}
                     = \eta(s,\,m_1^2,\,m_2^2)\,$.\,
Accordingly, the unitarity conditions
(\ref{eq:al-bound}) and (\ref{eq:UC-IEn}) become,
\beq
\label{eq:al-boundx}
\ba{rcccrcc}
\dis\left|{\RE}\,\ah^{\rm el}_j\right|
   &\leqq&  \dis\f{\,\varrho_{e}^{~}\,}{2} \,,
~~~&~~~
\dis\left|\,\ah^{\rm el}_j\right| &\leqq&  \dis\varrho_{e}^{~} \,,
\ea
\eeq
and
\beq
\label{eq:UC-IEnx}
\eta_{\rm in}^2\,
\sigma_{\rm inel}[2\to n] ~\leqq~ \dis\f{\,4\pi\varrho_e\,}{s} ~,
\eeq
where the total inelastic cross section is given by
$\,\dis\sigma_{\rm inel}[2\to n]=\f{1}{\,2s\eta_{\rm in}\,}
   \int_{{\rm PS}_n}^{~}\left|\TT_{\rm inel}[2\to n]\right|^2
$\,.\,
Similarly we rederive Eq.\,(\ref{eq:inel22a}) as
\beq
\label{eq:inel22ax}
\dis
\sum_j (2j+1)
\left\{
   \f{\varrho_{\rm e}^{~}}{4} - \f{1}{\varrho_{\rm e}^{~}}
   \[\( {\RE}\,\ah^{\rm el}_j\)^2
    +\( {\IM}\,\ah^{\rm el}_j - \f{\varrho_{\rm e}^{~}}{2}\)^2\]
\right\}
~>~ \sum_j (2j+1)\f{1}{\varrho_{\rm i}^{~}}
  | \ah_j^{\rm inel} |^2 ~,
\eeq
where
$~ \ah_j^{\rm inel} \equiv 
   \sqrt{\eta_{\rm in}^{~}\eta_{\rm f}^{\rm inel}\,}
 a^{\rm inel}_j\,$
and
$\,\eta_{\rm f}^{\rm inel}\equiv\eta (s,m_3^2,m_4^2)\,$.\, 
Thus, for $2\to 2$ inelastic scattering, we find that 
the unitarity condition (\ref{eq:a0-inel}) becomes
\beqa
\label{eq:a0-inelx}
|\ah_j^{\rm inel}| &<& \dis
\f{\,\sqrt{ \varrho_{\rm i}^{~}
            \varrho_{\rm e}^{~}\,}\,}{2} \,.
\eeqa
For the present study in Sec.\,3-6, we have
$\,s\gg m_j^2\,$ ($j=1,2,\cdots,n+2$), so that 
$\,\eta_{\rm in}^{~},\eta_{\rm f}^{~} \longrightarrow 1$\,,\,
under which all the formulas above reduce back to Sec.\,2.3.1.
In the traditional case of $2\to 2$ scattering, 
a similar effect of $\,\eta_{\rm in,f}^{~}$
due to the finite masses of the initial/final state
was noted before\,\cite{Ubook}.

%\vspace{8mm}
\newpage
\baselineskip 14.0pt  %13.84pt 

%\begin{center}
%{\bf {\Large References}}
%\end{center}
%\vspace*{-6mm}


\begin{thebibliography}{99}

\bibitem{SM1}
S. L. Glashow, Nucl. Phys. {\bf 22}, 579 (1961);
S. Weinberg, Phys. Rev. Lett. {\bf 19}, 1264 (1967);
A. Salam, in {\it Elementary Particle Theory,}
Nobel Symposium No.\,8, ed. N. Svartholm, pp.\,367 
(Almqvist and Wiksell, Stockholm, 1968).

  

\bibitem{SM2}
H. Fritzsch and M. Gell-Mann, in {\it Proceedings of 
XVI International Conference on High Energy Physics,}
eds. J. D. Jackson and A. Roberts
(Fermilab, Batavia, IL, 1972);
D. J. Gross and F. Wilczek, 
Phys. Rev. Lett. {\bf 30}, 1343 (1973);
H. D. Politzer,  
Phys. Rev. Lett. {\bf 30}, 1346 (1973).


\bibitem{CKM}
N. Cabibbo, Phys. Rev. Lett. {\bf 10}, 531 (1963);\\
M. Kobayashi and T. Maskawa, 
Prog. Theor. Phys. {\bf 49}, 652 (1973).


\bibitem{atm}
S. Fukuda, {\it et al.}, 
[Super-Kamiokande Collaboration], 
Phys. Rev. Lett. {\bf 85}, 3999 (2000);
{\bf 86}, 5656 (2001);
{\bf 82}, 1810 (1999);
{\bf 81}, 1562 (1998);
{\bf 81}, 1158 (1998); and  
 Y. Ashie, {\it et al.},
[Super-Kamiokande Collaboration], 
Phys. Rev. Lett. {\bf 93}, 101801 (2004) 
[{\tt hep-ex/0404034}];
M. Ishitsuka, {\tt hep-ex/0406076}.



\bibitem{sol}
S. Fukuda, {\it et al.}, 
[Super-Kamiokande Collaboration],
Phys. Rev. Lett. {\bf 86}, 5656 (2001); 
Q. R. Ahmad, {\it et al.,}
[SNO collaboration],
Phys. Rev. Lett. 
{\bf 87}, 071301 (2001).  %[nucl-ex/0106015].     
Q. R. Ahmad, {\it et al.,}
[SNO Collaboration], 
Phys. Rev. Lett. {\bf 89}, 011301 (2002)
[{\tt nucl-ex/0204008}] and
Phys. Rev. Lett. {\bf 89}, 011302 (2002)
[{\tt nucl-ex/0204009}]; 
B. Aharmim, {\it et al.}, 
[SNO collaboration],
{\tt hep-ex/0407029};
A. Bellerive, {\tt hep-ex/0401018}.


\bibitem{K2K}
S. H. Ahn, {\it et al.} [K2K Collaboration],
Phys. Rev. Lett. {\bf 93}, 051801 (2004)
[{\tt hep-ex/0402017}]
Phys. Rev. Lett. {\bf 90}, 041801 (2003);
Phys. Lett. B\,{\bf 511}, 178 (2001).

 


\bibitem{CHOOZ}
M. Apolloni, {\it et al.}, [CHOOZ Collaboration],
Phys. Lett. B{\bf 466}, 415 (1999);
Phys. Lett. B{\bf 420}, 397 (1998); 
F. Boehm, {\it et al.}, [Palo Verde Collaboration],
Phys. Rev. D{\bf 64}, 112001 (2001); 
Phys. Rev. Lett. {\bf 84}, 3764 (2000).
 
 

\bibitem{Kam}
K. Eguchi, {\it et al.},
[KamLAND Collaboration], Phys. Rev. Lett. {\bf 90}, 021802 (2003)
[{\tt hep-ex/0212021}];
T. Araki, {\it et al.},
[KamLAND Collaboration], Phys. Rev. Lett. {\bf 94}, 081801 (2005)
[{\tt hep-ex/0406035}].


\bibitem{MNS}
B. Pontecorvo, Sov. Phys. JETP {\bf 6}, 429 (1957);
Z. Maki, M. Nakagawa, and S. Sakata, 
Prog. Theor. Phys. {\bf 28}, 870 (1962).


\bibitem{nu-rev} 
For recent reviews, {\it e.g.,}
J. W. F. Valle, {\tt hep-ph/0410103};
G. Altarelli and F. Feruglio, 
New J. Phys. {\bf 6}, 106 (2004)
[{\tt hep-ph/0405048}] and [{\tt hep-ph/0206077}];
R. N. Mohapatra, New J. Phys. {\bf 6}, 82 (2004)
[{\tt hep-ph/0411131}];
V. Barger, {\it et al.,}
Int. J. Mod. Phys. E\,{\bf 12}, 569 (2003)
[{\tt hep-ph/0308123}];
M. C. Gonzalez-Garcia and  Y. Nir, 
Rev. Mod. Phys. {\bf 75}, 345 (2003)
[{\tt hep-ph/0202058}].



\bibitem{G4nu}
M. Maltoni, T. Schwetz, M. A. Tortola and J.\,W.\,F. Valle,
New J. Phys. {\bf 6}, 122 (2004)
[{\tt hep-ph/0405172}]; {\tt hep-ph/0305312};
Nucl. Phys. B\,{\bf 643}, 321 (2002) [{\tt hep-ph/0207157}];
and references therein.


\bibitem{BooNE}
J. Monroe, [MiniBooNE Collaboration],
[{\tt hep-ex/0406048}],   
in proceedings of the Moriond Electroweak 2004 Conference;
M. H. Shaevitz, [MiniBooNE Collaboration],
{\tt hep-ex/0407027};
A. Bazarko, {\it et al.},
[MiniBooNE Collaboration], 
Nucl. Phys. Proc. Suppl. {\bf 91}, 210 (2001).


\bibitem{LSND}
C. Athanassopoulos, {\it et al.,}
[LSND Collaboration],
Phys. Rev. Lett. {\bf 81}, 1774 (1998);
and A. Aguilar {\it et al.,} 
Phys. Rev. D{\bf 63}, 112001 (2001)
[{\tt hep-ex/0101039}].


\bibitem{WMAP}
D. N. Spergel, {\it et al.}, [WMAP Collaboration],
Astrophys. J. Suppl. {\bf 148}, 175 (2003)
[{\tt astro-ph/0302209}]. 


\bibitem{2dF}
O. Elgaroy, {\it et al.,}
Phys. Rev. Lett. {\bf 89}, 061301 (2002) 
[{\tt astro-ph/0204152}];
S. Hannestad, {\tt astro-ph/0205223};
O. Elgaroy and O. Lahav, JCAP {\bf 04}, 004 (2003)
[{\tt astro-ph/0303089}].


\bibitem{WMAP-2dF}
S. Hannestad, 
New J. Phys. {\bf 6}, 108 (2004) [{\tt hep-ph/0404239}]; 
JCAP {\bf 05}, 004 (2003) [{\tt astro-ph/0303076}];
and references therein.


\bibitem{Hannest}
S. Hannestad,  
{\tt hep-ph/0409108} and {\tt hep-ph/0412181};
O. Elgaroy and O. Lahav, New J. Phys. {\bf 7}, 61 (2005)
[{\tt hep-ph/0412075}].


\bibitem{higgs}
P. W. Higgs, Phys. Lett. {\bf 12}, 132 (1964);
             Phys. Rev. Lett. {\bf 13}, 508 (1964);
             Phys. Rev. {\bf 145}, 1156 (1966);
F. Englert and R. Brout,
             Phys. Rev. Lett. {\bf 13}, 321 (1964);
G. S. Guralnik, C. R. Hagen, and T. W. Kibble, 
             Phys. Rev. Lett. {\bf 13}, 585 (1964).


\bibitem{weinberg5}
S. Weinberg, Phys. Rev. Lett. {\bf 43}, 1566 (1979).  


\bibitem{nu-seesaw}
P. Minkowski, Phys. Lett. B{\bf 67}, 421 (1977); 
M. Gell-Mann, P. Ramond and R. Slansky,
in {\it Proceedings of the Workshop on Supergravity}, 
eds. F. van Nieuwenhuizen and D. Freedman, Amsterdam, p.315, 1979;
T. Yanagida, {\it Proceedings of the Workshop on Unified Theories
and Baryon Number in the Universe}, 
eds. O. Sawada and A. Sugamoto, KEK, Tsukuba, p.95, 1979;
R. N. Mohapatra and G. Senjanovic, 
Phys. Rev. Lett. {\bf 44}, 912 (1980).


\bibitem{nu-rad}
A. Zee, Phys. Lett. B{\bf 93}, 389 (1980);
        Phys. Lett. B{\bf 161}, 141 (1985);
L. Wolfenstein, Nucl. Phys. B{\bf 175}, 93 (1980);
S. T. Petcov, Phys. Lett. B{\bf 115}, 401 (1982).


\bibitem{DSB}
For an updated overview of dynamical symmetry
breaking and compositeness,   
C. T. Hill and E. H. Simmons, 
Phys. Rept. {\bf 381}, 235 (2003)
[{\tt hep-ph/0203079}];
and references therein.


\bibitem{seesaw}
E.g.,
B. A. Dobrescu and C. T. Hill, 
Phys. Rev. Lett. {\bf 81}, 2634 (1998)
[{\tt hep-ph/9712319}];
R. S. Chivukula, B. A. Dobrescu, H. Georgi, C. T. Hill,
Phys. Rev. D\,{\bf 59}, 075003 (1999)
[{\tt hep-ph/9809470}];\,
H.-J. He, C. T. Hill, T. Tait,
Phys. Rev. D{\bf 65}, 055006 (2002) 
[{\tt hep-ph/0108041}].


\bibitem{LH}
E.g., 
N.~Arkani-Hamed, A.~G.~Cohen and H.~Georgi,
Phys.\ Lett.\ B {\bf 513}, 232 (2001);
%%[arXiv:hep-ph/0105239]
%%CITATION = HEP-PH 0105239;%% 
N.~Arkani-Hamed, A.~G.~Cohen, T.~Gregoire and J.~G.~Wacker,
JHEP {\bf 08}, 020 (2002);
%hep-ph/0202089;
%%CITATION = HEP-PH 0202089;%%
N.~Arkani-Hamed, A.~G.~Cohen, E.~Katz, 
A.~E.~Nelson, T.~Gregoire and J.~G.~Wacker,
JHEP {\bf 08}, 021 (2002);
%hep-ph/0206020.
%%CITATION = HEP-PH 0206020;%%
N.~Arkani-Hamed, A.~G.~Cohen, E.~Katz, A.~E.~Nelson,
%``The littlest Higgs,''
JHEP {\bf 07}, 034 (2002);
%[hep-ph/0206021].
%%CITATION = HEP-PH 0206021;%%
I. Low, W. Skiba, D. Smith, 
Phys. Rev. D {\bf 66}, 072001 (2002);
D.\,E. Kaplan and M. Schmaltz, JHEP {\bf 10}, 039 (2003);
S. Chang and J.\,G. Wacker,
Phys. Rev. D {\bf 69}, 035002 (2004);
S. Chang, JHEP {\bf 12}, 057 (2003);
S. Chang and H.-J. He, Phys. Lett. B {\bf 586}, 95 (2004);
%[hep-ph/0311177]
%%CITATION = HEP-PH 0311177;%%
and references therein.


\bibitem{SUSY}
For reviews, {\it e.g.,}
P. Fayet and S. Ferrara,    Phys. Rept. {\bf 32}, 249 (1977);
H. P. Nilles,               Phys. Rept. {\bf 110}, 1 (1984);
H. E. Haber and G. L. Kane, Phys. Rept. {\bf 117}, 75 (1985);
and more recently,  H. E. Haber, {\tt hep-ph/0409008} and
{\tt hep-ph/0212136},
%Nucl. Phys. Proc. Suppl. {\bf 101}, 217 (2001)
%[{\tt hep-ph/0103095}]; 
and references therein.                             


\bibitem{SUSYhigh}
N. Arkani-Hamed and S. Dimopoulos,
{\tt hep-th/0405159}.


\bibitem{GUT}
H. Georgi and S. L. Glashow, Phys. Rev. Lett. {\bf 32}, 438 (1974);
H. Georgi, H. R. Quinn, and S. Weinberg, 
Phys. Rev. Lett. {\bf 33}, 451 (1974);
S. Dimopoulos and H. Georgi, Nucl. Phys. B{\bf 193}, 150 (1981);
S. Dimopoulos, S. Raby, and F. Wilczek, 
Phys. Rev. D{\bf 24}, 1681 (1981).


\bibitem{exd}
N. Arkani-Hamed, S. Dimopolous, G. R. Dvali,
Phys. Lett. B{\bf 429} (1998) 263;
I. Antoniadis, N. Arkani-Hamed, S. Dimopolous, G. R. Dvali,
Phys. Lett. B{\bf 436} (1998) 257;
K. R. Dienes, E. Dudas, T. Gherghetta,
Phys. Lett. B{\bf 436}, 55 (1998);
Nucl. Phys. B{\bf 537}, 47 (1999); 
L. Randall and R. Sundrum,
Phys. Rev. Lett. {\bf 83} (1999) 3370. 


\bibitem{Decons}
N. Arkani-Hamed, A. G. Cohen, and H. Georgi,
Phys. Rev. Lett. {\bf 86}, 4757 (2001); \\
C. T. Hill, S. Pokorski, and J. Wang,
Phys. Rev. D{\bf 64}, 105005 (2001).




\bibitem{string}
See:
L. Susskind, {\tt hep-th/0302219} and {\tt hep-th/0405189};  
J. Polchinski, [{\tt hep-th/0209105}], in Heisenberg Centennial
Symposium, Munich, 2001; and references therein.



\bibitem{Hill-topc}
C. T. Hill, Phys. Lett. B{\bf 345}, 483 (1995)
[{\tt hep-ph/9411426}].


\bibitem{xTC}
S. Weinberg, Phys. Rev. D{\bf 13}, 974 (1976);
L. Susskind, Phys. Rev. D{\bf 20}, 2619 (1979).\\
For an early review,
E. Farhi and L. Susskind, Phys. Rept. {\bf 74}, 277 (1981).


\bibitem{xETC}
S. Dimopoulos and L. Susskind, 
Nucl. Phys. B{\bf 155}, 237 (1979);
E. Eichten and K. Lane, 
Phys. Lett. B{\bf 90}, 125 (1980).
For recent reviews, K. Lane, {\tt hep-ph/0202255};
E. H. Simmons, {\tt hep-ph/0211335}.



\bibitem{Ebad1}
C. H.  Llewellyn Simth, Phys. Lett. B{\bf 46}, 233 (1973).


\bibitem{DM}
D. A. Dicus and V. S. Mathur, Phys. Rev. D{\bf 7}, 3111 (1973).


\bibitem{Ebad2}
J. M. Cornwall, D. N. Levin, and G. Tiktopoulos,
%Phys. Rev. Lett.  {\bf 30}, 1268 (1973);
Phys. Rev. D{\bf 10}, 1145 (1974).


\bibitem{LQT}
B. W. Lee, C. Quigg, and H. B. Thacker, 
Phys. Rev. D{\bf 16}, 1519 (1977);
Phys. Rev. Lett. D{\bf 38}, 883 (1977).


\bibitem{Vel}
M. Veltman,  Acta. Phys. Polon. B{\bf 8}, 475 (1977).


\bibitem{CG}
M. S. Chanowitz and M. K. Gaillard, 
Nuc. Phys. B{\bf 261}, 379 (1985).


\bibitem{LW}
M. L\"{u}scher and P. Weisz, 
Phys. Lett. B{\bf 212}, 472 (1988).


\bibitem{MVW}
W. Marciano, G. Valencia, and S. Willenbrock,
Phys. Rev. D{\bf 40}, 1725 (1989).


\bibitem{DJL}
S. Dawson and S. Willenbrock, 
Phys. Rev. Lett. {\bf 62}, 1232 (1989);
L. Durand, J. M. Johnson, and J. L. Lopez,
Phys. Rev. Lett. {\bf 64}, 1215 (1990).



\bibitem{AC}
T. Appelquist and M. S. Chanowitz, 
Phys. Rev. Lett. {\bf 59}, 2405 (1987).


\bibitem{scott}
F. Maltoni, J. M. Niczyporuk, and S. Willenbrock,
Phys. Rev. D {\bf 65}, 033004 (2002)
[{\tt hep-ph/0106281}].



\bibitem{nudy}
T. Appelquist and R. Shrock, 
Phys. Lett. B {\bf 548}, 204 (2002)
[{\tt hep-ph/0204141}];
Phys. Rev. Lett. {\bf 90}, 201801 (2003) [{\tt hep-ph/0301108}].
 
 

\bibitem{nudy1}
P. Sikivie, L. Susskind, M. Voloshin, and V. Zakharov,
Nucl. Phys. B{\bf 173}, 189 (1980);
B. Holdom, Phys. Rev. D{\bf 23}, 1637 (1981);
Phys. Lett. B{\bf 246}, 169 (1990);
T. Appelquist and J. Terning, Phys. Rev. D{\bf 50}, 2116 (1994).  

\bibitem{CCWZ}
C. G. Callen, S. Coleman, J. Wess, and B. Zumino,
Phys. Rev. {\bf 177}, 2247 (1969).


\bibitem{weinberg79}             
S. Weinberg, Physica A{\bf 96}, 327 (1979).


\bibitem{ABL}
T. Appelquist and C. Bernard, Phys. Rev. D{\bf 22}, 200 (1980);
A. C. Langhitano, Nucl. Phys. B{\bf 188}, 118 (1981).

 
\bibitem{GB}
J. Goldstone, Nuovo Cim. {\bf 19}, 154 (1961).   


\bibitem{He-p1}
H.-J. He, Y.-P. Kuang and C.-P. Yuan,
Phys. Rev.  D{\bf 55}, 3038 (1997) [{\tt hep-ph/9611316}];\\
Phys. Lett. B{\bf 382}, 149 (1996) [{\tt hep-ph/9604309}].



\bibitem{He-p2}
For comprehensive reviews,  
H.-J. He, {\tt hep-ph/9804210},
invited review in the Proceedings of 
``{\it Physics at the First Muon Collider}'', 
pp. 685-700, FermiLab, 
Batavia, IL, Nov.\,6-9, 1997, USA;
H.-J. He, Y.-P. Kuang and C.-P. Yuan, 
DESY-97-056 [{\tt hep-ph/9704276}],
invited lectures in the Proceedings of 
the Workshop on ``{\it Physics at the TeV Energy Scale} '',
CCAST (World Laboratory), vol.\,72, pp.\,119-234, 
July 15-26, 1996. 



\bibitem{C5DU}
R.\,S. Chivukula, D.\,A. Dicus,  H.-J. He,
Phys. Lett. B\,{\bf 525}, 175 (2002) 
[{\tt hep-ph/0111016}];
%%CITATION = HEP-PH 0111016;%%
R.\,S. Chivukula and H.-J. He,
Phys. Lett. B\,{\bf 532}, 121 (2002)
[{\tt hep-ph/0201164}];
%%CITATION = HEP-PH 0201164;%%
R.\,S. Chivukula, D.\,A. Dicus, H.-J. He, S. Nandi,
Phys. Lett. B\,{\bf 562}, 109 (2003) 
[{\tt hep-ph/0302263}];
%%CITATION = HEP-PH 0302263;%%
%\bibitem{HeSUSY03}
R.\,S. Chivukula, D.\,A. Dicus, H.-J. He, S. Nandi,
{\tt hep-ph/0402222}, in the proceedings of the international
conference of SUSY-2003, June\,5-10, 2003, Arizona, USA.
%%CITATION = HEP-PH 0402222;%%


\bibitem{HeDPF04}
H.-J. He, 
{\tt hep-ph/0412113}, in the proceedings of DPF-2004:
Annual Meeting of the Division of Particles and Fields, 
American Physical Society, Riverside, CA,
USA, August 26-31, 2004.              
%%CITATION = HEP-PH 0412113;%%




\bibitem{Csaki}
C. Csaki, C. Grojean, H. Murayama, L. Pilo and J. Terning, 
Phys. Rev. D{\bf 69}, 055006 (2004) [{\tt hep-ph/0305237}]. 
%... 
% R.S. Chivukula, E.H. Simmons, H.-J. He, M. Kurachi,
% M. Tanabashi, Phys. Rev. D70 (2004) 075008, 
% hep-ph/0406077, %%CITATION = HEP-PH 0406077;%%
% Phys. Lett. B603 (2004) 210, 
% hep-ph/0408262  %%CITATION = HEP-PH 0408262;%%
% Phys. Rev. D71 (2005) 035007, 
% and hep-ph/0410154; %%CITATION = HEP-PH 0410154;%%
% N. Evans and P. Membry, hep-ph/0406285;
%%CITATION = HEP-PH 0406285;%%
% H. Georgi, hep-ph/0408067;  %%CITATION = HEP-PH 0408067;%%
% and references therein.
    


\bibitem{Eold}
C. E. Vayonakis, Lett. Nuovo. Cimento {\bf 17}, 383 (1976);
G. J. Gounaris, R. K\"{o}gerler, H. Neufeld, 
Phys. Rev. D{\bf 34}, 3257 (1986);
Y.-P. Yao and C.-P. Yuan, Phys. Rev. D{\bf 38}, 2237 (1988);
J. Bagger and C. Schmidt, Phys. Rev. D{\bf 41}, 264 (1990).


\bibitem{HKL}
H.-J. He, Y.-P. Kuang and X. Li, 
Phys. Rev. Lett. {\bf 69}, 2619 (1992);
Phys. Rev. D{\bf 49}, 4842 (1994);
Phys. Lett. B{\bf 329}, 278 (1994) [{\tt hep-ph/9403283}].
%%CITATION = HEP-PH 9403283;%%


\bibitem{HKY94}
H.-J. He, Y.-P. Kuang and C.-P. Yuan,
Phys. Rev. D{\bf 51}, 6463 (1995) [{\tt hep-ph/9410400}].
%%CITATION = HEP-PH 9410400;%%


\bibitem{HK96}
H.-J. He and W.B. Kilgore,
Phys. Rev. D{\bf 55}, 1515 (1997) [{\tt hep-ph/9609326}].
%%CITATION = HEP-PH 9609326;%%


\bibitem{xet}
For related papers, {\it e.g.,}
H. Veltman, Phys. Rev. D{\bf 41}, 2294 (1990);
W. B. Kilgore, Phys. Lett. B{\bf 294}, 257 (1992);
A. Dobado and J. R. Pelaez, Nucl. Phys. B{\bf 425}, 110 (1994),
Erratum B{\bf 434}, 475 (1995);  
J. F. Donoghue and J. Tandean, Phys. Lett. B{\bf 361}, 69 (1995);
T. Torma, Phys. Rev. D{\bf 54}, 2168 (1996);
A. Denner and S. Dittmaier, Phys. Rev. D{\bf 54}, 4499 (1996).


\bibitem{PRL-KK-WZ}
C. Balazs, D.\,A. Dicus, H.-J. He, W.\,W. Repko, C.-P. Yuan, 
Phys. Rev. Lett.  {\bf 83}, 2112 (1999) [{\tt hep-ph/9904220}].
%%CITATION = HEP-PH 9904220;%%


\bibitem{PDG}
S. Eidelman, {\it et al.,}
[Particle Data Group], Phys. Lett. B {\bf 592}, 1 (2004)
%Phys. Rev. D{\bf 66}, 010001 (2002)
[{\tt http://pdg.lbl.gov}].



\bibitem{CGG}
M. S. Chanowitz, H. Georgi, and M. Golden,
Phys. Rev. Lett. {\bf 57}, 2344 (1986);
Phys. Rev. D{\bf 36}, 1490 (1987).


\bibitem{Don}
J. F. Donoghue, C. Ramirez, and G. Valencia,
Phys. Rev. D{\bf 38}, 2195 (1988).


\bibitem{FFVV}
M. S. Chanowitz, M. A. Furman and I. Hinchliffe,
Nucl. Phys. B{\bf 153}, 402 (1979);
Phys. Lett. B{\bf 78}, 285 (1978).


\bibitem{scott-PRL}
F. Maltoni, J. M. Niczyporuk, and S. Willenbrock,
Phys. Rev. Lett. {\bf 86}, 212 (2001)
[{\tt hep-ph/0006358}].
%%CITATION = HEP-PH 0006358;%%


\bibitem{CDF-D0mt}
V. Abazov {\it et al.}, [D0 Collaboration],
Nature {\bf 429}, 638 (2004) and {\tt hep-ex/0407005}; \\
P. A. M. Fernandez [CDF Collaboration],
{\tt hep-ex/0409001}. 


\bibitem{mt-world}
P. Azzi, {\it et al.,}
[Tevatron Electroweak Working Group],
{\tt hep-ex/0404010}; 
%[http://tevewwg.fnal.gov];
L. Cerrito [CDF \& D0 Collaborations],
{\tt hep-ex/0405046},
in the proceedings of XXXIXth Rencontres de Moriond, 
``QCD and High Energy Hadronic Interactions'', March 2004.



\bibitem{HDN}
E.g., H.-J. He, D. A. Dicus and J. N. Ng, 
Phys. Lett. B {\bf 536}, 83 (2002) [{\tt hep-ph/0203237}];
%%CITATION = HEP-PH 0203237;%%
H.\,S. Goh, R.\,N. Mohapatra and S.\,P. Ng,
Phys. Lett. B {\bf 542}, 116 (2002) [{\tt hep-ph/0205131}];
and references therein.


 
\bibitem{0nu2Ba}
For comprehensive reviews of the neutrinoless double $\beta$-decay,
S. R. Elliott and P. Vogel, 
Ann. Rev. Nucl. Part. Sci. {\bf 52}, 115 (2002) 
[{\tt hep-ph/0202264}];
S. R. Elliott and J. Engel,
J. Phys. G {\bf 30}, R183 (2004)
[{\tt hep-ph/0405078}].


\bibitem{0nu2Bb}
H. V. Klapdor-Kleingrothaus, 
Part. Nucl. Lett. {\bf 104}, 20 (2001)
[{\tt hep-ph/0102319}];
H. V. Klapdor-Kleingrothaus, {\it et al.,}
Eur. Phys. J. A{\bf 12}, 147 (2001) [{\tt hep-ph/0103062}];
Mod. Phys. Lett. A{\bf 16}, 2409 (2002) [{\tt hep-ph/0201231}]
and [{\tt hep-ph/0205228}];
C. E. Aalseth, {\it et al.,} 
Mod. Phys. Lett. A{\bf 17}, 1475 (2002)
[{\tt hep-ph/0202018}];
H. V. Klapdor-Kleingrothaus, {\it et al.,}
Phys. Lett. B\,{\bf 586}, 198 (2004)
[{\tt hep-ph/0404088}].



\bibitem{MTY}
V. A. Miransky, M. Tanabashi and K. Yamawaki,
Mod. Phys. Lett. A{\bf 4}, 1043 (1989). 


\bibitem{BHL}
W. A. Bardeen, C. T. Hill, and M. Lindner,
Phys. Rev. D{\bf 41}, 1647 (1990).


\bibitem{TOPC-Rev}
For a review, G. Cvetic, 
Rev. Mod. Phys. {\bf 71}, 513 (1999)
[{\tt hep-ph/9702381}], and references therein.


\bibitem{DEWSBnu}
S. Antusch, J. Kersten and M. Lindner,
Nucl. Phys. B {\bf 658}, 203 (2003)
[{\tt hep-ph/0211385}].


\bibitem{4F-taumu}
D. Black, T. Han, H.-J. He, M. Sher,
Phys. Rev. D {\bf 66}, 053002 (2002) 
[{\tt hep-ph/0206056}]; %%CITATION = HEP-PH 0206056;%%
\\
S. Chen {\it et al.,}
[CLEO Collaboration], 
Phys. Rev. D\,{\bf 66}, 071101 (2002);
[{\tt hep-ex/0208019}]; \\ 
K. K. Gan [CLEO Collaboration],
Nucl. Phys. Proc. Suppl.{\bf 123}, 121 (2003)
[{\tt hep-ex/0211027}];
Y. Enari {\it et al.,} [Belle Collaboration],
Phys. Lett. B (2005) [{\tt hep-ex/0503041}].



\bibitem{nu-susy}
N. Arkani-Hamed, L. Hall, H. Murayama, D. Smith, N. Weiner,
Phys. Rev. D {\bf 64}, 115011 (2001)
[{\tt hep-ph/0006312}].



\bibitem{HeDicus2}
D. A. Dicus and H.-J. He, work in progress.



\bibitem{2Loopnu}
A. Zee,     Nucl. Phys. B{\bf 264}, 99 (1986);
K. S. Babu, Phys. Lett. B{\bf 203},  132 (1988).      


\bibitem{Ma}
E. Ma, Phys. Rev. Lett. {\bf 86}, 2502 (2001);
E. Ma, M. Raidal, U. Sarkar, 
Nucl. Phys. B {\bf 615}, 313 (2001); and references therein.


\bibitem{nu-susy2}
F. Borzumati, K. Hamaguchi, Y. Nomura, T. Yanagida,
{\tt hep-ph/0012118};
J. March-Russell and S. West, {\tt hep-ph/0403067};
and references therein.


\bibitem{DCT}
T. Appelquist and J. Carrazone, Phys. Rev. D{\bf 11}, 2856 (1975).      


\bibitem{susy-f}
For a comprehensive analysis, see, {\it e.g.},
S. Dimopoulos and D. Sutter, Nucl. Phys. B{\bf 452}, 496 (1995)
[{\tt hep-ph/9504415}];
and references therein.



\bibitem{LEPG}
M. Fukugita and T. Yanagida,  Phys. Lett. B{\bf 174}, 45 (1986).


\bibitem{2nuLEPG}
E.g., P. H. Frampton, S. L. Glashow, and T. Yanagida,
Phys. Lett. B {\bf 548}, 119 (2002)
[{\tt hep-ph/0208157}];
V. Barger, D. A. Dicus, H.-J. He, and T. Li,
Phys.  Lett. B {\bf 583}, 173 (2004)
[{\tt hep-ph/0310278}]; and references therein.
%%CITATION = HEP-PH 0310278;%%


\bibitem{2HDM-alpha}
E.g., J. F. Gunion and H. E. Haber, 
Nucl. Phys. B\,{\bf 272}, 1 (1986);
B\,{\bf 278}, 449 (1986).



\bibitem{scr}
M. J. Veltman, Acta. Phys. Polon. B{\bf 12}, 437 (1981).


\bibitem{rLPG}
T. Hambye, J. March-Russel, S. M. West,
JHEP {\bf 07}, 070 (2004) [{\tt hep-ph/0403183}];
S. M. West,  Phys. Rev. D\,{\bf 71}, 013004 (2005) 
[{\tt hep-ph/0408318}].


\bibitem{VLHC}
P. Limon, ``VLHC Accelerator'', talk given at
International Conference on {\it Future Hadron Colliders, 
Physics, Detectors and Machines,}
Fermilab, Batavia, IL, USA,
Oct.\,16-18, 2003. 


\bibitem{DH-PRL}
D. A. Dicus and H.-J. He, 
Phys. Rev. Lett. {\bf 94}, 221802 (2005) 
[{\tt hep-ph/0502178}];
%%CITATION = HEP-PH 0502178;%%
and see also, H.-J. He and D. A. Dicus,
{\tt hep-ph/0411024\,},\,
%%CITATION = HEP-PH 0411024;%%
presentation at the DPF-2004 Meeting of 
American Physical Society,  
Riverside, California, USA, Aug.\,26-31, 2004. 


\bibitem{Ubook}
E.g., G. Barton, {\it Introduction to dispersion techniques
in field theory,}  
Lecture Notes and Supplements in Physics series, 
W. A. Benjamin, Inc., NY, 1965. 


\end{thebibliography}
\end{document}